\documentclass[aps,pra,twocolumn,groupedaddress,showpacs]{revtex4-1}
\usepackage{comment}
\usepackage{lipsum}
\usepackage[version=3]{mhchem}
\usepackage{amsmath,bm}
\usepackage{xcolor}
\usepackage{graphicx}
\usepackage{appendix}
\usepackage[latin1]{inputenc}   
\usepackage[T1]{fontenc}      
\usepackage[tmargin=2 cm, bmargin=2.5 cm, lmargin=1.8 cm, rmargin=1.8 cm ]{geometry}         
\usepackage[english]{babel}  
\usepackage[%
  colorlinks=true,
  urlcolor=blue,
  linkcolor=blue,
  citecolor=blue
]{hyperref}

\begin{document}

\title{Time-dependent restricted-active-space self-consistent-field theory for bosonic many-body systems}
\author{Camille L\'ev\^eque}
\email[]{camille.leveque@phys.au.dk}
\affiliation{Department of Physics and Astronomy, Aarhus University, 8000 Aarhus C, Denmark}
\author{Lars Bojer Madsen}
\affiliation{Department of Physics and Astronomy, Aarhus University, 8000 Aarhus C, Denmark}

\date{\today}

\begin{abstract}
We describe the time-dependent restricted-active-space self-consistent-field (TD-RASSCF) method for a system of interacting bosons. We provide the theory of the method  and discuss its numerical implementation. The method provides a general wavefunction based approach to solve the time-dependent and time-independent  Schr\"odinger equation for a system of bosons. It is based on the time-dependent variational principle to optimize at each instant of time a set of time-dependent coefficients and time-dependent orbitals used to describe the total wavefunction. Including the concept of a restricted-active-space, the exponential growth of the configurational space, resulting from all possible distributions of $N$ bosons in $M$ orbitals, can be controlled trough a specific excitation scheme. We show, by illustrative time-independent and time-dependent examples, that the method provides an accurate description of the system with a substantially smaller configurational space  than the one required in the multi-configurational time-dependent Hartree method for bosons (MCTDHB).  The TD-RASSCF method can also tackle problems beyond the reach of the MCTDHB method when a large number of orbitals are required. 
\end{abstract}


\maketitle

\section{Introduction}

	Since the first realizations of Bose-Einstein condensates (BEC) \cite{Anderson95, Bradley95, Davis95}, the experimental and theoretical investigation of trapped cold atoms has attracted much attention. It is nowadays experimentally possible to design and control systems with a specific number of atoms \cite{Preiss15, Kaufman14} trapped in various potential shapes \cite{Bloch05, Jotzu14} and dimensions \cite{Greiner02}, with tunable inter-particle interactions \cite{Courteille98, Inouye98}, and to provide a controllable transition from a few- to a many-particle system. Such a detailed control of cold atom systems has opened the possibility to simulate various physical systems \cite{Bloch12} from solid-state physics \cite{Anderson98} to black-holes analogs \cite{Steinhauer16} through matter-light interaction \cite{Sala13} and electrons dynamics in molecules \cite{Sengstock15}.
	
	Various theoretical models \cite{Bloch08} have been used so far to describe static and dynamical properties of many-boson systems, among which, a handful are exactly solvable. One of the most prominent models was introduced by Lieb and Liniger \cite{Liniger63, Lieb63}, to describe a system of spinless bosons interacting through a two-body contact interaction: using the Bethe \textit{Ansatz} and periodic boundary conditions the resulting Schr\"odinger equation can be solved exactly for any interaction strength and an arbitrary number of bosons. Unfortunately, this model is exactly solvable only without a trapping potential. In the limit of infinite interaction strength, the Tonks-Girardeau model use the Fermi-Bose mapping to map the wavefunction of bosons into a fermonic wavefunction of non-interacting fermions with frozen parallel spins \cite{Girardeau60}. This mapping provides the exact solution for the ground-state of the system for arbitrary trapping potentials and remains valid also for the excited states, as well as non-equilibrium solutions also for any external potential \cite{Yukalov05}. In the case of non-interacting bosons or more generally in the Gross-Pitaevskii (GP) limit, i.e., $N \rightarrow \infty $ and $N\lambda = constant$, with $N$ the number of bosons and $\lambda$ the interaction strength, the GP equation or its time-dependent (TD-GP) analog provides the exact description of the system. In this situation, the exact wavefunction of the system is described by a single product of single-particle functions and the interactions between the particles are correctly described by the mean-field approach. The above models assume that the bosons interact through a pair-wise contact potential. Considering other types of interaction potentials between the particles, other models can be solved exactly with an external potential. One model uses an inverse-harmonic interaction between the particles and can be solved exactly with a harmonic trapping potential \cite{Calogero69}, while an other model considers a harmonic interaction potential \cite{Cohen85,Yan03}. The latter model has the peculiarity that it can be solved exactly numerically also for time-dependent Hamiltonians with a time-dependent trapping potential or a time-dependent interaction potential \cite{Lode12}.
	
	These exact models, unfortunately, do not cover the large variety of interaction or trapping potentials that are encountered in experiments. Nonetheless, they are of primary interest  as they provide a unique way to benchmark numerical methods and approximations. The GP equation can be simplified when the potential and interaction energies are much larger than the kinetic energy, giving rise to the Thomas-Fermi approximation when the kinetic energy is neglected \cite{Baym96}. On an other hand, to overcome the lack of correlation in the GP theory and to take into account a small depletion of the BEC, i.e., to account for atoms which are not in the condensate, a perturbative expansion of the particle number in the condensate leads to Bogoliubov theory \cite{Bogoliubov47, Lee57, Lee57_2}. In the specific case of periodic trapping potentials, such as optical lattices \cite{Greiner02}, for weak contact interactions and deep lattices the Bose-Hubbard model (BHM) \cite{Fisher89} is obtained by expanding the Bose field operator in term of the Wannier functions of the lowest Bloch band and neglecting the tunneling between nonconsecutive sites and interactions between different sites. The BHM and its various extensions have been extensively and successfully used to describe the ground state of  trapped atoms in optical lattices and their dynamics \cite{Dutta15}. A more general and efficient numerical approach to deal with optical lattices is the density-matrix renormalization group (DMRG) method \cite{White92, White92_2, White93} based on the matrix product states \textit{Ansatz} \cite{Schollwock05}. The method has been used to provide accurate results for ground and exited states of the system, and more recently has been used to investigate time-dependent systems \cite{Vidal04, Daley04, White04}. The second wide-spread and promising numerical method to study trapped atoms is the quantum Monte Carlo (QMC) approach. It includes, among others, the variational Monte Carlo (VMC) \cite{McMillan65} and the diffusion Monte Carlo (DMC) \cite{Anderson75, Reynolds82, Blume2001, Astrakharchik04} methods, which used a Bijl-Jastrow decomposition of the wavefunction \cite{Bijl40, Jastrow55}, but are, however, not applicable to time-dependent systems.
	
	Along with the above theory developments it has been a long standing idea to explore quantum chemistry methodologies to describe a \textit{time-independent} system of trapped cold atoms. This idea was, to the best of our knowledge, introduced by the work of Ersy \cite{Esry97}, applying the mean-field Hartree-Fock (HF) theory and the configuration interaction (CI) method up to double excitations (CISD) to harmonically trapped bosons. The HF method for bosons can be viewed has a variant of the GP theory but has the advantage that it provides a set of optimized virtual orbitals, i.e., non-occupied orbitals, that can be subsequently used in a CI expansion of the wavefunction. The CI expansion corrects the lack of correlation between the particles, not included at the HF level. The CI method is in principle exact but requires a severe truncation of the CI expansion to be numerically tractable. Later, Streltsov et \textit{al} \cite{Streltsov06} introduced the multiconfigurational Hartree theory for bosons (MCHB), which is an extension of the multiconfiguration \textit{self-consistent} field (MCSCF) method introduced for fermions and widely used in electronic-structure calculations in atoms and molecules \cite{Lowdin55}. The MCHB method uses a CI expansion \textit{Ansatz} for the many-body wavefunction in which both the coefficients of the expansion and the orbitals are variationally optimized, providing better accuracy with substantially less configurations and orbitals. The coupled-cluster (CC) method was originally introduced in nuclear physics \cite{Coester58, Coester60} and subsequently extended to describe electronic wavefunctions in atoms and molecules \cite{Cizek66}. This framework was also extended to bosons up to double excitations (CCSD) by Cederbaum et \textit{al}, and successfully applied to various particle numbers and interaction strengths \cite{Cederbaum06}.

	Over the past decade, numerous numerical methods have been developed \cite{Lysaght09,Hochstuhl12, Pabst12,Kvaal12,Sato13,Bauch14} to tackle the problem of \textit{time-dependent} multi-electron dynamics induced by laser pulses that are strong or short or both \cite{Popmintchev10, Calegari14, Kraus15}. In short, the various successful methods used so far to investigate static properties of atoms and molecules have been extended to solve the time-dependent Schr\"odinger equation including a time-dependent operator.  Among these methods,  the multiconfigurational time-dependent Hartree-Fock method \cite{Zanghellini03, Kato04, Nest05,Haxton11} variationally optimizes a set of time-dependent orbitals and CI coefficients, following the idea of the multiconfigurational time-dependent Hartree (MCTDH) method \cite{Meyer90,Beck00}, originally introduced to describe molecular dynamics. The MCTDHF method has been extended to identical bosons, within the framework of the MCTDH for bosons MCTDHB \cite{Alon08}, in which the indistinguishability is taken into account using permanents instead of Slater determinants. Further development includes the case of particle mixtures of different type of bosons and fermions \cite{Alon07, Alon12}. The fundamental concept of using a set of \textit{time-dependent} single-particle functions or orbitals to expand the total wavefunction offers the possibility to use substantially less orbitals than in the case of time-independent orbitals, because the former basis optimally adapts during the evolution of the system. Recently, the framework of the multi-layer (ML) MCTDH method \cite{Wang03,Manthe08,Vendrell11} was extended to systems of bosons and mixtures of them \cite{Kronke13, Cao13}. The method uses a ML expansion to reduce the size of the wavefunction in comparison to the MCTDHB method for multi-species or multi-dimensional systems. It is particularly effective for systems which can be subdivided in strongly interacting subsystems while the individual subsystems interact only weakly with each other. In the case of a one-dimensional system consisting of only a single type of particles, the ML-MCTDHB and MCTDHB wavefunctions are identical \cite{Schmitz13}.

	The MCTDHB method shed new light on the dynamics of trapped cold atoms, especially when fragmentation occurs and more than one orbital is populated - a situation which can not be describe by the TD-GP theory. Fragmentation occurs in different systems such as during the dynamics at a Josephson junction \cite{Sakmann09}, which is a universal phenomenon \cite{Sakmann14}, and can not be described, even qualitatively, using the TD-GP or BH theories. In double-well trapping potentials, fragmentation of the BEC is also obtained for the ground-state \cite{Sakmann08} for large barrier height between the two wells and the GP theory fails to describe the variance of position and momentum operators \cite{Klaiman15}. Multiconfigurational methods are also required to accurately describes the formation and dynamics of fragmented states with repulsive or attractive interactions between the particles \cite{Streltsov08, Streltsov09, Streltsova14} and tunneling of a many boson system to open space \cite{Lode12_2} or tunneling of trapped vortices \cite{Beinke15}. Using time-dependent orbitals reduce the number of orbitals and thus the number of configurations required to describe accurately time-evolving systems in comparison with methods with time-independent orbitals. Nevertheless, simulations using such full-configurational wavefunctions remain a difficult task due to the exponential scaling of the configurational space, i.e., the dimensionality determined by the number of ways to arrange $N$ particles in $M$ orbitals, especially for bosons.
	
	This challenge leads us to the quest for a method which maintains the appealing properties of the time-dependent orbitals based methods mentioned above, but is free from the exponential scaling problem. One such method uses the concept of a restricted active-space (RAS), well-known in quantum chemistry, where it has been applied with time-independent molecular orbitals \cite{Olsen88}. The RAS based method was successfully extended to time-dependent orbitals in the time-dependent restricted active-space self-consistent-field method (TD-RASSCF) to deal with electron dynamics in atoms \cite{Haru13, Haru14_1}. Introducing a RAS scheme by fixing the promotion of the electrons between three sets of orbital spaces can considerably reduced the number of configurations. In addition, the theory has the specificity to include, as limiting cases, the TD Hartree-Fock (HF), the TD complete active space self-consistent field (TD-CASSCF) \cite{Sato13} and the MCTDHF frameworks, as a particular RAS schemes are applied to the MCTDHF wavefunction. Successful applications of the TD-RASSCF method include calculations of the ground-states (GS) of atoms, and time-dependent dynamics in the presence of strong laser fields to describe, for instance, high-order harmonic generation \cite{Haru13, Haru14_1}. The aim of this work is to extend the TD-RASSCF method to systems of spinless interacting bosons. To follow the generic naming introduced for the MCTDH methods, we call this method TD-RASSCF-B where the additional B stands for bosons and we will refer to the original TD-RASCSF method for fermions \cite{Haru13, Haru14_1, Haru14_2} as TD-RASSCF-F to avoid any confusion concerning the particles considered. As a main finding, we derive the working equations of the TD-RASSCF-B method and we present the general set of working equations for the TD-RASSCF method where the type of particles plays a role in the symmetry of the creation and annihilation operators, only. The applications of the method to compute the GS energy of trapped bosons show that the TD-RASSCF-B theory provides accurate results, in comparison to MCTDHB, while the expansion of the wavefunction is considerably reduced. Moreover, the MCTDHB accuracy can be overtaken by using large numbers of time-dependent orbitals while the number of configurations remains small thanks to the RAS schemes. The investigation of the breathing dynamic of a BEC illustrates how the TD-GP theory fails to describe the time-evolution of the system, while various examples of the TD-RASSCF-B method \textit{qualitatively} or \textit{quantitatively} reproduce the exact dynamics obtained using the MCTDHB method, depending of the choice of the excitation scheme.
	
      The paper is organized as follows. In Sec. \ref{wavefunc} we introduce the TD-RASSCF-B \textit{Ansatz} for the wavefunction and in Sec. \ref{EOM_TDRAS} we derive the equations of motion for the set of coefficients and orbitals. In Sec. \ref{time_indpdt} the method is applied and compared to the MCTDHB method to study the static properties of a system consisting of $N=100$ bosons trapped in a harmonic potential. The applicability of the method to time-dependent systems is illustrated by two examples of a breathing dynamics following a sudden quenching of the two-body interaction in Sec. \ref{time_dpdt}. Finally, in Sec. \ref{conclusion} we conclude and provide perspectives to future work. In the Appendices \ref{wf_representation} to \ref{efficiency_TDRAS}, we provide the key ingredients for the numerical implementation of the method and discuss the numerical effort in comparison to the MCTDHB method.
       
\section{Theoretical framework} \label{theory}

\subsection{Ansatz for the many-body wavefunction} \label{wavefunc}

\begin{figure}[h!]
\centering 
\includegraphics[scale=0.5]{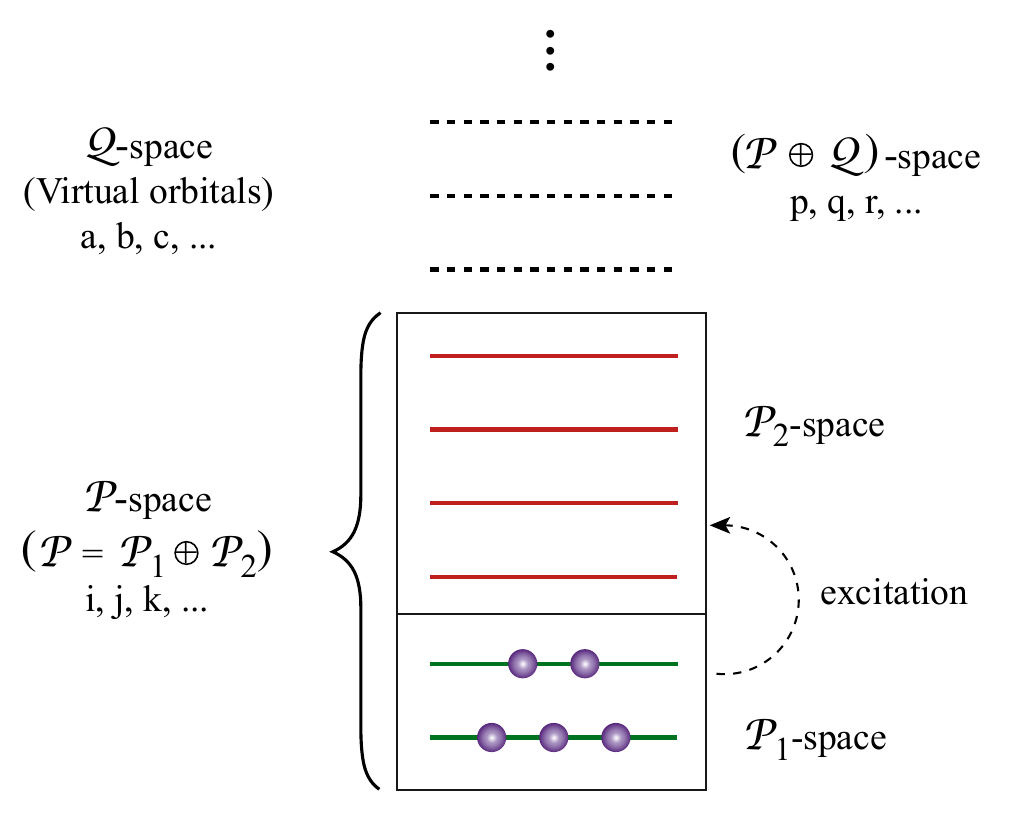}
\caption{ \small{Division of the single-particle Hilbert space within the TD-RASSCF-B framework. The $\cal{P}$-space orbitals, used to expand the wavefunction, are divided in a ${\cal P}_{1}$ and ${\cal P}_{2}$ space between which the particles can be excited through a specific scheme chosen at will. The orthogonal complement of virtual (i.e., unoccupied) orbitals is referred to as the $\cal{Q}$-space. The indexes $i, j, k, \cdots$ are used to label the orbitals of the $\cal{P}$-space, the indexes $a, b, c, \cdots$ to label the orbitals of the $\cal{Q}$-space and the indexes $p, q, r, \cdots$ are used for orbitals in either the $\cal{P}$- or $\cal{Q}$-space.}}
   \label{General_orbtial_space}
 \end{figure}

For the energy regime of interest, the time evolution of a system composed of $N$ bosons is governed by the time-dependent Schr\"odinger equation:
\begin{equation}
i \hbar \frac{\partial}{\partial t}|\Psi(t)\rangle= \hat{H}(t)|\Psi(t)\rangle
\end{equation}
with $\hat{H}(t)$ the many-body Hamiltonian of the system and $|\Psi(t)\rangle$ the $N$-particle wavefunction. Hereafter we set $\hbar=1$, unless  explicitly specified. We can approximate the wavefunction using linear combinations of suitably symmetrized sets of products of time-dependent single-particle functions $\{\phi_{i}(\bf{r},t)\}$. In the following, the single-particle functions are denoted \textit{orbitals}. To take into account the indistinguishability of the bosons, the total wavefunction is expressed in terms of permanents. For a given number of bosons and orbitals the multi-configurational wavefunction is constructed by taking into account all the possible arrangement of the particles in the given orbitals, each arrangements being called a configuration $|\Phi_{I}(t)\rangle$,  
\begin{equation}
|\Psi(t)\rangle=\sum_{I\in {\cal V}_{\text{FCI}}}{ C_{I}(t)|\Phi_{I}(t)}\rangle.
\end{equation}
This \textit{Ansatz} converges to the exact wavefunction when the number of orbitals increases to infinity. The configurational space increases exponentially with respect to the number of orbitals and often makes a numerical treatment impossible, even for a small number of orbitals. In the case of a system of $N$ bosons and $M$ orbitals, the dimension of the full-configurational Fock space ${\cal V}_{\text{FCI}}$ can be evaluated as,
\begin{equation}\label{dim_wf_MCTDHB}
dim({\cal V}_{\text{FCI}})=\begin{pmatrix}
   N+M-1  \\
   N  
\end{pmatrix} 
=    \frac{(N+M-1)!}{N!(M-1)!}. 
\end{equation}
In the case of the TD-RASSCF-B method, we introduce two orbital spaces, ${\cal P}_{1}$ and ${\cal P}_{2}$, such that $M_{1}+M_{2} = M$, with $M_{1}$ and $M_{2}$ the number of orbitals in ${\cal P}_{1}$ and ${\cal P}_{2}$, respectively (see Fig. \ref{General_orbtial_space}). The ${\cal P}_{1}$ subspace must include enough orbitals such that it can accommodate all the particles. For bosons one orbital, i.e.,  $M_{1}=1$ is the lower bound, and there is no restriction concerning the upper bound. In this subspace all the configurations are used to construct the total wavefunction. Concerning the ${\cal P}_{2}$-space, particles are promoted from ${\cal P}_{1}$ to ${\cal P}_{2}$ according to a specific excitation scheme, which is based on the highest  number of particles that can be promoted. This number is chosen at will and the \textit{restriction} of the configurational space provides a way to constrain its size, defining the \textit{Ansatz} for the TD-RASSCF method as,
\begin{equation}\label{RAS_wf}
|\Psi(t)\rangle=\sum_{I\in {\cal V}} C_{I}(t)|\Phi_{I}(t)\rangle,
\end{equation}
where the configurations are drawn from the space $\cal{V}$ subject to restrictions. To evaluate the size of the $\cal{V}$, we introduce $N_{\text{max}}$ the highest number of bosons in ${\cal P}_{2}$ and consider a RAS scheme allowing all occupations of ${\cal P}_{2}$ from $0$ to $N_{\text{max}}$. The dimension of the $\cal{V}$ of Eq. (\ref{RAS_wf}), is then given by
  \begin{equation}\label{config_general_RAS}
  \begin{split} 
dim({\cal V})& =\begin{pmatrix}
   N+M_{1}-1  \\
   N  
\end{pmatrix} + \\
 & \sum_{k=1}^{N_{\text{max}}} 
\begin{pmatrix}
   k+M_{2}-1  \\
   k 
\end{pmatrix}
\begin{pmatrix}
   (N-k)+M_{1}-1  \\
   N-k
\end{pmatrix} .  
\end{split}
\end{equation} 
 The first term is the total number of configurations obtained with the $N$ bosons in the $M_{1}$ orbitals and no particle in ${\cal P}_{2}$. The sum takes into account the configurations resulting from the excitation of $k$ bosons in ${\cal P}_{2}$, with $1\le k\le N_{\text{max}}$. The total number of configurations with $k$ bosons in ${\cal P}_{2}$ is obtained as a product of the possible arrangements of $k$ bosons in $M_{2}$ orbitals and $(N-k)$ bosons in $M_{1}$ orbitals, see also Appendix \ref{wf_representation}. 
 
	The TD-RASSCF-B \textit{Ansatz} holds some interesting specificities. First, if only ${\cal P}_{1}$ orbitals are used, i.e., $M_{1}=M$ and $M_{2}=0$, then the TD-RASSCF-B and MCTDHB \textit{Ans\"atze} are equivalent with the same number of configurations, as seen be replacing $M_{1}$ by $M$ in Eq. (\ref{config_general_RAS}). Note that this is also true for $M_{2}\ne 0$ and $N_{\text{max}}=N$. Moreover, if only a single time-dependent orbital is considered, i.e.,  $M_{1}=1$ and $M_{2}=0$, the RAS wavefunction includes a single configuration with all particles in one orbital, which is equivalent the to time-dependent GP wavefunction. Thus the theoretical framework of TD-RASSCF-B is very general and holds, as limiting cases, the GP and MCTDHB theories. The TD-RASSCF-B wavefunction is built from a set of time-dependent coefficients $\{C_{I}(t)\}$ and orbitals $\{|\phi_{i}(t)\rangle\}$. To describe its dynamics, we need a set of equations of motion (EOM), which provides the time-evolution of the coefficients and orbitals through their time-derivatives $\{\dot{C}_{I}(t)\}$ and $\{|\dot{\phi}_{i}(t)\rangle\}$. The $\cal{P}$-space is a subset of the total single-particle Hilbert space and we can define its orthogonal complement, $\cal{Q}$, collecting the virtual orbitals, as depicted in Fig. \ref{General_orbtial_space}. While in the case of time-independent orbitals these two subspaces remain fixed, in the case of time-dependent orbitals the $\cal{P}$-space is variationally optimized at each time and both $\cal{P}$- and $\cal{Q}$-space are time-dependent. We can define $\hat{P}$ and $\hat{Q}$, the time-dependent projectors onto the subspaces $\cal{P}$ and $\cal{Q}$, respectively, with the property $\hat{P}+\hat{Q}=\hat{1}$, the identity operator. The role of the $\cal{Q}$-space emanates from the time-derivative of the $\cal{P}$-space orbitals, that can be written as,
\begin{equation}\label{general_time_deriv_Porb}
|\dot{\phi}_{i}(t)\rangle=(\hat{P}+\hat{Q})|\dot{\phi}_{i}(t)\rangle = \hat{P}|\dot{\phi}_{i}(t)\rangle+\hat{Q}|\dot{\phi}_{i}(t)\rangle,
\end{equation}
 with one contribution from the $\cal{P}$-space and one contribution from the $\cal{Q}$-space. In the following we establish the EOM of the TD-RASSCF-B theory, providing the time-derivative of the expansion coefficients in Sec. \ref{Sec_EOM_Coeff} and the time-derivative of the orbitals (Sec. \ref{Sec_EOM_Orb}) through the $\cal{Q}$- and $\cal{P}$-space contributions in Secs. \ref{Sec_EOM_OrbQ} and \ref{Sec_EOM_OrbP}, respectively. 
 
 \subsection{Derivation of the working equations} \label{EOM_TDRAS}
 
The EOM for the TD-RASSCF-F theory have been already established in Refs. \cite{Haru13,Haru14_1}. In the following we provide the derivation of the EOM in the case of the TD-RASSCF-B method, and highlight the differences with respect to the TD-RASSCF-F theory. Starting from the Lagrangian formulation of the time-dependent Schr\"odinger equation \cite{Kramer81}, we define the action functional using the TD-RASSCF-B \textit{Ansatz}, Eq. (\ref{RAS_wf}), as,
\begin{equation}\label{action_func}
\begin{split} 
S [\{C_{I}(t)\},&\{ |\phi_{i}(t)\rangle\}, \{\epsilon_{j}^{i}(t)\}] =  \int_{t_{1}}^{t_{2}} \Bigg[ \langle\Psi(t)| \hat{K} |\Psi(t)\rangle \\
 &  \left.+ \sum_{ij}\epsilon_{j}^{i}(t)\bigg( \langle\phi_{i}(t)|\phi_{j}(t)\rangle - \delta_{ij} \bigg)\right] dt,
\end{split}
\end{equation}
with $\hat{K}\equiv i\partial/\partial t -\hat{H}$ and $\delta_{ij}$ the Kronecker delta function. The Lagrange multipliers, $\epsilon_{j}^{i}(t)$, ensure that the orbitals remain orthonormal for all time $t$. In the following the indexes $i, j, k, \cdots$ are used to denote the orbitals of the $\cal{P}$-space, the indexes $a, b, c, \cdots$ denote the orbitals of the $\cal{Q}$-space and $p, q, r, \cdots$ are used for either $\cal{P}$- or $\cal{Q}$-space orbitals, see also Fig. \ref{General_orbtial_space}. For our purpose, we consider only one- and two-body operators, such that the Hamiltonian can be expressed in the framework of second quantization as 
\begin{equation}\label{Hamiltonian}
\hat{H}(t)=\sum_{pq}h_{q}^{p}(t)\hat{b}_{p}^{\dag}\hat{b}_{q}+\frac{1}{2}\sum_{pqrs}{v_{qs}^{pr}(t) \hat{b}_{p}^{\dag}\hat{b}_{r}^{\dag}\hat{b}_{s}\hat{b}_{q}},
\end{equation}
with $\hat{b}_{p}$  ($\hat{b}_{p}^{\dag}$) the annihilation (creation) operator of a particle in the orbital $|\phi_{p}(t)\rangle$ [see also Appendix \ref{apply_op}]. These operators satisfy the commutation relation, $[ \hat{b}_{p},\hat{b}_{q}^{\dag} ] = \hat{b}_{p}\hat{b}_{q}^{\dag}- \hat{b}_{q}^{\dag}\hat{b}_{p} = \delta_{qp}$, for bosons and the anti-commutation relation, $\{\hat{b}_{p},\hat{b}_{q}^{\dag} \}= \hat{b}_{p}\hat{b}_{q}^{\dag}+\hat{b}_{q}^{\dag}\hat{b}_{p} = \delta_{qp}$, for fermions, see for instance Ref. \cite{Cederbaum16}. The matrix elements of the one-body and two-body operators in the basis of the time-dependent orbitals, are expressed as
\begin{equation} \label{one-bod}
h_{q}^{p}(t)=\int \phi_{p}^{*}(\textbf{r},t) h(\textbf{r},t) \phi_{q}(\textbf{r},t) d\textbf{r},
\end{equation}
and 
\begin{equation} \label{two-bod}
\begin{split}
v_{qs}^{pr} (t) =\int \int &\phi_{p}^{*}({\bf r},t) \phi_{r}^{*}({\bf r'},t)  \\
&\times W({\bf r},{\bf r'},t) \phi_{q}({\bf r},t)\phi_{s}({\bf r'},t) d{\bf r}d{\bf r'},
\end{split}
\end{equation}
respectively. In the following, the explicit time dependence of the operators, coefficients and orbitals is dropped for brevity.

According to the time-dependent variational principle \cite{Kramer81,Broeckhove88}, the best approximation using the wavefunction \textit{Ansatz} is obtained by seeking stationarity of the action $S$, i.e., $\delta S = 0$, for any variation of the parameters and with the boundary condition $|\delta\Psi(t_{1})\rangle = |\delta\Psi(t_{2})\rangle=0$. The variation of the action gives, 
\begin{equation}\label{var_action_func}
\begin{split}
\delta &S =  \int_{t_{1}}^{t_{2}} \Bigg[ \langle\delta\Psi| \hat{K} \Psi\rangle + \langle \hat{K}\Psi| \delta\Psi\rangle \\
&+ \sum_{ij}\left[\epsilon_{j}^{i}\left( \langle\delta\phi_{i}|\phi_{j}\rangle + \langle\phi_{i}|\delta\phi_{j}\rangle\right) + \delta\epsilon_{j}^{i} (\langle\phi_{i}|\phi_{j}\rangle-\delta_{ij}) \right]  \Bigg] dt,
\end{split}
\end{equation}
where the boundary condition is used to remove the additional term $i\partial_{t} \langle\Psi|\delta\Psi\rangle$ resulting from the action of $\hat{K}$ on $\langle\Psi|$ instead of $|\delta\Psi\rangle$, see Ref. \cite{Broeckhove88}. The variation of the wavefuntion is explicitly written as \cite{Haru13},
 \begin{equation}\label{var_wavefunc}
|\delta\Psi\rangle  =  \sum_{I\in {\cal V}} \delta C_{I}|\Phi_{I}\rangle + \sum_{pq}  \hat{b}_{p}^{\dag}\hat{b}_{q}|\Psi\rangle\langle\phi_{p}|\delta\phi_{q}\rangle.
\end{equation}
We can now proceed with the stationarity condition of the action with respect to the parameters $\{C_{I}\}$, $\{|\phi_{i}\rangle\}$ and $\{\epsilon_{j}^{i}\}$ to obtain the EOM of the TD-RASSCF-B method.  
 
\subsubsection{Equations of motion for the coefficients}  \label{Sec_EOM_Coeff}

The variation w.r.t. the Lagrange multipliers leads to the conservation of the orthonormality of the orbitals. We then consider the variation of the action functional with respect to the expansion coefficients. The action $S$ depends on the expansion coefficient $C_{I}^{*}$ only through the \textit{bra} $\langle \delta \Psi|$ of the first expectation value in Eq. (\ref{var_action_func}). Thus, the stationarity condition, $\delta S/\delta C_{I}^{*}=0$, readily leads to $\langle\Phi_{I}|i\partial_{t}-\hat{H}|\Psi\rangle = 0$. Moreover, the derivative of the wavefunction with respect to time reads,
\begin{equation}\label{TD_WF_RAS}
\frac{\partial}{\partial t}|\Psi\rangle=\sum_{I\in {\cal V}} \dot{C}_{I}|\Phi_{I}\rangle + \left[ \sum_{pq}\eta_{q}^{p} \hat{b}_{p}^{\dag}\hat{b}_{q} \right]|\Psi\rangle,
\end{equation}
with  $\eta_{q}^{p}\equiv\langle \phi_{p}|\dot{\phi}_{q}\rangle$, which results from the time-derivative of the orbitals used to build the configurations in $|\Psi\rangle$. Hereafter, the operator in bracket in Eq. (\ref{TD_WF_RAS}), $\sum_{pq}\eta_{q}^{p} \hat{b}_{p}^{\dag}\hat{b}_{q}$, will be called $\hat{D}$, for brevity. We can now rewrite the stationary condition $\delta S/\delta C_{I}^{*}=0$ using the explicit form of $\partial_{t}|\Psi\rangle$, Eq. (\ref{TD_WF_RAS}), as
\begin{equation}
 i\dot{C}_{I}+\langle\Phi_{I}|(i\hat{D}-\hat{H})|\Psi\rangle=0, \forall I \in {\cal V} \label{EOM_C_general},
\end{equation}
or equivalently, using the expressions of $\hat{H}$ and $\hat{D}$,
\begin{equation}\label{EOM_C_general_expand}
\begin{split}
i\dot{C}_{I}=\sum_{ij}&\left(h_{j}^{i}-i\eta_{j}^{i}\right)\langle\Phi_{I}|\hat{b}_{i}^{\dag}\hat{b}_{j}|\Psi\rangle\\
 &+\frac{1}{2}\sum_{ijkl}v_{jl}^{ik}\langle\Phi_{I}|\hat{b}_{i}^{\dag}\hat{b}_{k}^{\dag}\hat{b}_{l}\hat{b}_{j}|\Psi\rangle.
\end{split}
\end{equation}  
The indexes in the summations are now restricted to the $\cal{P}$-space. It is clear that if either annihilation or creation operators  act on an orbital of $\cal{Q}$, the inner product with all RAS configurations $\langle\Phi_{I}|$ vanishes. The EOM for the expansion coefficients, the amplitude equations (\ref{EOM_C_general_expand}),  are identical to those obtained for fermions in the TD-RASSCF-F theory, see Refs. \cite{Haru13, Haru14_1}, and those of the MCTDHB \cite{Alon08} and MCTDHF \cite{Caillat05} theories. It is worthwhile to keep in mind that the action of the creation and annihilation operators differs for fermions and bosons. The $\eta_{j}^{i}$ matrix elements in the amplitude equations [Eq. (\ref{EOM_C_general_expand})] describe the rotation of the orbitals into one another and are also present in the EOM of the MCTDHB/F methods. In these latter cases, besides to be elements of an anti-Hermitian matrix, there are no constraints on the $\eta_{j}^{i}$ and their values are usually set to zero. The same is true in the TD-RASSCF-B/F methods for equivalent orbitals, i.e., for pairs of orbitals which belong to the same ${\cal P}_{i}$-space (i=1, 2). For orbitals which do not belong to the same ${\cal P}_{i}$-space, the $\eta_{j}^{i}$ matrix elements must be evaluated, as discussed in Refs. \cite{Haru13, Haru14_1} and in Sec. \ref{even_exci} and \ref{all_exci}. 

\subsubsection{Equations of motion for the orbitals} \label{Sec_EOM_Orb}
 
Seeking stationarity of $S$ with respect to a variation of an orbital $\langle\phi_{i}|$, i.e., $\delta S/ \delta\langle\phi_{i}|=0$, gives 
 \begin{equation}\label{EOM_orb}
 \begin{split}
 \sum_{q} |\phi_{q}\rangle \langle\Psi_{i}^{q}| &\left[\sum_{I\in {\cal V}} i\dot{C}_{I}Ê|\Phi_{I}\rangle + ( i\hat{D}-\hat{H})|\Psi\rangle \right] \\
 &+ \sum_{j}\epsilon_{j}^{i}|\phi_{j}\rangle= 0,
 \end{split}
\end{equation}
with $\langle\Psi_{i}^{q}| \equiv \langle\Psi|\hat{b}_{i}^{\dag}\hat{b}_{q}$. The index $q$ in the above equation runs over all the orbitals, i.e., the orbitals of the ${\cal P}$-space and the ${\cal Q}$-space, see Fig. \ref{General_orbtial_space}. The EOM for the orbitals of the ${\cal P}$- and ${\cal Q}$-space are obtained by projecting Eq. (\ref{EOM_orb}) on either an orbital of the ${\cal P}$-space, $\langle\phi_{j}|$, or of the ${\cal Q}$-space, $\langle\phi_{a}|$, as done in the following. \\

\paragraph{Equations of motion for the ${\cal Q}$-space orbitals} \hspace{0pt} \label{Sec_EOM_OrbQ} \\

Starting with the EOM for the ${\cal Q}$-space orbitals, we multiply Eq. (\ref{EOM_orb}) from the left  with an orbital $\langle\phi_{a}|$ belonging to the ${\cal Q}$-space and obtain, 
 \begin{equation}\label{EOM_orb_Q}
 \sum_{I\in {\cal V}} i\dot{C}_{I} \langle\Psi_{i}^{a}Ê|\Phi_{I}\rangle +  \langle\Psi_{i}^{a}Ê| (i\hat{D}-\hat{H})|\Psi\rangle= 0,
\end{equation}
where we used the orthogonality between the orbitals of the ${\cal P}$ and ${\cal Q}$ spaces to get rid of the Lagrange multipliers. Moreover the inner product $\langle\Psi_{i}^{a}Ê|\Phi_{I}\rangle = \langle\Psi|\hat{b}_{i}^{\dag}\hat{b}_{a}|\Phi_{I}\rangle$ vanishes because in all configurations $|\Phi_{I}\rangle$ the orbital $|\phi_{a}\rangle$ is unoccupied. Using the explicit expression of the Hamiltonian, Eq. (\ref{Hamiltonian}), and for the operator $\hat{D}$, Eq. (\ref{TD_WF_RAS}), we obtain
\begin{equation}\label{EOM_with_H_Q}
\begin{split}
\sum_{pq}\left( i\eta_{q}^{p}-h_{q}^{p} \right) &\langle\Psi|\hat{b}_{i}^{\dag}\hat{b}_{a}\hat{b}_{p}^{\dag}\hat{b}_{q}|\Psi\rangle \\
&=\frac{1}{2}\sum_{pqrs}v_{qs}^{pr}\langle\Psi|\hat{b}_{i}^{\dag}\hat{b}_{a}\hat{b}_{p}^{\dag}\hat{b}_{r}^{\dag}\hat{b}_{s}\hat{b}_{q}|\Psi\rangle.
\end{split}
\end{equation}  
Using the commutation relation for the creation/annihilation operators for bosons (fermions), we can reestablish the normal ordering of the chains of operators, 
\begin{align}
\hat{b}_{i}^{\dag}\hat{b}_{a}\hat{b}_{p}^{\dag}\hat{b}_{q}&=\hat{b}_{i}^{\dag}\hat{b}_{q}\delta_{pa} \pm \hat{b}_{i}^{\dag}\hat{b}_{p}^{\dag}\hat{b}_{q}\hat{b}_{a} \label{four_op} \\ 
\hat{b}_{i}^{\dag}\hat{b}_{a}\hat{b}_{p}^{\dag}\hat{b}_{r}^{\dag}\hat{b}_{s}\hat{b}_{q}&= \hat{b}_{i}^{\dag}\hat{b}_{r}^{\dag}\hat{b}_{s}\hat{b}_{q}\delta_{pa} \pm \hat{b}_{i}^{\dag}\hat{b}_{p}^{\dag}\hat{b}_{s}\hat{b}_{q}\delta_{ra}+\hat{b}_{i}^{\dag}\hat{b}_{p}^{\dag}\hat{b}_{r}^{\dag}\hat{b}_{s}\hat{b}_{q}\hat{b}_{a}, \label{six_op} 
\end{align} 
with the upper sign holding for bosons and the lower for fermions. The chain of four operators in Eq. (\ref{four_op}) and six operators in Eq. (\ref{six_op}) both annihilate a particle in orbital $|\phi_{a}\rangle$, from the ${\cal Q}$-space, which is not include in $|\Psi\rangle$ and thus vanish. The l.h.s. of Eq. (\ref{EOM_with_H_Q}) now reads,
\begin{equation}
\sum_{q}\left( i\eta_{q}^{a}-h_{q}^{a} \right) \langle\Psi|\hat{b}_{i}^{\dag}\hat{b}_{q}|\Psi\rangle = \sum_{j}\left( i\eta_{j}^{a}-h_{j}^{a} \right) \langle\Psi|\hat{b}_{i}^{\dag}\hat{b}_{j}|\Psi\rangle, 
\end{equation}
where we restrict the summation over $j\in {\cal P}$, the summation over the ${\cal Q}$-space orbitals being zero. In the same way, inserting Eq. (\ref{six_op}) in the r.h.s. of Eq. (\ref{EOM_with_H_Q}) simplifies its expression to
\begin{equation}
 \sum_{j}\left( i\eta_{j}^{a}-h_{j}^{a} \right) \langle\Psi|\hat{b}_{i}^{\dag}\hat{b}_{j}|\Psi\rangle =\sum_{jkl}v^{ak}_{jl}\langle\Psi|\hat{b}_{i}^{\dag}\hat{b}_{k}^{\dag}\hat{b}_{l}\hat{b}_{j}|\Psi\rangle.
\end{equation}
Here we used that $v^{ak}_{jl}=v^{ka}_{lj}$ [Eq. (\ref{two-bod})]. Interestingly, this equation is exactly the same for bosons and fermions. The time-derivative of the orbitals, included in the term $\eta_{j}^{a}$, requires the explicit consideration of the ${\cal Q}$-space orbitals. This issue is circumvented by using the projector $\hat{Q}$ onto the subspace spanned by the ${\cal Q}$-space orbitals, 
\begin{eqnarray}
\hat{Q}&=&\sum_{a}|\phi_{a}\rangle\langle\phi_{a}| \nonumber \\
   &=& \hat{1}-\sum_{i}|\phi_{i}\rangle\langle\phi_{i}|  \nonumber \\
   &=&  \hat{1}-\hat{P},
\end{eqnarray}
with $\hat{P}$ the projector onto the ${\cal P}$-space. Introducing the one-body density matrix, $\rho_{i}^{j}= \langle\Psi|\hat{b}_{i}^{\dag}\hat{b}_{j}|\Psi\rangle$ and the two-body density matrix $\rho_{ik}^{jl}=\langle\Psi|\hat{b}_{i}^{\dag}\hat{b}_{k}^{\dag}\hat{b}_{l}\hat{b}_{j}|\Psi\rangle$, we obtain,
\begin{equation} \label{EOM_Qspace}
 i\sum_{j} \hat{Q}|\dot{\phi}_{j}\rangle \rho_{i}^{j} = \hat{Q}\left[\sum_{j} \hat{h}|\phi_{j}\rangle \rho_{i}^{j} + \sum_{jkl} \hat{W}_{l}^{k}|\phi_{j}\rangle \rho_{ik}^{jl}\right],
\end{equation}
with,
\begin{equation}
 \hat{W}_{l}^{k}({\bf r})=\int{\phi_{k}^{*}({\bf r'})W({\bf r},{\bf r'})\phi_{l}({\bf r'})d{\bf r'}},
\end{equation}
the mean-field operator, which describes the interaction between the particles. The role of the $\cal{Q}$-space appears in the time-derivative of the orbitals of the $\cal{P}$-space through the term $\hat{Q}|\dot{\phi}_{i}\rangle$, see Eq. (\ref{general_time_deriv_Porb}). We rearrange Eq. (\ref{EOM_Qspace}), see appendix \ref{Ap_Num_imple}, to uncouple the contribution of each $\hat{Q}|\dot{\phi}_{i}\rangle, \forall i\in\cal{P}$, and obtain
\begin{equation}\label{final_Qspace}
\hat{Q}|\dot{\phi}_{i}\rangle = -i(\hat{1}-\hat{P})\left[ \hat{h}|\phi_{i}\rangle + \sum_{jklm} (\underline{\underline{\bm{\rho}}}^{-1})_{i}^{m} \rho_{mk}^{jl}\hat{W}_{l}^{k}|\phi_{j}\rangle\right],
\end{equation}             
with $\underline{\underline{\bm{\rho}}}^{-1}$ the inverse of the one-body density matrix. The MCTHB theory leads also to Eq. (\ref{final_Qspace}), see Ref. \cite{Alon08}, but the l.h.s. is subsequently simplified thanks to the choice of the matrix elements $\eta_{j}^{i}=0$ and using $\hat{Q}=\hat{1}-\hat{P}$, see Eq. (\ref{time_deriv_Porb}) below. As discussed in \ref{Sec_EOM_Coeff}, such a fixed choice of $\eta_{j}^{i}$ is not possible in the TD-RASSCF theory. The derivation of $\cal{Q}$-space EOM differ slightly for bosons and fermions, see Eqs. (\ref{four_op}) and (\ref{six_op}), but the final result, Eq. (\ref{final_Qspace}), is the same for both types of particles.  \\

\paragraph{Equations of motion for the ${\cal P}$-space orbitals} \hspace{0pt} \label{Sec_EOM_OrbP} \\

Going back to the stationary condition for the variation of the action functional with respect to an orbital, Eq. (\ref{EOM_orb}), we multiply this latter on the left by an orbital of the ${\cal P}$-space, $\langle\phi_{j}|$, leading to,
\begin{equation}\label{EOM_orb_P_1}
\sum_{I\in {\cal V}} i\dot{C}_{I} \langle\Psi_{i}^{j}Ê|\Phi_{I}\rangle +  \langle\Psi_{i}^{j}Ê| ( i\hat{D}-\hat{H})|\Psi\rangle + \epsilon_{j}^{i} = 0.
\end{equation}
This equation still contains the Lagrange multiplier $ \epsilon_{j}^{i}$. A variation of $S$ with respect to the orbital $|\phi_{j}\rangle$ and its projection onto the orbital $\langle\phi_{i}|$, leads to an equation containing the same Lagrange multiplier, 
\begin{equation}\label{EOM_orb_P_2}
\sum_{I\in  {\cal V}} -i\dot{C}_{I}^{*} \langle\Phi_{I}|\Psi_{j}^{i}Ê\rangle +  \langle\Psi | ( i\hat{D}-\hat{H})|\Psi_{j}^{i}Ê\rangle + \epsilon_{j}^{i} = 0,
\end{equation}
and subtracting Eq. (\ref{EOM_orb_P_1}) and Eq. (\ref{EOM_orb_P_2}) gives the EOM for the $\cal{P}$-space orbitals, i.e., 
 \begin{equation} \label{P_space_general}
 \langle\Psi | ( i\hat{D}-\hat{H})|\Psi_{j}^{i}Ê\rangle -  \langle\Psi_{i}^{j}Ê| ( i\hat{D}-\hat{H})|\Psi\rangle = i\dot{\rho}_{i}^{j},
\end{equation}      
where we have introduced $\dot{\rho}_{i}^{j}\equiv\sum_{I\in  {\cal V}}(\dot{C}^{*}_{I}\langle\Phi_{I}|\Psi_{j}^{i}\rangle+\langle\Psi_{i}^{j}|\Phi_{I}\rangle\dot{C}_{I})$. The $\cal{P}$-space EOM provide the contribution of the $\cal{P}$-space in the time-derivative of the orbitals, see Eq. (\ref{general_time_deriv_Porb}), 
\begin{equation}\label{time_deriv_Porb}
\hat{P}|\dot{\phi}_{i}\rangle = \sum_{j}|\phi_{j}\rangle\eta_{i}^{j},
\end{equation}
through the evaluation of the matrix elements $\eta_{i}^{j}$ included in the operator $\hat{D}$. Nonetheless, solving Eq. (\ref{P_space_general}) is not a trivial task because of the presence of $\dot{\rho}_{i}^{j}$, which couples the amplitude and ${\cal P}$-space orbitals equations. In the case of the wavefunction based on the RAS \textit{Ansatz}, a freedom in the choice of the elements $\eta_{i}^{j}$ is still possible for pairs of orbitals which belong to the same ${\cal P}_{i}$-space ($i=1,2$), and we use $\eta_{i}^{j}=0, \forall \{i,j\}\in{\cal P}_{1}\text{ or }{\cal P}_{2}$. The $\cal{P}$-space equation [Eq. (\ref{P_space_general})] remains to be solved only for pairs of orbitals $\{i',j''\}$, which belong to different ${\cal P}_{i}$-spaces,
 \begin{equation}\label{P_space_pair}
 \langle\Psi | ( i\hat{D}-\hat{H})|\Psi_{j''}^{i'}Ê\rangle -  \langle\Psi_{i'}^{j''}Ê| ( i\hat{D}-\hat{H})|\Psi\rangle = i\dot{\rho}_{i'}^{j''},
\end{equation}      
but remains coupled to the amplitude equations through $\dot{\rho}_{i'}^{j''}$. In the meantime it is noted that \textit{if} Eq. (\ref{P_space_pair}) is solved for $\eta_{i}^{j}$, the r.h.s. of Eq. (\ref{time_deriv_Porb}) can be constructed. Moreover Eq. (\ref{final_Qspace}) can be solved, and hence $|\dot{\phi}_{i}(t)\rangle$ of Eq. (\ref{general_time_deriv_Porb}) can be evaluated. In the derivation of the TD-RASSCF-F method, a way to circumvent the difficulty of solving Eq. (\ref{P_space_general}) was proposed \cite{Haru13,Haru14_1}. This approach will be used in the following also for bosons. \\

\paragraph{Even excitation RAS scheme} \hspace{0pt} \label{even_exci} \\
 
\begin{figure}[h!]
\centering 
\includegraphics[scale=0.32]{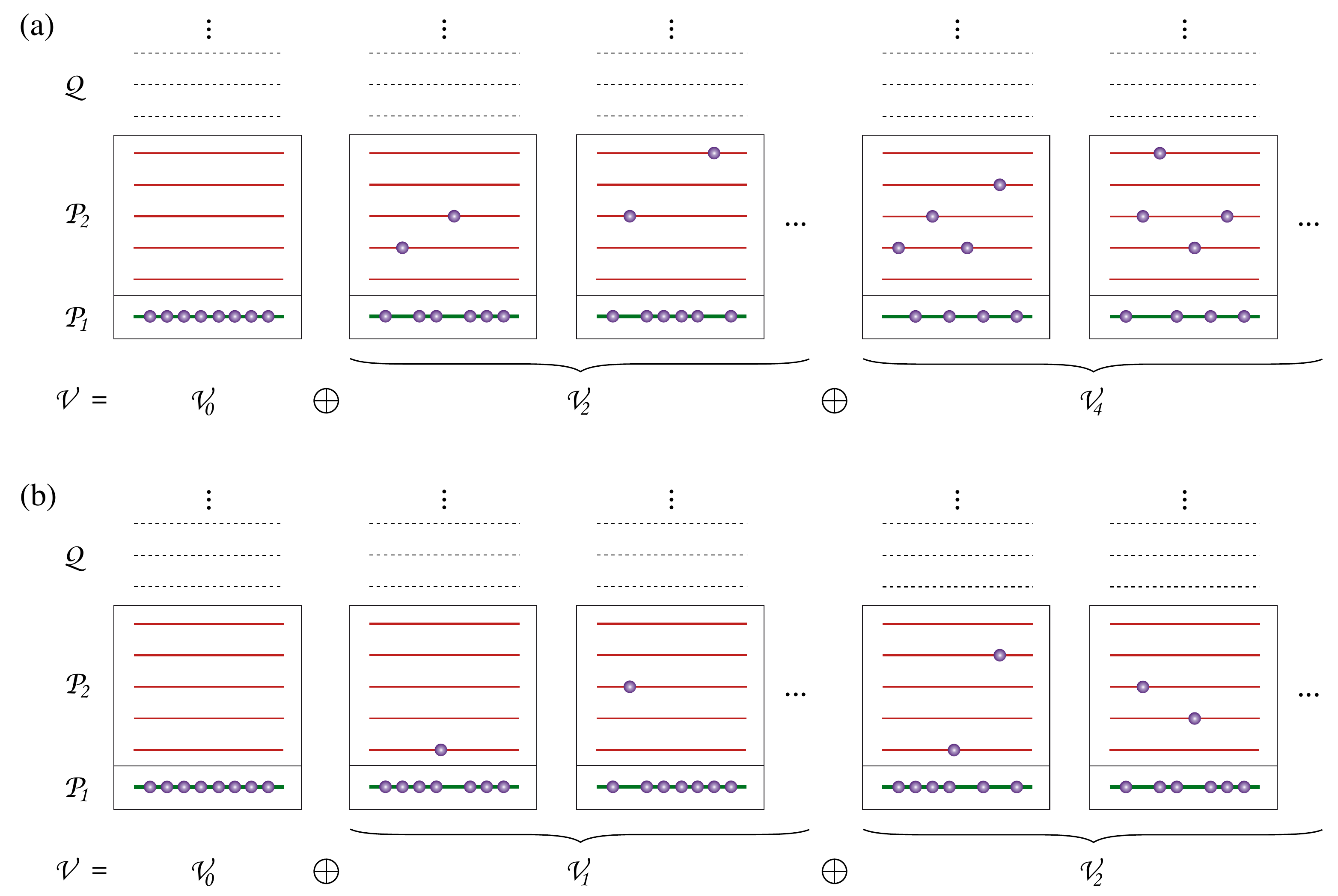}
   \caption{ \small{Illustration of the decomposition of the Fock-space used in the TD-RASSCF-B method with $N=8$ bosons, a single ${\cal P}_{1}$ orbital, $M_{1}=1$ (thick (green) line) and $M_{2}=5$ ${\cal P}_{2}$ orbitals (thin (red) lines). (a) For a RAS scheme including only even excitations, the RAS Fock-space $\cal{V}$ is decomposed into the direct sum of subspaces ${\cal V}_{n}$ with $n=0,2,4, \cdots,N_{\text{max}}$. Each subspace ${\cal V}_{n}$ includes \textit{all} the configurations with $n$ bosons in ${\cal P}_{2}$ and $N-n$ bosons in ${\cal P}_{1}$. (b) The decomposition of the RAS Fock-space for the general excitation scheme is also written as a direct sum of subspaces ${\cal V}_{n}$ except that in that case $n=0,1,2, \cdots,N_{\text{max}}$.}}   \label{Ras_Schemes}
 \end{figure}

First we suggest to consider the case in which only an even number of particles is promoted from ${\cal P}_{1}$ to ${\cal P}_{2}$, see Fig. \ref{Ras_Schemes}a. In this case, $\dot{\rho}_{i'}^{j''}$ explicitly reads,
\begin{equation}
\dot{\rho}_{i'}^{j''}=\sum_{I\in {\cal V}}\left(\dot{C}^{*}_{I}\langle\Phi_{I}|\hat{b}_{i'}^{\dag}\hat{b}_{j''}|\Psi\rangle+\langle\Psi|\hat{b}_{i'}^{\dag}\hat{b}_{j''}|\Phi_{I}\rangle\dot{C}_{I}\right).
\end{equation} 
The action of $\hat{b}_{i'}^{\dag}\hat{b}_{j''}$ on the wavefunction $|\Psi\rangle$ annihilates one particle in ${\cal P}_{2}$ and creates one in ${\cal P}_{1}$. Since only an even number of particles is present in ${\cal P}_{2}$, $\hat{b}_{i'}^{\dag}\hat{b}_{j''}|\Psi\rangle$ would contain only configurations with an odd number of particles in ${\cal P}_{2}$, which makes the inner product with $\langle\Phi_{I}|, \forall I \in \text{RAS}$ vanish. In the same manner $\hat{b}_{i'}^{\dag}\hat{b}_{j''}$ acting on $|\Phi_{I}\rangle$ is either zero, if $|\phi_{j''}\rangle$ is unoccupied in the configuration $|\Phi_{I}\rangle$, or gives an odd number of particles in ${\cal P}_{2}$. In this specific excitation scheme, $\dot{\rho}_{i'}^{j''}=0$, for all pairs of orbitals $\{i',j''\}$, leaving the amplitudes and the ${\cal P}$-space orbitals equations uncoupled. Using the explicit expressions of the Hamiltonian [Eq. (\ref{Hamiltonian})] and the operator $\hat{D}$, [Eq. (\ref{TD_WF_RAS})], Eq. (\ref{P_space_pair}) reads,
\begin{equation}\label{eta_eq_even}
\begin{split}
\sum_{pq}&\left(h_{q}^{p}-i\eta_{q}^{p}\right)\left[\langle\Psi|\hat{b}_{i'}^{\dag}\hat{b}_{j''}\hat{b}_{p}^{\dag}\hat{b}_{q}-\hat{b}_{p}^{\dag}\hat{b}_{q}\hat{b}_{i'}^{\dag}\hat{b}_{j''}|\Psi\rangle\right]+\\
&\frac{1}{2}\sum_{pqrs}v_{qs}^{pr}\left[\langle\Psi|\hat{b}_{i'}^{\dag}\hat{b}_{j''}\hat{b}_{p}^{\dag}\hat{b}_{r}^{\dag}\hat{b}_{s}\hat{b}_{q}-\hat{b}_{p}^{\dag}\hat{b}_{r}^{\dag}\hat{b}_{s}\hat{b}_{q}\hat{b}_{i'}^{\dag}\hat{b}_{j''}|\Psi\rangle\right]=0.
\end{split}
\end{equation} 
We can simplify this expression, starting with 
\begin{align}
\langle\Psi|&\hat{b}_{i'}^{\dag}\hat{b}_{j''}\hat{b}_{p}^{\dag}\hat{b}_{q}-\hat{b}_{p}^{\dag}\hat{b}_{q}\hat{b}_{i'}^{\dag}\hat{b}_{j''}|\Psi\rangle\nonumber \\ 
&= \langle\Psi|\hat{b}_{i'}^{\dag}\hat{b}_{q} \delta_{pj''} \pm \hat{b}_{i'}^{\dag}\hat{b}_{p}^{\dag}\hat{b}_{j''}\hat{b}_{q}-\hat{b}_{p}^{\dag}\hat{b}_{j''}\delta_{i'q}\mp \hat{b}_{i'}^{\dag}\hat{b}_{p}^{\dag}\hat{b}_{j''}\hat{b}_{q}|\Psi\rangle \nonumber \\
&=\langle\Psi|\hat{b}_{i'}^{\dag}\hat{b}_{q} \delta_{pj''}  -\hat{b}_{p}^{\dag}\hat{b}_{j''}\delta_{i'q}|\Psi\rangle \nonumber \\  
&=\rho_{i}^{q}\delta_{pj''}  - \rho_{p}^{j''}\delta_{i'q}\nonumber \\
&\equiv A_{pi'}^{qj''}.       
\end{align}
Now we turn to the chains of six operators in the last term in Eq. (\ref{eta_eq_even}). The first product of operators is expressed as
\begin{equation}
\begin{split}
\hat{b}_{i'}^{\dag}\hat{b}_{j''}\hat{b}_{p}^{\dag}\hat{b}_{r}^{\dag}\hat{b}_{s}\hat{b}_{q} =\ &\hat{b}_{i'}^{\dag}\hat{b}_{r}^{\dag}\hat{b}_{s}\hat{b}_{q}\delta_{pj''} \\
&\pm \hat{b}_{i'}^{\dag}\hat{b}_{p}^{\dag}\hat{b}_{s}\hat{b}_{q}\delta_{rj''}+\hat{b}_{i'}^{\dag}\hat{b}_{p}^{\dag}\hat{b}_{r}^{\dag}\hat{b}_{j''}\hat{b}_{s}\hat{b}_{q} \label{simply_1}
\end{split}
\end{equation}
and the second product of operators as
\begin{equation}
\begin{split}
 -\hat{b}_{p}^{\dag}\hat{b}_{r}^{\dag}\hat{b}_{s}\hat{b}_{q}\hat{b}_{i'}^{\dag}\hat{b}_{j''}= &-\hat{b}_{p}^{\dag}\hat{b}_{r}^{\dag}\hat{b}_{s}\hat{b}_{j''}\delta_{i'q} \\ 
 &\mp\hat{b}_{p}^{\dag}\hat{b}_{r}^{\dag}\hat{b}_{q}\hat{b}_{j''}\delta_{i's}-\hat{b}_{i'}^{\dag}\hat{b}_{p}^{\dag}\hat{b}_{r}^{\dag}\hat{b}_{j''}\hat{b}_{s}\hat{b}_{q}. \label{simply_2}
 \end{split}
\end{equation}
The sum of Eqs. (\ref{simply_1}) and (\ref{simply_2}) enters Eq. (\ref{eta_eq_even}), and we see that only chains of four operators remain.  Using the fact that $v^{ik}_{jl}=v^{ki}_{lj}$ [Eq. (\ref{two-bod})] and that $\eta_{q}^{p}$ must be evaluated for orbitals which belong to different ${\cal P}_{i}$ space, $\{l',k''\}$, Eq. (\ref{eta_eq_even}) can be rewritten,
\begin{equation}\label{eta_eq_even_final}
\sum_{k''l'}(h_{l'}^{k''}-i\eta_{l'}^{k''})A_{k''i'}^{l'j''}+\sum_{klm}(v_{kl}^{j''m}\rho_{i'm}^{kl}-v_{i'm}^{kl}\rho_{kl}^{j''m})=0.
\end{equation} 
This equation, used to determine $\eta_{l'}^{k''}$, is identical for fermions and bosons, only the evaluation of the one- and two-body reduced density matrices depends on the kind of particles. The coefficients and the ${\cal P}$-space orbitals equations are separable and can now be solved. The $\eta_{l'}^{k''}$ are obtained using Eq. (\ref{eta_eq_even}), and their values are used to determine the time-derivative of the coefficients from Eq. (\ref{EOM_C_general_expand}) and the time-derivative of the ${\cal P}$-space orbitals, Eq. (\ref{general_time_deriv_Porb}), is obtained from Eq. (\ref{time_deriv_Porb}) in addition to the $\cal{Q}$-space equations [Eq. (\ref{final_Qspace})].\\

\paragraph{General RAS scheme} \hspace{0pt} \label{all_exci} \\

Considering only even excitations provides an efficient and simple way to uncouple the equations of the TD-RASSCF-B method. Nonetheless, it is also possible to consider both even and odd excitations in the configurational space. In the following, we specifically consider a RAS scheme with all successive numbers of particles occupying ${\cal P}_{2}$ from $0$ to $N_{\text{max}}$, where $N_{\text{max}}$, defined in Sec. \ref{wavefunc}, is the highest number of particles allowed in ${\cal P}_{2}$, see Fig. \ref{Ras_Schemes}b. Note that  $N_{\text{max}}$ must fulfill the condition $N_{\text{max}}\le N$. For instance, taking $N_{\text{max}}=4$, we consider the promotion of 0,1,2,3 and 4 particles from ${\cal P}_{1}$ to ${\cal P}_{2}$. In this way, the configurational space is span by the direct sum of $N_{\text{max}}+1$ subspaces, 
\begin{equation} \label{Fock_decomp}
\mathcal{V} = \mathcal{V}_{0}\oplus\mathcal{V}_{1}\oplus \cdots \oplus \mathcal{V}_{N_{\text{max}}}.
\end{equation}
Using the expression of the time derivative of the $\{C_{I}\}$ coefficients, Eq. (\ref{EOM_C_general}), the time derivative of the one-body density matrix, present in Eq. (\ref{P_space_general}), can be expressed as,
\begin{equation} 
\begin{split}
i\dot{\rho}_{i'}^{j''} = \sum_{I\in\mathcal{V}} &\left(\langle\Psi|(i\hat{D}-\hat{H})|\Phi_{I}\rangle\langle\Phi_{I}|\Psi_{j''}^{i'}\rangle\right.\\
 &\left.-\langle\Psi_{i'}^{j''}|\Phi_{I}\rangle\langle\Phi_{I}|(i\hat{D}-\hat{H})|\Psi\rangle\right).
\end{split}
\end{equation}
We introduce the projector onto the RAS space $\mathcal{V}$ as $\hat{\Pi} = \sum_{I\in\mathcal{V}} |\Phi_{I}\rangle\langle\Phi_{I}|$. Using the above expression of $\dot{\rho}_{i}^{j}$, we obtain a new formulation of the ${\cal P}$-space orbital equation,
\begin{equation} \label{P_space_EOM_proj}
 \langle\Psi|(i\hat{D}-\hat{H})(\hat{1}-\hat{\Pi})|\Psi_{j''}^{i'}\rangle-\langle\Psi_{i'}^{j''}|(\hat{1}-\hat{\Pi})(i\hat{D}-\hat{H})|\Psi\rangle=0.
\end{equation}
 For $|\phi_{i'}\rangle\in P_{1}$ and $|\phi_{j''}\rangle\in P_{2}$, we note that $|\Psi_{j''}^{i'}\rangle$ belongs to $\mathcal{V}$, with one particle from ${\cal P}_{2}$ being annihilated and one particle in ${\cal P}_{1}$ created, leading to $(\hat{1}-\hat{\Pi})|\Psi_{j''}^{i'} \rangle= 0$. On the other hand,  $\langle\Psi_{i'}^{j''}| = \langle\Psi|\hat{b}_{i'}^{\dag}\hat{b}_{j''}$, provides configurations with a creation of an additional particle in ${\cal P}_{2}$, which may lie in $\mathcal{V}_{N_{\text{max}}+1}$, not included in $\mathcal{V}$. In this case, $\langle\Psi_{i'}^{j''}|(\hat{1}-\hat{\Pi})\ne 0$ and  Eq. (\ref{P_space_EOM_proj}) simplifies to,
 \begin{equation}\label{Eq_P_space_proj_RAS}
\begin{split} 
\langle\Psi_{i'}^{j''}|(\hat{1}&-\hat{\Pi})(i\hat{D}-\hat{H})|\Psi\rangle=0, \textnormal{ with } \\
& Ê\langle\Psi_{i'}^{j''}|(\hat{1}-\hat{\Pi})= \sum_{I\in \mathcal{V}_{N_{\text{max}}}} C_{I}^{*}\langle\Phi_{I}|\hat{b}_{i'}^{\dag}\hat{b}_{j''}.
\end{split}
 \end{equation}   
Using the expression of the Hamiltonian, [Eq. (\ref{Hamiltonian})], of the operator $\hat{D}$, [Eq. (\ref{TD_WF_RAS})], and keeping in mind that $\eta_{l}^{k}$ has only to be determined for pairs of orbitals $\{l',k''\}$ which belong to different ${\cal P}_{i}$ subspaces, Eq. (\ref{Eq_P_space_proj_RAS}) is equivalent to 
 \begin{equation}
\sum_{k''l'}(i\eta_{l'}^{k''}-h_{l'}^{k''}) \zeta_{k''i'}^{l'j''} = \frac{1}{2}\sum_{klmn}v_{ln}^{km}\zeta_{kmi'}^{lnj''},Ê\label{Pspace_General_RAS}
 \end{equation}    
where the fourth- and sixth-order tensors are defined by 
 \begin{eqnarray}
\zeta_{k''i'}^{l'j''} &=& \langle\Psi_{i'}^{j''}|(\hat{1}-\hat{\Pi})\hat{b}_{k''}^{\dag}\hat{b}_{l'}|\Psi\rangle \label{four_order_tens} \\
\zeta_{kmi'}^{lnj''}&=&\langle\Psi_{i'}^{j''}|(\hat{1}-\hat{\Pi})\hat{b}_{k}^{\dag}\hat{b}_{m}^{\dag}\hat{b}_{n}\hat{b}_{l}|\Psi\rangle. \label{six_order_tens}
 \end{eqnarray}    
Here again, the $\cal{P}$-space EOM for the determination of the $\eta_{l'}^{k''}$, Eq. (\ref{Pspace_General_RAS}), are identical for bosons and fermions \cite{Haru13,Haru14_1}. These equations are solved to determine the $\eta_{l'}^{k''}$ for each pairs of orbitals belonging to different ${\cal P}_{i}$-space. The value of  $\eta_{l'}^{k''}$ is subsequently used to solve the amplitudes equations [Eq. (\ref{EOM_C_general_expand})] and to evaluate the time-derivative of the $\cal{P}$-space orbitals from Eqs. (\ref{time_deriv_Porb}) and (\ref{final_Qspace}), as for the case of the even excitation scheme.

The \textit{general} excitations scheme and the \textit{only even} excitations schemes were originally introduced in the case of fermions in Refs. \cite{Haru13,Haru14_1}. We mention that recently Haxton et \textit{al}.  \cite{Haxton15} derived a general RAS scheme for fermions, in the sense that the configurational space can be build from of any arbitrary configurations. For both excitation schemes presented in this work, the time-derivative of the coefficients and orbitals are obtained by solving the amplitudes equations, Eq. (\ref{EOM_C_general_expand}), the $\cal{Q}$-space equations, Eq. (\ref{final_Qspace}) and the $\cal{P}$-space equations Eq. (\ref{Pspace_General_RAS}) and Eq. (\ref{eta_eq_even_final}) for the general RAS scheme and the only even excitation scheme, respectively. In the case of the MCTDHB method, the amplitude [Eq. (\ref{EOM_C_general_expand})] and the $\cal{Q}$-space equations [Eq. (\ref{final_Qspace})] are also solved to obtain the time-derivative of the wavefunction, see Appendix \ref{Ap_Num_imple}. The numerical efficiency to solve these equations scale differently with the number of configurations and the number of orbitals, as detailed in Appendix \ref{efficiency_TDRAS}. For a given number of orbitals, the TD-RASSCF-B method is more efficient to solve Eqs. (\ref{EOM_C_general_expand}) and (\ref{final_Qspace}), irrespectively of the excitation scheme used. Nonetheless, in the TD-RASSCF-B framework one additional system of equations needs to be solved, namely the $\cal{P}$-space equations  [Eq. (\ref{eta_eq_even_final}) or (\ref{Pspace_General_RAS})]. For only even excitations, the number of operations required to obtain the time-derivative of the wavefunction is always smaller in the case of the TD-RASSCF-B method than in the MCTDHB method. In the case of the general excitation scheme, the evaluation of the sixth-order tensor, Eq. (\ref{six_order_tens}), requires a significantly large number of operations. Thus, the TD-RASSCF-B method may require more operations than MCTDHB for large values of $N_{\text{max}}$ and large numbers of orbitals. As shown in Appendix \ref{efficiency_TDRAS}, this happens only for large values of $N_{\text{max}}$, for instance for $N_{\text{max}} > 38$ with $N=50$ or $N_{\text{max}} > 909$ for $N=1000$ bosons. Except for these high excitation schemes, the TD-RASSCF-B method is numerically more efficient than the MCTDHB method, but more importantly the exponential grows of the configurational space with respect to the number of orbitals can be controlled thanks to the RAS $\textit{Ansatz}$. In addition, we have shown that the TD-RASSCF equations of motion are the same for bosons and fermions, which means that the TD-RASSCF theory is a general framework including as limiting cases the TD-GP (TD-HF) and the MCTDHB (MCTDHF) theories for bosons (fermions). This result is reminiscent to the work of Alon et \textit{al} \cite{Alon07_2} where a unified set of EOM for the MCTDH theory for both bosons and fermions was derived.     

\section{Application to a time-independent system: Ground state energy} \label{time_indpdt}

	In this section, we consider a system of $N=100$ bosons trapped in a 1-dimensional (1D) harmonic potential. Experimentally, quasi-1D systems have been obtained by using a tight confinement in the transversal coordinates, freezing in that way the transversal dynamics of the system \cite{Gorlitz01, Moritz03, Laburthe04, Kinoshita05, Hofferberth07}. In the following, we consider an anisotropic harmonic trap such that the longitudinal frequency ($\omega_{x}$) is much smaller than the transversal frequency ($\omega_{\perp}$), i.e., $\omega_{\perp}\gg \omega_{x}$, such that the transverse part of the wavefunction can be assumed to be energetically frozen to the ground state and be integrated out. The resulting 1D Hamiltonian for the $N$ boson system reads,
\begin{equation}\label{H_relax}
\hat{H}=\frac{1}{2}\sum_{i=1}^{N} \left( -\frac{\partial^2}{\partial x_{i}^{2}}+x_{i}^{2}\right) + \lambda\sum_{i<j}\delta(x_{i}-x_{j}),
\end{equation}  
using the unit of length $l_{0}=\sqrt{\hbar(m\omega_{x})^{-1}}$ and the unit of energy $E_{0}=\hbar\omega_{x}$, with $m$ the mass of the particles. Assuming no confinement induced resonances \cite{Olshanii98}, the interaction strength, $\lambda$, is related to the 3D s-wave scattering length of the particles, $a_{s}$, through $\lambda=2a_{s}l_{0}l^{-2}_{\perp}$, with $l_{\perp}$ the transversal harmonic oscillator length. Experimentally, the 1D interaction strength can be tuned either by controlling the longitudinal and transversal frequencies or using an external magnetic field \cite{Courteille98, Inouye98}. 

	To solve numerically the EOM of the MCTDH and TD-RASSCF-B theories, the time-dependent orbitals are expanded on a time-independent basis or \textit{primitive} basis, which consists of a sine discrete variable representation (DVR), see Ref. \cite{Beck00}. We use $101$ basis functions in a box $[-8,8]$ and compare the results with larger basis sets to ensure the convergence of the energies presented in Tables I  and II. We numerically integrate the EOM using different integration algorithms, namely the 4th order runge-kutta (RK), the adaptive time-step 5th order RK \cite{NumRecipe} and the Adams-Bashforth-Moulton (ABM) predictor-corrector as implemented in the Heidelberg MCTDH package \cite{MCTDH}. The different integration schemes were tested against each other and we report the results obtained using the ABM integrator to the 7th order, as it is the most efficient. We calculate the GS energy using imaginary time propagation \cite{Kosloff86} of the EOM and give its energy in units of $E_{0}$.

	To assess the accuracy of the GP, MCTDHB and TD-RASSCF-B methods we compare the GS energies and by virtue of the variational principle (see for instance Ref. \cite{Szabo96}), the lower the energy the higher the accuracy. First, as a general remark, for any value of $\lambda \ne 0$ the energy obtained with the MCTDHB method systematically decreases with increasing number of orbitals and subsequently for increasing number of configurations, see for instance the first line of Table I where the numbers of configurations are indicated in parentheses. Concerning the TD-RASSCF-B method, for a given excitation scheme the energy also decreases when the number of orbitals is increased. In addition, for a given number of orbitals the energy decreases when we increase the highest number of allowed particles in ${\cal P}_{2}$, $N_{\text{max}}$. To simplify the following discussion, we introduce some quantities to help the comparison between the MCTDHB and TD-RASSCF-B methods. Firstly, we define the correlation energy as the difference between the energy obtained with a given method and the mean-field GP energy,
\begin{equation}\label{Ecorr}
{\cal E}_{\text{corr}}=E_{\text{GP}}-E_{\text{method}},
\end{equation}  
where $E_{\text{method}}$ designates the energy obtained with a given method. By definition, the GP correlation energy is equal to zero and is considered as uncorrelated. We use as a reference, ${\cal E}_{\text{ref}}$, the correlation energy obtained for the MCTDHB method with $5$ orbitals, i.e.,
\begin{equation}\label{Eref}
{\cal E}_{\text{ref}}=E_{\text{GP}}-E^{5}_{\text{MCTDHB}},
\end{equation}  
where the superscript $5$ denotes for the number of orbitals. Using this reference, we can easily compare the results obtained for different numbers of orbitals by expressing the correlation energy in percent of ${\cal E}_{\text{ref}}$. Secondly, we define the \textit{relative} correlation energy, ${\cal E}_{\text{rel}}^{X}$, as the difference between the energy obtained from a MCTDHB calculation with $X$ orbitals and the GP energy, i.e.,	
\begin{equation}\label{Erel}
{\cal E}^{X}_{rel}=E_{\text{GP}}-E^{X}_{\text{MCTDHB}}.
\end{equation}  
This quantity is particularly useful to compare the results of different RAS schemes within a given number of orbitals. Indeed, the TD-RASSCF-B \textit{Ansatz}, with restrictions on the active space, is an approximation to the MCTDHB wavefunction. Thus, when the correlation energy of a RAS scheme with $X$ orbitals is equal to ${\cal E}_{\text{rel}}^{X}$ the calculation is converged.

	We first focus on the results obtained with $\lambda=0.01$, the weakest interaction strength considered and we report the results in Table I. The reference for the correlation energy is ${\cal E}_{\text{ref}}=-2.9 \times 10^{-2}$ (the GP result is obtained from MCTDHB with a single orbital). Increasing the number of orbitals in the MCTDHB calculations from $2$ to $5$ allows us to account for more and more of ${\cal E}_{\text{ref}}$. Specifically we obtain $51\%$, $78\%$ and $91\%$ of ${\cal E}_{\text{ref}}$ for 2, 3, and 4 orbitals, respectively. The $\simeq 10\%$ variation of the correlation energy between $4$ and $5$ orbitals indicates that the results are not fully converged with respect to the number of orbitals, and more than 5 orbitals are required to converge the energy below $10^{-3}$, see Table I. Unfortunately, the MCTDHB wavefunction with $5$ orbitals includes already $\sim 4.6 \times 10^{6}$ configurations and using 6 (8) orbitals leads to $\sim 96 \times 10^{6}$ ($\sim 26 \times 10^{9}$) configurations, far beyond the scope of any practical numerical implementation. 

	The TD-RASSCF-B method provides more flexibility to describe the wavefunction in the sense that we can choose different RAS schemes, different numbers of orbitals and their partitions into ${\cal P}_{1}$ and ${\cal P}_{2}$ spaces. In Table I we report the results obtained for $M=2$ to $8$ orbitals with a single ${\cal P}_{1}$ orbital, i.e., $M_{1}=1$ and $M_{2}=M-M_{1}$ ${\cal P}_{2}$ orbitals, and a few specific cases of the \textit{general} RAS scheme. We indicate the excitation schemes with the usual notation. For example -SD denotes that single and double excitations are allowed from ${\cal P}_{1}$ to ${\cal P}_{2}$. We follow this notation up to -SDTQ56789 and for larger excitations, we just indicate the value of $N_{\text{max}}$ (e.g., "-10" means that all excitations from ${\cal P}_{1}$ to ${\cal P}_{2}$ up to $10$ are included). For each number of orbitals, when we increase the excitation scheme the energy becomes closer to the MCTDHB result and converges to this latter for the -SDTQ5678 RAS scheme, as indicated by the underlined digits in Table I. Thus, ${\cal E}_{\text{ref}}$ is recovered for the -SDTQ5678 scheme with 5 orbitals, but the expansion of wavefunction includes \textit{only} $495$ configurations, i.e.,  $9 \times 10^{3}$ times fewer configurations than the MCTDHB expansion for $5$ orbitals. It is worthwhile to note that this RAS scheme converged for all number of orbitals, and always for much fewer configurations than with the MCTDHB. The least accurate TD-RASSCF-B calculation, presented in Table I, consists of $2$ orbitals and the -SD RAS scheme. The correlation energy includes $98.7\%$ of ${\cal E}_{\text{rel}}^{2}$ and interestingly when we increase  the number of orbitals from $3$ to $5$, a similar amount of correlation is obtained ($98.8\%$ for all values) in comparison to the respective ${\cal E}_{\text{rel}}^{X=3,4,5}$ correlation energies. Thus, using the -SD scheme with $5$ orbitals $98.8\%$ of ${\cal E}_{\text{ref}}$ is obtained but the TD-RASSCF-B wavefunction includes only $15$ configurations while the MCTDHB wavefunction includes more than $4.5Ê\times 10^{6}$ configurations, i.e., $\sim 3 \times 10^{5}$ times more configurations. Moreover, the energy difference between the -SD scheme and MCDTHB method is systematically below $5 \times 10^{-4}$, lower than the convergence obtained with respect to the number of orbitals. Concerning the correlation energy of the -SDTQ and -SDTQ56 schemes, we find that they include $99.99\%$ and $99.9999\%$ of ${\cal E}_{\text{rel}}^{X}$, with $X = 2$ to $5$. It is remarkable that the correlation energy remains almost constant while the difference between the number of configurations between the MCTDHB and TD-RASSCF-B increases exponentially with the number of orbitals. These results show that the correlation energy does not strongly depend on the number of configurations used in the wavefunction expansion, as the configurational space of the -SD RAS scheme increases only from $3$ to $15$ configurations for $M=2$ to $M=5$ but captures $98.8\%$ of ${\cal E}_{\text{rel}}^{X}$, with $X=2$ to $5$. Thus, the correlation depends more critically on the number of orbitals than the number of configurations used in the calculation. To illustrate this point, we compute the GS energy with $6$ to $8$ orbitals with the TD-RASSCF-B method, see Table I, and we obtain energies lower than the energy of the MCTDHB with $5$ orbitals for all excitation schemes used here. It means that the -SD scheme with $8$ orbitals and $36$ configurations is more accurate than the MCTDHB method with $5$ orbitals and $\sim 4.6 \times 10^{6}$ configurations. Moreover, comparing the energies obtained for the -SDTQ5678 and -10 excitation schemes, we can conclude that the GS energy has converged with respect to the number of excitations. Thus, the TD-RASSCF-B method, thanks to the restriction imposed on the configurational space, can provide more accurate results than the MCTDHB method, whoes practical applicability is limited by the exponential growth of the number of configurations.

	We also consider RAS schemes with only even excitations (see Table II), for which the numerical effort is always reduced in comparison to the MCTDHB method, see Appendix \ref{efficiency_TDRAS}. The energy difference between the -D and the -SD schemes is below $1.3 \times 10^{-6}$ for all numbers of orbitals, indicating that more than $98.7 \%$ of the relative correlation energy is obtained with slightly fewer configurations. The same conclusion holds for the comparison of the -DQ and -SDTQ schemes with an energy difference below $1.5 \times 10^{-5}$, including more than $99.95 \%$ of the relative correlation energy. The number of configurations is slightly smaller in the case of the RAS schemes with only even excitations but the numerical efficiency is better as the $\cal{P}$-space EOM, Eq. (\ref{eta_eq_even_final}), does not require the update of a sixth-order tensor at each time-step as it is the case of the general RAS schemes, see Eq. (\ref{six_order_tens}). For values of $N_{\text{max}}\ge6$ and $M\ge3$, the energy converges with respect to $N_{\text{max}}$, as the energy does not change by increasing $N_{\text{max}}$ further, but with energy slightly larger than the MCTDHB ones. The energy difference between the -DQ68 scheme and the MCTDHB calculation, with $5$ orbitals for both methods, is $\sim1.1 \times 10^{-5}$ and includes $99.96 \%$ of ${\cal E}_{\text{ref}}$. In the case of only even excitations, we also find that for $M>5$ the energy for all schemes is below the energy of the best MCTDHB calculation performed. The comparison of the converged -DQ68 and -SDTQ5678 RAS schemes, show that the energy difference remains below $1.5 \times 10^{-5}$ for $M>5$, which is two orders of magnitude small then the convergence obtain with respect to the number orbitals $\sim 10^{-3}$.
	   	         	
	We perform the same analysis for an interaction strength $\lambda=0.1$ and we obtain, as a reference for the correlation energy, ${\cal E}_{\text{ref}}=-1.34$, see Table I. This value is much larger than the one obtained previously and can be explained by the stronger interaction between the particles. Indeed, for a stronger interaction strength, the energy of the system is lowered by allowing the particles to occupied higher orbitals, i.e., orbitals leading to higher kinetic and potential energies, such that the interaction energy is reduced. In the mean-field GP theory, this possibility is not possible as only one orbital is used to describe the wavefunction. The orbitals that diagonalized the reduced density matrix and their respective eigenvalues, or \textit{population}, can be used to characterized the system. If the largest eigenvalue is of the same order as $N$, the system is condensed \cite{Penrose56}. As the GP wavefunction includes a single orbital, it can only describe condensed systems. If more than one eigenvalue is of the order of $N$, then the system is fragmented (see Ref. \cite{Nozieres82} and the discussion in Ref. \cite{Sakmann08}). We find that, indeed, the occupation of the lowest natural orbital in the MCTDHB calculation using $5$ orbitals decreases from a population of $99.987 \%$ for $\lambda=0.01$ to a population of $99.465 \%$ for $\lambda=0.1$. This slightly larger depletion of the condensate has a strong impact on the correlation energy, as the mean-field GP theory provides a less accurate description of the system. We point out that increasing the number of orbitals from $4$ to $5$ in the MCTDHB calculations gives an energy difference $\sim 10^{-1}$, see Table I, which means that the energy does not converge below this value. The relative correlation energy ${\cal E}_{\text{rel}}^{2}$, ${\cal E}_{\text{rel}}^{3}$ and ${\cal E}_{\text{rel}}^{4}$ include $40.1\%$, $68.8\%$ and $86.7\%$ of ${\cal E}_{\text{ref}}$, respectively.       
	Starting the discussion with the \textit{general} RAS schemes, see Table I, we note that to converge to the MCTDHB energies and thus include $100\%$ of the relative correlation energies, ${\cal E}_{\text{rel}}^{X}$ with $X=2$ to $5$, large values of $N_{\text{max}}$ are required. For the -SDTQ5 RAS scheme we find that the correlation energy includes $\sim 97\%$ of ${\cal E}_{\text{rel}}^{X}$,  the -10 RAS scheme includes  $\sim 99.9\%$ of ${\cal E}_{\text{rel}}^{X}$, the -15 RAS scheme includes  $\sim 99.997\%$ of ${\cal E}_{\text{rel}}^{X}$ and the -20 RAS scheme is converged with more than $99.9999\%$ of ${\cal E}_{\text{rel}}^{X}$. These results are obtained irrespectively of the number of orbitals, i.e., $X=2$ to $5$. Even if large values of $N_{\text{max}}$ are used, the expansion of the wavefunction using the -20 RAS scheme includes $\sim 10^{4}$ configurations for $5$ orbitals while the MCTDHB wavefunction includes $\sim 4.6\times 10^{6} $ configurations. We also use RAS schemes with \textit{only} even excitations and we report the results in Table II. Similarly to the case with $\lambda=0.01$, we find that, except for $2$ orbitals, the energy does not converge to the MCTDHB energy, irrespectively to the value of $N_{\text{max}}$ used. Thus the energies obtained with $3$, $4$ and $5$ orbitals include $99.7\%$, $99.0\%$ and $98.8\%$ of the respective ${\cal E}_{\text{rel}}^{X}$ with the -20 RAS scheme, which is converged. For similar numbers of configurations the \textit{general} RAS scheme provides more accurate results but remains more demanding in term of computation, see Appendix \ref{efficiency_TDRAS}. It is important to keep in mind that the convergence with respect to the number of orbitals is $\sim 10^{-1}$ and a convergence one order of magnitude below is achieved with the -SDTQ5 excitation scheme (Table I) and the -DQ excitation scheme (Table II).          	     
 	As previously, the configurational space of the MCTDHB wavefunction becomes unworkable for more than $5$ orbitals, but the TD-RASSCF-B method can include more orbitals. At the level of the -SDT scheme and $8$ orbitals and for higher excitation schemes with $6$ to $8$ orbitals, the GS energy is always below the energy obtained with the MCTDHB method with $5$ orbitals, see Table I. In the same way, using only even excitations provides more accurate results for excitation schemes higher than -D. As a remark, we obtain an energy $\sim 0.3$ below the energy of the MCTDHB method with $5$ orbitals by using the -15 RAS scheme with 8 orbitals, see Table I.
	
	These preceding examples show that the TD-RASSCF-B method provides an efficient approach for computing the GS energy of trapped cold atoms. This wavefunction based approach gives access to quantities  of interest such as the one and two-body reduced densities and the fragmentation using the population analysis of the natural orbitals. The accuracy was compared with the MCTDHB results and we showed that the TD-RASSCF-B method converges for relatively low excitation schemes. Moreover, the possibility to constrain the growth of the configurational space gives the possibility to use more orbitals than in the MCTDHB calculations and better results were systematically obtained using the TD-RASSCF-B method. This result can be understand as follows, the MCTDHB wavefunction for a small number of orbitals generates a large number of configurations, as all orbitals can be equally populated. The main part of these configurations, however, do not contribute to lower the energy of the system as they describe states with many particles occupying the same spatial orbital, which induced a large interaction energy for a large value of $\lambda$. As a limiting case, we know that in the Tonks-Girardeau model \cite{Girardeau60}, obtained for an infinite value of $\lambda$, each boson occupies a different orbital. Thus, using a larger number of orbitals in the TD-RASSCF-B method introduces configurations for which a small number of particles occupy a larger number of different orbitals, and thus describes more efficiently the system. This flexibility of the TD-RASSCF-B method of choosing more orbitals opens a new possibility to explore the static properties of trapped cold atoms in systems with hundreds of particles and large numbers of orbitals, which are for the moment beyond the possibility of the MCTDHB method.       
    
\section{Application to a time-dependent system: Dynamics of bosons with harmonic interaction} \label{time_dpdt}

	As an illustration of an application of the TD-RASSCF-B method to a truly time-dependent problem, we simulate the dynamics of an ensemble of $N=10$ bosons  in a 1D harmonic trap interacting through a harmonic interaction potential. We consider an initial system of non-interacting bosons, for which the Hamiltonian $\hat{H}_{0}$ reads,
\begin{equation}\label{H0_dyn}
\hat{H}_{0}=\frac{1}{2}\sum_{i=1}^{N} \left( -\frac{\partial^2}{\partial x_{i}^{2}}+x_{i}^{2}\right),
\end{equation}  
where we use the units described in Sec. \ref{time_indpdt} and the time is expressed in units of $t_{0}=\omega_{x}^{-1}$. The analytical ground state wavefunction and energy are used to ensure the convergence of the imaginary time propagation and, as expected, are the same for all methods. The dynamics is initiated at $t=0$ by quenching instantaneously the strength of the two-body interaction, as performed in Ref. \cite{Lode12}, leading to the evolution of the system under the action of the following Hamiltonian,
\begin{equation}\label{H_quench}
\hat{H}=\frac{1}{2}\sum_{i=1}^{N} \left( -\frac{\partial^2}{\partial x_{i}^{2}}+x_{i}^{2}\right) + \sum_{i<j}\lambda (x_{i}-x_{j})^2,
\end{equation}
with $\lambda$ the strength of the two-body interaction. This sudden change in the interaction between the bosons leads to a breathing dynamics of the BEC with frequencies $\Omega_{n}=2n\sqrt{\omega_{x}^{2}+2N\lambda}$, with $\omega_{x}$ the frequency of the harmonic trap, see Ref. \cite{Lode12}. For positive values of $\lambda$ the two-body interaction is \textit{attractive}, while for negative values the interaction is repulsive and leads to unbound dynamics for $\lambda<-\omega_{x}/2N$. Note that we use the same parameters than in Sec. \ref{time_indpdt} for the numerical resolution of the EOM.          

\subsection{Breathing dynamics with $\lambda = 0.1$}

\begin{figure}[h!]
\centering 
\includegraphics[scale=0.25]{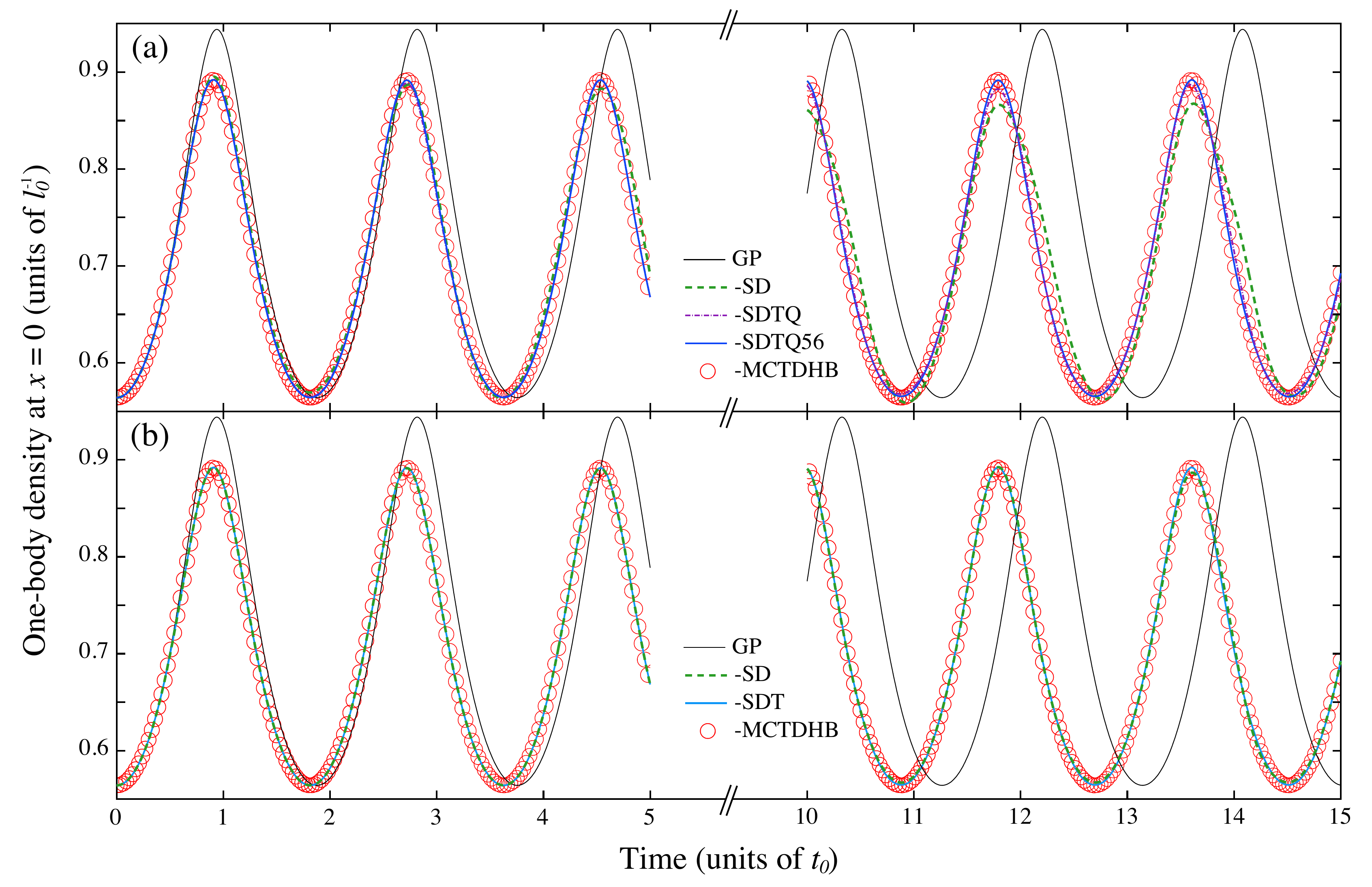}
   \caption{ \small{Time evolution of the one-body density at the center of the harmonic trap, $\rho(x=0,t)$ after an instantaneous quench of the interparticle interaction strength from $\lambda=0$ to $\lambda=0.1$ [Eqs. (\ref{H0_dyn}) and (\ref{H_quench})]. The result obtained using the mean-field GP theory (thin (black) line) strongly disagrees with the numerically exact result ((red) open circles) obtained using MCTDHB with $4$ orbitals in both panels. Note that the results are plotted for $0\le t\le 5$ on the left and $10 \le t \le 15$ on the right. (a) Results of the TD-RASSCF-B method using a single ${\cal P}_{1}$ orbital, $M_{1}=1$, and $M_{2}=3$ ${\cal P}_{2}$ orbitals for different RAS schemes, namely -SD (dashed (green)), -SDTQ (dash-dot (purple)) and -SDTQ56 (thick (blue) line). For short time, the lines are on top of the MCTDHB result. (b) Results of the TD-RASSCF-B method using $M_{1}=2$ and $M_{2}=2$ keeping the total number of orbitals constant for the -SD (dashed (green)) and -SDT (thick (blue) line) schemes. The results are on top of the MCTDHB curve, which show the good accuracy of the method.}}   \label{gene_RAS_0_1}
 \end{figure}

\begin{figure}[h!]
\centering 
\includegraphics[scale=0.25]{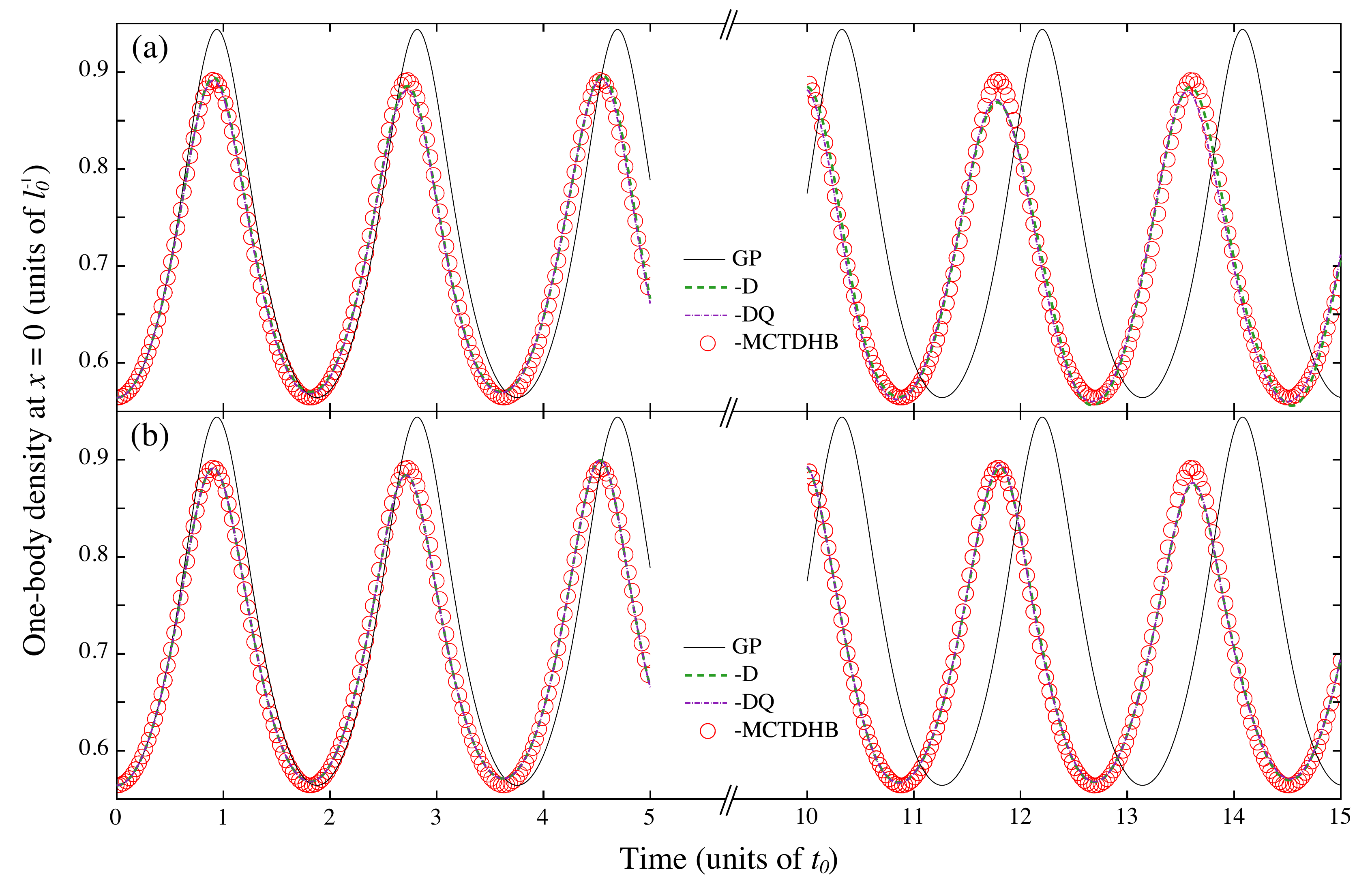}
   \caption{ \small{Same as Fig. \ref{gene_RAS_0_1} but using the TD-RASSCF-B method with RAS schemes including only even excitations. (a) Results using a single ${\cal P}_{1}$ orbital, $M_{1}=1$, and $M_{2}=3$ ${\cal P}_{2}$ orbitals for -D (dashed (green)) and  -DQ (dash-dot (purple)) RAS schemes. (b) Results of the TD-RASSCF-B method using $M_{1}=2$ and $M_{2}=2$ keeping the total number of orbitals constant.}}   \label{even_RAS_0_1}
 \end{figure}

We first consider the dynamics following a quenching of the interaction strength from $\lambda=0$ to $\lambda=0.1$ [Eq. (\ref{H_quench})]. We find that the MCTDHB method with $M=4$ orbitals and $286$ configurations is numerically exact for the propagation time considered here, i.e., $0\le t\le 15$, see Fig. \ref{gene_RAS_0_1}. The time evolution of the system is characterized by the one-particle density, $\rho(x=0,t)$, at the center of the trap $x=0$ and exhibits a periodic evolution with a frequency $\omega_{\text{MCTDHB}}=3.46$. This value agrees perfectly with the analytical prediction $\Omega_{n=1}=3.46$ meaning that the first excited state is mainly responsible for the dynamics. Nonetheless, the discrepancy with a pure cosine function $\sim \cos(\Omega_{1}t)$ indicates the role of higher excited states with higher harmonic frequencies \cite{Lode12}. The mean-field GP fails to describe the system evolution, even at short time ($t<1$) as depicted in Fig. \ref{gene_RAS_0_1} (a) and we obtain a lower frequency $\omega_{\text{GP}} = 3.35$. First, we perform TD-RASSCF-B simulations using a single ${\cal P}_{1}$ orbital ($M_{1}=1$) and $M_{2}=3$ ${\cal P}_{2}$ orbitals for different RAS schemes reported in Fig. \ref{gene_RAS_0_1} (a). For a short time, i.e., $0\le t\le 5$, all RAS schemes describe accurately the dynamics of the system, in contrast to the GP result. On the scale of the figure, the convergence to the MCTDHB result is obtained by using the -SDTQ excitation scheme including $35$ configurations,  a reduction by a factor of $\sim 8$. For a longer time, $10\le t\le 15$, the -SD RAS scheme substantially differs from the MCTDHB result with a shift in the frequency and a smaller amplitude of the oscillations. The -SDTQ scheme only slightly differs by a smaller amplitude for $t>11.5$ and convergence is achieved for the -SDTQ56 excitation scheme with $84$ configurations. We also investigate the role of the ${\cal P}_{1}$ orbitals on the accuracy of the computations by considering $M_{1}=2$ and $M_{2}=2$, such as the total number of orbitals, $M=4$, remains unchanged. The -SD excitation scheme converged for short time, see Fig. \ref{gene_RAS_0_1} (b), and provides a better description of long time dynamics than the -SDTQ scheme used previously [Fig.  \ref{gene_RAS_0_1} (a)] but includes $58$ configurations. This number of configurations is similar to the $56$ configurations obtained by using the -SDTQ5 RAS scheme with a single ${\cal P}_{1}$ orbital, which differs from the MCTDHB results for $t>11.5$ (not shown) while the -SD scheme with $2$ ${\cal P}_{1}$ orbitals differs for $t>13.5$, see Fig. \ref{gene_RAS_0_1} (b). We obtain converged results for the -SDT RAS scheme with $90$ configurations. In the previous section, we showed that the RAS schemes with \textit{only} even excitations provide accurate results for GS energy and reduce the numerical effort (see Appendix \ref{efficiency_TDRAS}). We apply the -D and -DQ schemes with $M_{1}=1$ and $M_{1}=2$ Fig. \ref{even_RAS_0_1} (a) and (b), respectively, and keep $M=4$. For both $M_{1}=1$ and $M_{1}=2$, the results do not converge to the MCTDHB results and do not significantly improve for larger excitation schemes. For $M_{1}=1$, Fig. \ref{even_RAS_0_1} (a), the results obtained with the -D and the -DQ schemes are in very good agreement with the MCTDHB results concerning both the frequency and the amplitude and start to deviate only for $t>11$. In both cases, the number of configurations used to expand the wavefunction is substantially smaller than the expansion of the MCTDHB wavefunction with $7$ and $22$ configurations, respectively. When we use $2$ orbitals in the ${\cal P}_{1}$-space both -D and -DQ provide the same results for the dynamics, see Fig. \ref{even_RAS_0_1} (b). For short time dynamics, the results are similar to the ones obtained previously with $M_{1}=1$, but for a longer time, the oscillations remain in phase with the MCTDHB result, only the amplitude deviates for $t>13$. The wavefunction of the -D scheme includes $38$ configurations. Thus, the TD-RASSCF-B method provides an access to describe accurately the dynamics of the interacting system, while the mean-field GP theory failed even for short time. We obtain a good agreement in comparison to the MCTDHB method with a substantial reduction of the configurational space, using few tens instead of the few hundreds of configurations with the MCTDHB method. Moreover, the different parameters of TD-RASSCF-B method which  define the wavefunction can be used to converge the results to the MCTDHB calculations. The implications of this reduction on the CPU time are discussed at the end of this Section.  

\subsection{Breathing dynamics with $\lambda = 0.5$}

\begin{figure}[h!]
\centering 
\includegraphics[scale=0.25]{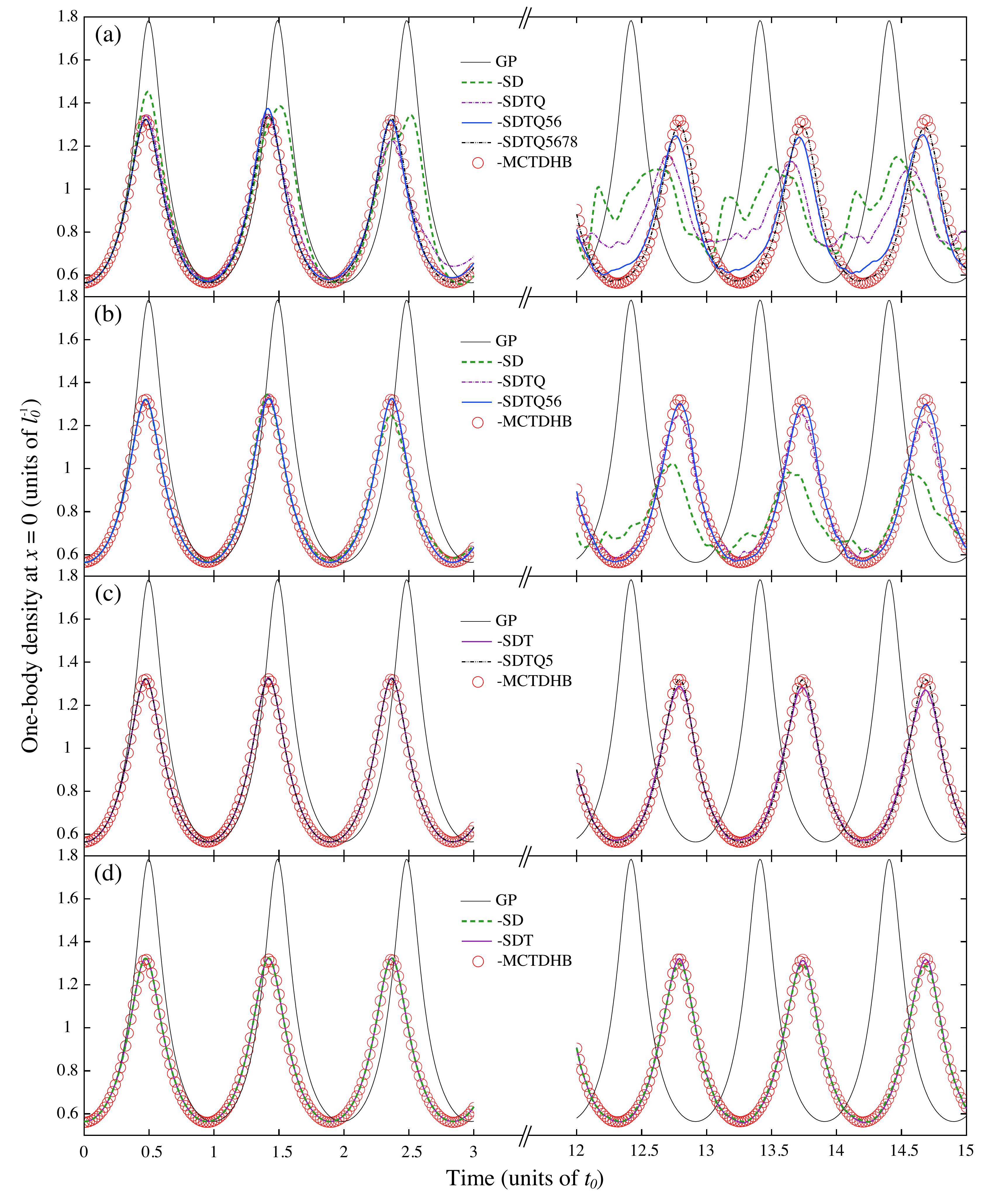}
   \caption{ \small{Time evolution of the one-body density at the center of the harmonic trap, $\rho(x=0,t)$ after an instantaneous quench of the interparticle interaction strength from $\lambda=0$ to $\lambda=0.5$ [Eqs. (\ref{H0_dyn}) and (\ref{H_quench})]. The result obtained using the mean-field GP theory (thin (black) line) strongly disagrees with the numerically exact result ((red) open circles) obtained using MCTDHB with $8$ orbitals in each panel. Note that the results are plotted for $0\le t\le 3$ on the left and $12 \le t \le 15$ on the right. (a) Results of the TD-RASSCF-B method using a single ${\cal P}_{1}$ orbital, $M_{1}=1$, and $M_{2}=7$ ${\cal P}_{2}$ orbitals for different RAS schemes, namely -SD (dashed (green)), -SDTQ (dash-dot (purple)), -SDTQ56 (thick (blue) line) and  -SDTQ5678 (dash-dot (black)). (b) Results using $M_{1}=2$ and $M_{2}=6$ keeping constant the total number of orbitals. For short time, $0 \le t\le 3$, all RAS schemes are on top of the MCTDHB result. (c) TD-RASSCF-B simulations using $M_{1}=3$ and $M_{2}=5$ orbitals, using the -SDT (solid purple) and -SDTQ5 (dash-dot black) RAS schemes. A very good agreement is obtained in comparison to the MCTDHB result. (d) Results of the TD-RASSCF-B method using $M_{1}=4$ and $M_{2}=4$ orbitals using the -SD and -SDT schemes. This latter converges to the MCTDHB results but includes $\sim 4$ times fewer configurations than the MCTDHB wavefunction.}}   \label{RAS_0_5}
 \end{figure}

We pursue the illustration of the TD-RASSCF-B method by considering a quenching from $\lambda=0$ to $\lambda=0.5$ [Eq. (\ref{H_quench})]. This interaction strength was used in Ref. \cite{Lode12} to benchmark the MCTDHB method. For the time interval considered here, $0\le t\le 15$, we find that the result obtained with the MCTDHB method using $8$ orbitals and $19448$ configurations is numerically exact, in agreement with Ref. \cite{Lode12}. As previously, the one-body density exhibits oscillations as a function of the time with a period $\omega_{\text{MCTDHB}} = 6.63$, in perfect agreement with the analytical frequency $\Omega_{n=1}=6.63$. In comparison to the previous results, the shape of the oscillations indicates that the role of higher excited states is stronger as a large deviation from a simple cosine function, $\sim \cos(\Omega_{1}t)$, is obtained. This is not surprising since the value of $\lambda$ is now $5$ times larger than the one of the previous example. Along with the MCTDHB result we report, for comparison, the result obtained using the mean-field GP theory, see Fig. \ref{RAS_0_5} (a), which strongly deviates from the MCTDHB result for $t>0.3$ with a larger amplitude in the oscillations and gives a lower frequency for the oscillations, $\omega_{\text{GP}}=6.32$. We start the TD-RASSCF-B simulations using $M=8$ orbitals with one single ${\cal P}_{1}$ orbital and $M_{2}=7$ ${\cal P}_{2}$ orbitals, Fig. \ref{RAS_0_5} (a). For times between $0$ and $2.5$, the excitation schemes larger or equal to -SDTQ provide an accurate description of the dynamics, while the -SDTQ scheme includes $330$ configurations in the wavefunction, i.e., a factor of $\sim 60$ less than the MCTDHB. For longer times, the -SDTQ5678 RAS scheme with $6435$ configurations is required to accurately describe the MCTDHB results, the lower excitation schemes give substantially different results. To improve the results of the TD-RASSCF-B method, we increase the number of orbital in ${\cal P}_{1}$ and keep constant the total number of orbitals, $M=8$. In Fig. \ref{RAS_0_5} (b), (c) and (d) we report the results with $M_{1}=2$, $3$ and $4$ ${\cal P}_{1}$ orbitals, respectively. In the case of $M_{1}=2$, the -SDTQ56 RAS scheme, with $5412$ configurations, provides a very accurate description of the dynamics for $0\le t\le15$. For lower excitation schemes, the results are accurate for $0\le t\le 3$ and  using the -SDTQ scheme, the frequency of the oscillation at a longer time are correctly obtained, see Fig. \ref{RAS_0_5} (b). For $M_{1}=3$, the results in Fig. \ref{RAS_0_5} (c) show that the -SDTQ5 RAS scheme including $6882$ configurations converged to the MCTDHB results, while the -SDT scheme with $2276$ configurations gives the correct frequency for the oscillation but with a smaller amplitude. Finally, Fig. \ref{RAS_0_5} (d) report converged results for $M_{1}=4$ using the -SDT schemes, which includes $5216$ configurations. Nonetheless, the -SD scheme with $2816$ configurations is accurate for the considered time of propagation in comparison to the MCTDHB result. 

Using different numbers of ${\cal P}_{1}$ orbitals and different RAS schemes points out that the number of configurations required to accurately describe the evolution of the system change substantially. For instance, a similar accuracy is achieved for the -SDTQ5678 scheme with $M_{1}=1$, the -SDTQ56 scheme with $M_{1}=2$, the -SDT scheme with $M_{1}=3$ and the -SD scheme with $M_{1}=4$. For each case, a smaller amplitude of the oscillations is observed in comparison to the MCTDHB for $t >12$. But the numbers of configurations used in the wavefunction expansions are $6435$, $5412$, $2276$ and $2816$, respectively. Thus, for a comparable accuracy, the number of configurations can be divided by a factor $\sim 2$ by choosing adequately the size of the ${\cal P}_{1}$-space and a factor $\sim 10$ in comparison to the MCTDHB configurational space. The reduction of the configurational space impacts strongly the required CPU times of the simulations. For instance, the -SDTQ5678 RAS scheme with $M_{1}=1$ took $\sim 18.7$ CPU hours, while the -SDTQ56 scheme with $M_{1}=2$, the -SDT scheme with $M_{1}=3$ and the -SD scheme with $M_{1}=4$ took $\sim 19.8$, $\sim 6.9$ and $\sim 11.1$ CPU hours, respectively, on a 2.4 GHz Intel E5-2680 CPU. Using these RAS schemes, the CPU time is drastically reduced in comparison to the $\sim 113.4$ CPU hours on a 2.5 GHz Intel E5-2680 CPU needed to perform the MCTDHB simulation, but the dynamics is accurately described. Moreover, the -SDTQ5 RAS scheme with $M_{1}=3$ converged to the exact solution with $\sim 4$ times less CPU time. Albeit the drastic reduction in the size of configuration space, the CPU time needed to perform the TD-RASSCF-B calculations is also substantially reduced.

We briefly summarize the findings concerning the dynamical evolution of trapped atoms after a quenching of the interaction strength of an attractive harmonic interaction. In the case of $\lambda=0.1$, the MCTDHB theory converged to the numerically exact result for $4$ orbitals. The TD-RASSCF-B method using different size of the ${\cal P}_{1}$-space and different RAS schemes can accurately describe the dynamics of the system characterized by $\rho(x=0,t)$ with $\sim 4$ times fewer configurations. Moreover, converged TD-RASSCF-B results were obtained with substantially less configurations, for instance with $M_{1}=2$ and the -SDT excitation scheme. In the case of $\lambda=0.5$, the exact solution was obtained with $8$ orbitals using the MCTDHB method leading to $19448$ configurations. A larger number of orbitals is needed for the stronger interaction between the particles. Accurate results were obtained with the TD-RASSCF-B method, reducing by a factor $\sim 10$ the size of the configurational space and converged results were obtained with $4$ times less configurations, for instance considering $M_{1}=4$ and the -SDT RAS scheme. Moreover, all calculations performed with the TD-RASSCF-B method were better than the mean-field GP theory, which failed to describe both scenarios.                          

\section{Conclusion and outlook}\label{conclusion}

In this work, we presented a general formalism for the time-dependent restricted active-space self-consistent field (TD-RASSCF) method, which includes the first derivation obtained for fermions (TD-RASSCF-F) \cite{Haru13, Haru14_1,Haru14_2} and extended it for systems of spinless bosons (TD-RASSCF-B). This TD-RASSCF-B method includes, as limiting cases, the (TD)-GP and the MCTDHB theories and provides a way to tackle the exponential growth of the configurational space in the MCTDHB method. The EOM were derived for two families of RAS schemes, which give the possibility to restrict the full-configurational description of the MCTDHB wavefunction. Through a set of numerical examples, we have shown that the method can provide an accurate description of the static properties of the ground-state of the system. In the case of hundreds of particles, the method can lead to results beyond the reach of the MCTDHB method, providing a better accuracy by including more orbitals while constraining the number of configurations. In this sense, the TD-RASSCF-B method paves the way for numerical investigation of intermediate system sizes with a few tens to hundreds of bosons with a better accuracy than what was possible with the MCTDHB method. We also provided a comparison between the MCTDHB and TD-RASSCF-B method in the case of breathing dynamics induced by a sudden quenching of the interaction strength with two different initial conditions. As for the MCTDHB method, the TD-RASSCF-B method does not have any restriction on the choice of the two-body interaction potential used, as was illustrated by the use of a non-contact harmonic interaction between the bosons. Using as a reference the numerically exact result obtained from the MCTDHB method, we showed that the TD-RASSCF-B method is always more accurate than the mean-field TD-GP theory to describe the dynamics of the system. Moreover, using different RAS schemes and partitions of the ${\cal P}$-space we obtained very accurate results for substantially less configurations and thus less CPU time than with the MCTDHB method. This reduction of the configurational space can be efficiently exploited to solve numerically the TDSE beyond the mean-field approach. Dynamical effects such as the four wave mixing (FWM) process \cite{Hilligsoe05} used to produce correlated atoms beams \cite{Bonneau13} or the dynamics of bright \cite{Khaykovich02,Strecker02} and dark \cite{Burger99, Denschlag97} solitons can be investigated \textit{ab-initio} beyond the mean-field TD-GP theory. In addition, the dynamics induced by a time-dependent Hamiltonian can also be explored using the TD-RASSCF-B method, such as in the case of periodically driven optical lattices \cite{Aidelsburger13,Goldman14}. \\
   
\section*{Acknowledgments}
The authors are indebted to Dr. Haruhide Miyagi for useful discussions. This work was supported by the ERC-StG (Project No. 277767-TDMET), and the VKR center of excellence, QUSCOPE.

\appendix

\section*{Appendices}

In these Appendices we provide a brief description of the implementation the TD-RASSCF-B method. The implementation is rather similar for bosons and fermions in the sense that the set of equations that we have to solve, i.e. Eqs. (\ref{EOM_C_general}), (\ref{EOM_Qspace}) and (\ref{eta_eq_even_final}) or (\ref{Pspace_General_RAS}), only depend on the type of particles trough the creation and annihilation operators and the set of configurations $\{|\Phi_{I}\rangle\}$.    

\section{Compact representation of the wavefunction}\label{wf_representation}
Our implementation is based on the general mapping of bosonic operators in Fock space introduced in Ref. \cite{Streltsov10} and implemented, for instance, for bosons \cite{Streltsov11} and fermions \cite{Fasshauer16} in the framework of multi-configurational TD methods. Assuming $M$ orbitals and $N$ bosons, the configurations are expressed using the occupation number formalism, where $|n_{1},n_{2},\cdots,n_{M}\rangle$, represents a configuration with $n_{1}$ bosons in the orbital $|\phi_{1}\rangle$, $n_{2}$ bosons in the orbital $|\phi_{2}\rangle$, etc. Such a configuration is indexed by an unique integer $J$  defined as,
 \begin{equation} \label{indexing_conf}
J =1+\sum_{k=1}^{M}
   \begin{pmatrix}
   N+M -1 -k - \sum_{l=1}^{k} n_{l}  \\
   M-k  
\end{pmatrix}.
 \end{equation}   
Thus, for each configuration we store its complex coefficient $C_{J}$ in an array according to the index $J$ provided by the above mapping. In Ref. \cite{Streltsov11}, this mapping was employed to avoid the storage of the configuration vectors $|n_{1},n_{2},\cdots,n_{M}\rangle$, which can be prohibitively memory consuming in the case of the MCTDHB method. Thus, to access the coefficient of the configurations, a set of M-nested loops over the occupation number $n_{i}$ is used to span the full configurational space and to compute, using Eq. (\ref{indexing_conf}), the respective indexes. In the case of the TD-RASSCF-B method only a selected number of configurations are used to expand the wavefunction, and the scheme of Ref. \cite{Streltsov11} is not readily applicable in the sense that we want to avoid explicit use of the full configurational space. Instead we follow a different strategy for indexing the RAS configurations. We introduce $M_{1}$ and $M_{2}$  the number of orbitals in the ${\cal P}_{1}$ and ${\cal P}_{2}$ spaces, respectively. For $\mathcal{V}_{0}$, i.e., configuration with particles only in ${\cal P}_{1}$, see Eq. (\ref{Fock_decomp}), we enumerate all possible configurations, evaluate their index, using Eq. (\ref{indexing_conf}), and store them. Then for the excited configurations, i.e., configurations with one or more particles in ${\cal P}_{2}$, we introduce the excitation number $n_{exc}$ which is equivalent to the number of particles in ${\cal P}_{2}$. The number of remaining particles in ${\cal P}_{1}$ is $N-n_{exc}$. For each excitation we enumerate the  configurations of $N-n_{exc}$ particles in the ${\cal P}_{1}$ orbitals and compute their index. The same is performed for the configurations with $n_{exc}$ in ${\cal P}_{2}$ orbitals. Then the permanents for the total number of bosons are obtained by combining them,
 \begin{equation}
 \begin{split}
|\textbf{n}^{n_{exc}}\rangle = &\left(\sum_{i=1}^{N_{1}^{n_{exc}}}|n_{1},n_{2},\cdots,n_{M_{1}}\rangle \right) \\
 &\otimes  \left(\sum_{i=1}^{N_{2}^{n_{exc}}}|n_{M_{1}+1},\cdots,n_{M}\rangle\right), 
 \end{split}
 \end{equation}   
 where $|\textbf{n}^{n_{exc}}\rangle$ is an array of dimension $M \times (N_{1}^{n_{exc}}\times N_{2}^{n_{exc}})$, with $N_{1}^{n_{exc}}$ ($N_{2}^{n_{exc}}$) the total number of configurations obtained from arranging the $N_{b}-n_{exc}$ ($n_{exc}$) particles in the ${\cal P}_{1}$ (${\cal P}_{2}$) orbitals. We evaluate the indexes of the configurations resulting from the ${\cal P}_{1}$ subsystem, $J_{P_{1}}^{n_{exc}}$, and from the ${\cal P}_{2}$ subsystem, $J_{P_{2}}^{n_{exc}}$, applying Eq. (\ref{indexing_conf}) for the subsystems, separately. The index of the configuration with the total number of bosons is build as a three components array defined by $\{J_{P_{1}}^{n_{exc}},J_{P_{2}}^{n_{exc}},n_{exc}\}$, which stores the position of the configuration in the configurational vector. This scheme is thus applied for all excitations, $n_{exc}$, included in the RAS scheme. With this storage or construction of the wavefunction, for a given configuration we have access to its coefficient in the following way. (i) We evaluate the index $J_{P_{1}}^{n_{exc}}$  for the ${\cal P}_{1}$ subsystem, (ii) we evaluate the index $J_{P_{2}}^{n_{exc}}$ for the ${\cal P}_{1}$ subsystem, (iii) we know or evaluate the excitation, $n_{exc}$, of the configuration (iv) we access to the index of the coefficient which is stored and the three component array at position $\{J_{P_{1}}^{n_{exc}},J_{P_{2}}^{n_{exc}},n_{exc}\}$. This scheme has, as a draw back, the requirement to store the list of occupation numbers and the indexes to be efficient for numerical evaluation. But the idea behind the TD-RASSCF-B method is to reduced the size of the configurational space, which makes such a storage manageable for applications done so far.      
 
 \section{Applying operators in second quantization}\label{apply_op}

In second quantization, the action of the Hamiltonian of Eq. (\ref{Hamiltonian}) on the wavefunction requires the application of annihilation and creation operators and multiplication by the matrix elements of the one- and two-body operators. To know the action of the Hamiltonian, we first need to know the action of the creation-annihilation operators \cite{Streltsov11}. Concerning the one-body term, we have, 
 \begin{equation}
  \begin{split}
b_{i}^{\dag}b_{j} &|n_{1},\cdots,n_{j},\cdots,n_{i},\cdots,n_{M}\rangle \\
&= \sqrt{n_{j}}\sqrt{n_{i}+1}|n_{1},\cdots,n_{j}-1,\cdots,n_{i}+1,\cdots,n_{M}\rangle. 
 \end{split}
 \end{equation}      
For a full-configurational wavefunction, the resulting configuration belongs to the configurational space and its index can be determined as described in Appendix \ref{wf_representation}. Now, considering the action on the wavefunction,
\begin{equation}\label{reord_conf}
b_{i}^{\dag}b_{j} \left[ \sum_{I\in {\cal V}}C_{I}|\Phi_{I}\rangle \right]= \sum_{I\in {\cal V}}C_{I}\sqrt{n_{j}}\sqrt{n_{i}+1}|\Phi_{I'}\rangle, 
\end{equation}     
with $|\Phi_{I'}\rangle$ the new configuration resulting from the action of $b_{i}^{\dag}b_{j}$ on the initial configuration $|\Phi_{I}\rangle$. The result can be interpreted as a reordering of the configuration in the wavefunction, as in Eq. (\ref{reord_conf}) or inversely to a reordering of the coefficients with a new factor ($\sqrt{n_{j}}\sqrt{n_{i}+1}$) if the configuration are reorganized in the initial order, i.e., 
\begin{equation}
b_{i}^{\dag}b_{j} \left[ \sum_{I\in {\cal V}}C_{I}|\Phi_{I}\rangle \right]= \sum_{I\in {\cal V}}C_{I'}|\Phi_{I}\rangle.
\end{equation}     
In the basis of the configurational states, $\{|\Phi_{I}\rangle\}$, the wavefunction is characterized by its coefficients only, and is stored as a vector. The new set of coefficients, $\{C'_{I}\}$, resulting from the action of $b_{i}^{\dag}b_{j}$, is obtained as,
\begin{equation}
C'_{I} = \langle \Phi_{I}| b_{i}^{\dag}b_{j} \left[ \sum_{I\in {\cal V}}C_{I}|\Phi_{I}\rangle \right].
\end{equation}   
In practice, we apply $b_{i}^{\dag}b_{j}$  on the bra $ \langle \Phi_{I}|$, which provides a new configurational state $\langle \Phi_{J}|$. The configurational vectors are orthonormal, and only the configuration $|\Phi_{J}\rangle$ in the sum remains from the projection. Thus, we evaluate the index of $\langle \Phi_{J}|$ to directly access the coefficient $C_{J}$, i.e.,
\begin{equation}
C'_{I} =C_{J}\sqrt{n_{i}}\sqrt{n_{j+1}}.
\end{equation} 
The action on the total wavefunction is obtained by repeating these operations for each configuration, providing a new coefficient vector. In the case of the RAS wavefunction, the configuration obtained from the successive application of the annihilation and creation operators may not belong to the configurational space. Nonetheless, the scheme applied above can be applied with, in addition, a test to check if the resulting configuration remains in the RAS space. This naive approach can be easily improved thanks to the representation of the wavefunction used and its indexing (see Appendix \ref{wf_representation}). The orbitals $\{i,j\}$, on which the operators $b_{i}^{\dag}b_{j}$ act, can (i) belong to ${\cal P}_{1}$ only $\{i',j'\}$, (ii) belong to ${\cal P}_{2}$ only $\{i'',j''\}$, or belong to ${\cal P}_{1}$ and ${\cal P}_{2}$, (iii) $\{i',j''\}$ and (iv) $\{i'',j'\}$. For (i) and (ii) the excitation, or the number of particle in ${\cal P}_{2}$ orbital, do not change in the resulting configuration. For the situation (iii) one particle is removed from ${\cal P}_{2}$ and added in ${\cal P}_{1}$ and the opposite happens for (iv). The excitation of the final configuration is thus known without counting the number of particles in ${\cal P}_{2}$, which is required to determine the index of the configuration. Moreover, to always remain in the RAS configurational space, the case (iv) is never applied to the configuration with the maximum excitation allowed for the general RAS scheme (Sec. \ref{all_exci}), and both (iii) and (iv) are not used for the scheme with only even excitations (Sec. \ref{even_exci}). The action of the one-body operator of the Hamiltonian is now straightforward, the coefficient vectors obtained by applying the $b_{i}^{\dag}b_{j}$ operators are multiplied by the corresponding matrix element $h_{j}^{i}$ [Eq. (\ref{one-bod})] and summed for each couple of $\{i,j\}$, with the restriction mentioned above for the RAS wavefunction. The two-body operator, see Eq. (\ref{two-bod}), included in the Hamiltonian of Eq. (\ref{Hamiltonian}) and the four- and six-order tensors specific to the RAS schemes [Eqs. (\ref{four_order_tens}) and (\ref{six_order_tens})] can be evaluated using the same strategy as the one detailed for the one-body operator. We mention that using the commutation relation for bosonic creation and annihilation operators can substantially reduced the numerical cost. For instance, if we consider the chain of operators $b^{\dag}_{i}b^{\dag}_{j}b_{k}b_{l}$, we have the equalities $b^{\dag}_{i}b^{\dag}_{j}b_{k}b_{l}=b^{\dag}_{i}b^{\dag}_{j}b_{l}b_{k}=b^{\dag}_{j}b^{\dag}_{i}b_{k}b_{l}=b^{\dag}_{i}b^{\dag}_{j}b_{l}b_{k}$.
  
 \section{Numerical implementation for the TD-RAS equations}\label{Ap_Num_imple}  
 
The EOM for the TD-RASSCF-B and F methods, Eqs. (\ref{EOM_C_general}), (\ref{EOM_Qspace}) and (\ref{eta_eq_even_final}) or (\ref{Pspace_General_RAS}), are solved to obtain the time derivative of the coefficients and orbitals. The main difference with the MCTDHB and F methods results from the evaluation of the matrix elements $\eta_{i'}^{j''}$. These elements are evaluated from Eqs. (\ref{eta_eq_even_final}) or (\ref{Pspace_General_RAS}) depending of the RAS scheme used, but both are solved in the same way. We recall that the matrix $\underline{\underline{\bm{\eta}}}$, with elements $\eta_{i'}^{j''}$, is anti-hermitian and thus $\eta_{j''}^{i'} = -(\eta_{i'}^{j''})^{*}$, and $i'$ and $j''$ hold for orbitals of the ${\cal P}_{1}$ and ${\cal P}_{2}$ subspace, respectively. Writing down Eqs. (\ref{eta_eq_even_final}) or (\ref{Pspace_General_RAS}) for any set of orbitals $\{i',j''\}$, provides a system of $M_{1}\times M_{2}$ linear equations with $M_{1}\times M_{2}$ unknowns. Introducing a composite index for the couple of $\{i',j''\}$, this system can be written in a matrix form,
\begin{equation}\label{sol_eta}
\underline{\underline{\bf{A}}} . \underline{\bf{X}}=\underline{\bf{B}},
\end{equation} 
where the matrix $\underline{\underline{\bf{A}}}$, of dimension ($M_{1}\times M_{2}$, $M_{1}\times M_{2}$), contains the values of $A_{k''i'}^{l'j''}$ or $\zeta_{k''i'}^{l'j''}$, for all set of $\{i',j''\}$ and $\{l',k''\}$. The vector $\underline{\bf{B}}$, of dimension ($M_{1}\times M_{2}$), contains the r.h.s. of Eq. (\ref{eta_eq_even_final}) or (\ref{Pspace_General_RAS}) for each set of $\{i',j''\}$ and the vector $\underline{\bf{X}}$ with the same dimension as $\underline{\bf{B}}$ contains the unknown values of $\eta_{i'}^{j''}$ and the matrix elements of the one-body operator.  The system of linear equations, Eq. (\ref{sol_eta}), can be solved using a standard numerical routine included, for instance, in the LAPACK library \cite{lapack}. The values for $\eta_{i'}^{j''}$ are trivially obtained from the elements of the vector $\underline{\bf{X}}$,  
\begin{equation}
   \eta_{i'}^{j''} = \left\{
    \begin{split}
     i(X\{i',j''\}-h_{i'}^{j''}) \ \ \ \ \text{for Eq. (\ref{eta_eq_even_final}),}\\ 
     -i(X\{i',j''\}+h_{i'}^{j''})  \ \ \ \ \text{for Eq. (\ref{Pspace_General_RAS}).} 
    \end{split}
  \right.
\end{equation}
After evaluating the matrix elements $\eta_{i'}^{j''}$, the time derivative of the coefficients $\{\dot{C}_{I}\}$ can be computed from Eq. (\ref{EOM_C_general_expand}) and the contribution of the ${\cal P}$-space orbitals to the time derivative of the orbitals is obtained from,
\begin{equation}
\hat{P}|\dot{\phi_{i}}\rangle= \sum_{j}^{M}|\phi_{j}\rangle\eta_{i}^{j}.
\end{equation} 
It remains to evaluate the contribution from the orbitals of the ${\cal Q}$-space, i.e. $\hat{Q}|\dot{\phi_{i}}\rangle$ from Eq. (\ref{EOM_Qspace}). This latter can be expressed in a matrix form,

\begin{equation}\label{mat_form_Q_space}
i\hat{Q}\underline{\underline{\bm{\rho}}}\underline{\dot{\bm{X}}}= \hat{Q}\left[ \underline{\underline{\bm{\rho}}}\underline{\bm{\tilde{h}}} + \underline{\bm{\tilde{W}}}\right], 
\end{equation} 
with $\underline{\underline{\bm{\rho}}}$ the one-body reduced density matrix, $\underline{\dot{\bm{X}}}$ a vector collecting the time derivative of the orbitals, $\underline{\bm{\tilde{h}}}$ and $\underline{\bm{\tilde{W}}}$ are both vectors with elements $h(\bm{r},t)|\phi_{i}\rangle$ and $\sum_{jlk}\hat{W}_{l}^{k}|\phi_{j}\rangle\rho_{ik}^{jl}$, respectively. To obtain Eq. (\ref{mat_form_Q_space}), we used the fact that $\underline{\underline{\bm{\rho}}}$ commutes with the projector $\hat{Q}$, as easily seen from the equality $\hat{Q}=\hat{1}-\hat{P}$. Multiplying on the left by the inverse of the one-body density matrix, $\underline{\underline{\bm{\rho}}}^{-1}$, we have, for the orbital $|\phi_{i}\rangle$, 
\begin{align}
\hat{Q}|\dot{\phi}_{i}\rangle &= -i\hat{Q}\left[ \hat{h}|\phi_{i}\rangle + \sum_{jklm} (\underline{\underline{\bm{\rho}}}^{-1})_{i}^{m} \rho_{mk}^{jl}\hat{W}_{l}^{k}|\phi_{j}\rangle\right] \nonumber\\
                                                &=-i(\hat{1}-\hat{P})\left[ \hat{h}|\phi_{i}\rangle + \sum_{jklm} (\underline{\underline{\bm{\rho}}}^{-1})_{i}^{m} \rho_{mk}^{jl}\hat{W}_{l}^{k}|\phi_{j}\rangle\right].
\end{align} 

The right hand side of the above equation is similar to the one that is solved in the MCTDH-based methods \cite{Meyer90,Alon08,Haxton11} and we follow the numerical implementation used for the MCTDH method \cite{Beck97} to avoid singularities in the inverse of the one-body reduced density matrix and in Eq. (\ref{sol_eta}) for the matrix $\underline{\underline{{\bm{A}}}}$, as well as for the projector onto the ${\cal P}$-space orbitals.   

\section{Numerical efficiency of the method}  \label{efficiency_TDRAS}

\begin{figure}[h!]
\centering 
\includegraphics[scale=0.3]{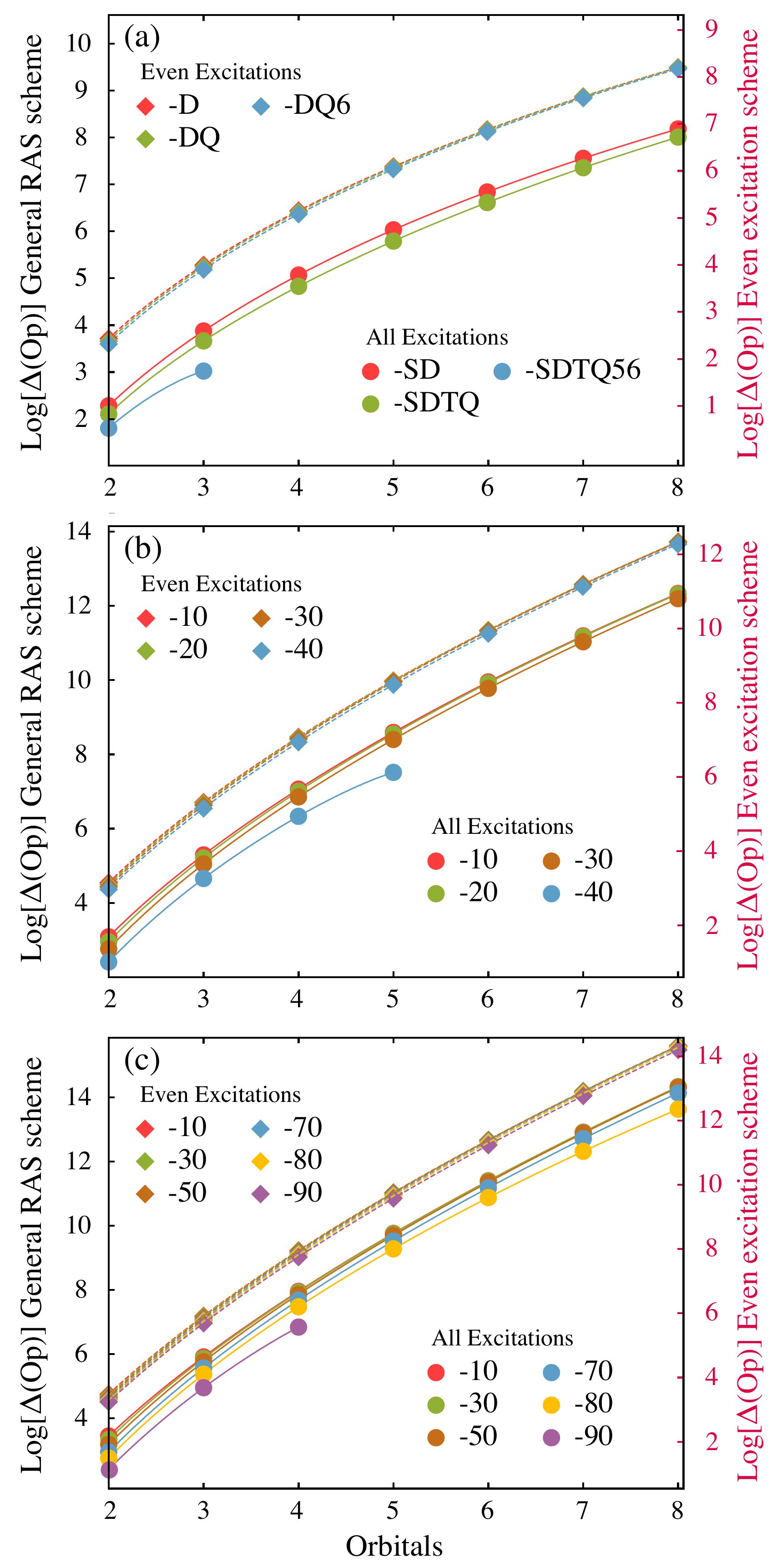}
   \caption{\small{Difference between the number of operations used to evaluate the time derivative of the wavefunction in the MCTDHB and TD-RASSCF-B method. The difference $\Delta(Op)$ is defined by Eq. (\ref{delta_Op}) in Appendix \ref{efficiency_TDRAS}. We report the value of $\Delta(Op)$ as a function of the number of orbitals for a system consisting of (a) $N=10$ particles, (b) $N=50$ particles and (c) $N=100$ particles. The diamond symbols depict the results for the case of RAS schemes with only even excitations and the number of operations is always fewer than for the MCTDHB method. The circles indicate the results for general RAS schemes, and for high excitation schemes and large numbers of orbitals number of operations can be larger than the one of the MCTDHB method. Note that the axis of the general RAS scheme is on the left while the axis of the RAS scheme with only even excitations in on the right.}}   \label{scaling_op}
 \end{figure}

Comparing the efficiency between different methods is a difficult task as it depends of the specific implementation and integration schemes used. Nonetheless, we can roughly estimate the number of operations required to evaluate the time derivative of the orbitals and coefficients and compare the MCTDHB and TD-RASSCF-B methods in this way. We denote by $N_{grid}$ the number of grid points that are used to describe the time-dependent orbital in the time-independent basis, usually a DVR \cite{Heather83}, which is the same for both methods. Starting with the MCTDHB method, at each evaluation of the time derivative the matrix elements of the two-body operator $v_{kl}^{ij}$ [Eq. (\ref{two-bod})] and the two-body reduced density matrix $\rho_{ik}^{jl}$ [see text above Eq. (\ref{EOM_Qspace})] are computed. These updates require $M^{4}N_{grid}^{2}$ and $M^{4}\mathcal{V}_{\text{FCI}}$ operations, respectively, where $\mathcal{V}_{\text{FCI}}$ is the size of the configurational space of the MCTDHB wavefunction evaluated from Eq. (\ref{dim_wf_MCTDHB}). Then computing the time derivative of the coefficients and the orbitals require $M^{4}\mathcal{V}_{\text{FCI}}$ and $M^{4}N_{grid}^{2}$ operations, respectively. The total cost is thus, approximatively, $2M^{4}(N_{grid}^{2}+\mathcal{V}_{\text{FCI}})$. Considering now the case of the TD-RASSCF-B method. The evaluation of the matrix elements of two-body operator and the calculation of the ${\cal Q}$-space equations for the time derivative of the orbitals require the same number of operations as with the MCTDHB method, i.e., $M^{4}N_{grid}^{2}$ operations for each. The evaluation of the time derivative of the coefficients and the matrix elements of the two-body reduced density matrix scale as $M^{4}\mathcal{V}$, with $\mathcal{V}$ the size of the configurational RAS space. In addition, we also need to solve the ${\cal P}$-space equations, which requires $M^{4}$ operations for excitation schemes with only even excitations and $M^{4}\mathcal{V}_{N_{\text{max}}}$ for the general RAS scheme, with $\mathcal{V}_{N_{\text{max}}}$ the number of configuration including $N_{\text{max}}$ particles in the ${\cal P}_{2}$-space. The total number of configurations included in the RAS wavefunction for the general excitation scheme is evaluated using Eq. (\ref{config_general_RAS}) and $\mathcal{V}_{N_{\text{max}}}$ is the last term of the summation. The dimension of the configurational space including only even excitations can be evaluated in a similar way,
\begin{equation}\label{config_even_RAS}
\begin{split}  
dim({\cal V})& =\begin{pmatrix}
   N+M_{1}-1  \\
   N  
\end{pmatrix} \\
&+  \sum_{k=1}^{N_{\text{max}}/2} 
\begin{pmatrix}
   2k+M_{2}-1  \\
   2k 
\end{pmatrix}
\begin{pmatrix}
   (N-2k)+M_{1}-1  \\
   N-2k
\end{pmatrix}.  
\end{split} 
\end{equation} 
Combining the results for the general RAS scheme, the number of operations required to evaluate the time derivative of the coefficients and orbitals scales as $2M^{4}(N_{grid}^2+\mathcal{V}+M^{2}\mathcal{V}_{\text{N}_{\text{max}}})$ and in the case of only even excitations it scales as $2M^{4}(N_{grid}^2+\mathcal{V}+1/2)$. To compare the numerical cost between the MCTDHB and TD-RASSCF-B methods, we can introduce $\Delta(Op)$, the difference between the MCTDHB and TD-RASSCF-B operations to remove the constant number of operation resulting from $N_{grid}$,
\begin{equation}\label{delta_Op}
\small{
   \Delta(Op)= \left\{
    \begin{split}
    & 2M^{4}(\mathcal{V}_{\text{FCI}}-\mathcal{V}-1/2)\quad \quad \quad \quad \text{\small{even excitation,}}\\ 
    & 2M^{4}(\mathcal{V}_{\text{FCI}}-\mathcal{V}-M^{2}\mathcal{V}_{\text{N}_{\text{max}}}/2)\ \  \text{\small{general scheme.}} 
    \end{split}
  \right.
  }
\end{equation}
 From the expression of $\Delta(Op)$, a positive value represents a computational gain with the TD-RASSCF-B in comparison to the MCTDHB method, while a negative value is obtained when the MCTDHB method is more efficient. In the case of a scheme with  only even excitations, $\Delta(Op)$ is proportional to the size difference of the MCTDHB and TD-RASSCF-B configurational spaces and is always positive, which means that the TD-RASSCF-B method is always more efficient. In the case of the general excitation scheme the six-order tensor of the ${\cal P}$-space equations, Eq. (\ref{six_order_tens}), can provide an overhead for the computation. To illustrate the computational efficiency we evaluate $\Delta(Op)$ for $10$, $50$ and $100$ bosons in $M=2$ to $8$ orbitals, see Fig. \ref{scaling_op}. For the TD-RASSCF-B, we consider the case of a single ${\cal P}_{1}$ orbital and $M-1$ orbitals in ${\cal P}_{2}$. The case with only even excitations reduces the computational cost  almost exponentially for increasing number of orbitals for any number of particles, which results from the efficiency of solving the ${\cal P}$-space equation. In the case of the general RAS scheme, there is always a value of $N_{\text{max}}$ which leads to more operations in the TD-RASSCF-B than in the MCTDHB method, due to the evaluation of the six-order tensor in the ${\cal P}$-space equation. But as shown in Fig. \ref{scaling_op}, this value is rather large, i.e. $N_{\text{max}}=6$ for 10, $N_{\text{max}}=40$ for 50 particles and $N_{\text{max}}=90$ for 100 particles for the schemes depicted in Figs.  \ref{scaling_op} (a), (b) and (c), respectively. \\

\bibliography{biblio}

\begin{thebibliography}{120}%
\makeatletter
\providecommand \@ifxundefined [1]{%
 \@ifx{#1\undefined}
}%
\providecommand \@ifnum [1]{%
 \ifnum #1\expandafter \@firstoftwo
 \else \expandafter \@secondoftwo
 \fi
}%
\providecommand \@ifx [1]{%
 \ifx #1\expandafter \@firstoftwo
 \else \expandafter \@secondoftwo
 \fi
}%
\providecommand \natexlab [1]{#1}%
\providecommand \enquote  [1]{``#1''}%
\providecommand \bibnamefont  [1]{#1}%
\providecommand \bibfnamefont [1]{#1}%
\providecommand \citenamefont [1]{#1}%
\providecommand \href@noop [0]{\@secondoftwo}%
\providecommand \href [0]{\begingroup \@sanitize@url \@href}%
\providecommand \@href[1]{\@@startlink{#1}\@@href}%
\providecommand \@@href[1]{\endgroup#1\@@endlink}%
\providecommand \@sanitize@url [0]{\catcode `\\12\catcode `\$12\catcode
  `\&12\catcode `\#12\catcode `\^12\catcode `\_12\catcode `\%12\relax}%
\providecommand \@@startlink[1]{}%
\providecommand \@@endlink[0]{}%
\providecommand \url  [0]{\begingroup\@sanitize@url \@url }%
\providecommand \@url [1]{\endgroup\@href {#1}{\urlprefix }}%
\providecommand \urlprefix  [0]{URL }%
\providecommand \Eprint [0]{\href }%
\providecommand \doibase [0]{http://dx.doi.org/}%
\providecommand \selectlanguage [0]{\@gobble}%
\providecommand \bibinfo  [0]{\@secondoftwo}%
\providecommand \bibfield  [0]{\@secondoftwo}%
\providecommand \translation [1]{[#1]}%
\providecommand \BibitemOpen [0]{}%
\providecommand \bibitemStop [0]{}%
\providecommand \bibitemNoStop [0]{.\EOS\space}%
\providecommand \EOS [0]{\spacefactor3000\relax}%
\providecommand \BibitemShut  [1]{\csname bibitem#1\endcsname}%
\let\auto@bib@innerbib\@empty
\bibitem [{\citenamefont {Anderson}\ \emph {et~al.}(1995)\citenamefont
  {Anderson}, \citenamefont {Ensher}, \citenamefont {Matthews}, \citenamefont
  {Wieman},\ and\ \citenamefont {Cornell}}]{Anderson95}%
  \BibitemOpen
  \bibfield  {author} {\bibinfo {author} {\bibfnamefont {M.~H.}\ \bibnamefont
  {Anderson}}, \bibinfo {author} {\bibfnamefont {J.~R.}\ \bibnamefont
  {Ensher}}, \bibinfo {author} {\bibfnamefont {M.~R.}\ \bibnamefont
  {Matthews}}, \bibinfo {author} {\bibfnamefont {C.~E.}\ \bibnamefont
  {Wieman}}, \ and\ \bibinfo {author} {\bibfnamefont {E.~A.}\ \bibnamefont
  {Cornell}},\ }\href {\doibase 10.1126/science.269.5221.198} {\bibfield
  {journal} {\bibinfo  {journal} {Science}\ }\textbf {\bibinfo {volume}
  {269}},\ \bibinfo {pages} {198} (\bibinfo {year} {1995})}\BibitemShut
  {NoStop}%
\bibitem [{\citenamefont {Bradley}\ \emph {et~al.}(1995)\citenamefont
  {Bradley}, \citenamefont {Sackett}, \citenamefont {Tollett},\ and\
  \citenamefont {Hulet}}]{Bradley95}%
  \BibitemOpen
  \bibfield  {author} {\bibinfo {author} {\bibfnamefont {C.~C.}\ \bibnamefont
  {Bradley}}, \bibinfo {author} {\bibfnamefont {C.~A.}\ \bibnamefont
  {Sackett}}, \bibinfo {author} {\bibfnamefont {J.~J.}\ \bibnamefont
  {Tollett}}, \ and\ \bibinfo {author} {\bibfnamefont {R.~G.}\ \bibnamefont
  {Hulet}},\ }\href {\doibase 10.1103/PhysRevLett.75.1687} {\bibfield
  {journal} {\bibinfo  {journal} {Phys. Rev. Lett.}\ }\textbf {\bibinfo
  {volume} {75}},\ \bibinfo {pages} {1687} (\bibinfo {year}
  {1995})}\BibitemShut {NoStop}%
\bibitem [{\citenamefont {Davis}\ \emph {et~al.}(1995)\citenamefont {Davis},
  \citenamefont {Mewes}, \citenamefont {Andrews}, \citenamefont {van Druten},
  \citenamefont {Durfee}, \citenamefont {Kurn},\ and\ \citenamefont
  {Ketterle}}]{Davis95}%
  \BibitemOpen
  \bibfield  {author} {\bibinfo {author} {\bibfnamefont {K.~B.}\ \bibnamefont
  {Davis}}, \bibinfo {author} {\bibfnamefont {M.~O.}\ \bibnamefont {Mewes}},
  \bibinfo {author} {\bibfnamefont {M.~R.}\ \bibnamefont {Andrews}}, \bibinfo
  {author} {\bibfnamefont {N.~J.}\ \bibnamefont {van Druten}}, \bibinfo
  {author} {\bibfnamefont {D.~S.}\ \bibnamefont {Durfee}}, \bibinfo {author}
  {\bibfnamefont {D.~M.}\ \bibnamefont {Kurn}}, \ and\ \bibinfo {author}
  {\bibfnamefont {W.}~\bibnamefont {Ketterle}},\ }\href {\doibase
  10.1103/PhysRevLett.75.3969} {\bibfield  {journal} {\bibinfo  {journal}
  {Phys. Rev. Lett.}\ }\textbf {\bibinfo {volume} {75}},\ \bibinfo {pages}
  {3969} (\bibinfo {year} {1995})}\BibitemShut {NoStop}%
\bibitem [{\citenamefont {Preiss}\ \emph {et~al.}(2015)\citenamefont {Preiss},
  \citenamefont {Ma}, \citenamefont {Tai}, \citenamefont {Lukin}, \citenamefont
  {Rispoli}, \citenamefont {Zupancic}, \citenamefont {Lahini}, \citenamefont
  {Islam},\ and\ \citenamefont {Greiner}}]{Preiss15}%
  \BibitemOpen
  \bibfield  {author} {\bibinfo {author} {\bibfnamefont {P.~M.}\ \bibnamefont
  {Preiss}}, \bibinfo {author} {\bibfnamefont {R.}~\bibnamefont {Ma}}, \bibinfo
  {author} {\bibfnamefont {M.~E.}\ \bibnamefont {Tai}}, \bibinfo {author}
  {\bibfnamefont {A.}~\bibnamefont {Lukin}}, \bibinfo {author} {\bibfnamefont
  {M.}~\bibnamefont {Rispoli}}, \bibinfo {author} {\bibfnamefont
  {P.}~\bibnamefont {Zupancic}}, \bibinfo {author} {\bibfnamefont
  {Y.}~\bibnamefont {Lahini}}, \bibinfo {author} {\bibfnamefont
  {R.}~\bibnamefont {Islam}}, \ and\ \bibinfo {author} {\bibfnamefont
  {M.}~\bibnamefont {Greiner}},\ }\href {\doibase 10.1126/science.1260364}
  {\bibfield  {journal} {\bibinfo  {journal} {Science}\ }\textbf {\bibinfo
  {volume} {347}},\ \bibinfo {pages} {1229} (\bibinfo {year}
  {2015})}\BibitemShut {NoStop}%
\bibitem [{\citenamefont {Kaufman}\ \emph {et~al.}(2014)\citenamefont
  {Kaufman}, \citenamefont {Lester}, \citenamefont {Reynolds}, \citenamefont
  {Wall}, \citenamefont {Foss-Feig}, \citenamefont {Hazzard}, \citenamefont
  {Rey},\ and\ \citenamefont {Regal}}]{Kaufman14}%
  \BibitemOpen
  \bibfield  {author} {\bibinfo {author} {\bibfnamefont {A.~M.}\ \bibnamefont
  {Kaufman}}, \bibinfo {author} {\bibfnamefont {B.~J.}\ \bibnamefont {Lester}},
  \bibinfo {author} {\bibfnamefont {C.~M.}\ \bibnamefont {Reynolds}}, \bibinfo
  {author} {\bibfnamefont {M.~L.}\ \bibnamefont {Wall}}, \bibinfo {author}
  {\bibfnamefont {M.}~\bibnamefont {Foss-Feig}}, \bibinfo {author}
  {\bibfnamefont {K.~R.~A.}\ \bibnamefont {Hazzard}}, \bibinfo {author}
  {\bibfnamefont {A.~M.}\ \bibnamefont {Rey}}, \ and\ \bibinfo {author}
  {\bibfnamefont {C.~A.}\ \bibnamefont {Regal}},\ }\href {\doibase
  10.1126/science.1250057} {\bibfield  {journal} {\bibinfo  {journal}
  {Science}\ }\textbf {\bibinfo {volume} {345}},\ \bibinfo {pages} {306}
  (\bibinfo {year} {2014})}\BibitemShut {NoStop}%
\bibitem [{\citenamefont {Bloch}(2005)}]{Bloch05}%
  \BibitemOpen
  \bibfield  {author} {\bibinfo {author} {\bibfnamefont {I.}~\bibnamefont
  {Bloch}},\ }\href {\doibase 10.1038/nphys138} {\bibfield  {journal} {\bibinfo
   {journal} {Nat Phys}\ }\textbf {\bibinfo {volume} {1}},\ \bibinfo {pages}
  {23} (\bibinfo {year} {2005})}\BibitemShut {NoStop}%
\bibitem [{\citenamefont {Jotzu}\ \emph {et~al.}(2014)\citenamefont {Jotzu},
  \citenamefont {Messer}, \citenamefont {Desbuquois}, \citenamefont {Lebrat},
  \citenamefont {Uehlinger}, \citenamefont {Greif},\ and\ \citenamefont
  {Esslinger}}]{Jotzu14}%
  \BibitemOpen
  \bibfield  {author} {\bibinfo {author} {\bibfnamefont {G.}~\bibnamefont
  {Jotzu}}, \bibinfo {author} {\bibfnamefont {M.}~\bibnamefont {Messer}},
  \bibinfo {author} {\bibfnamefont {R.}~\bibnamefont {Desbuquois}}, \bibinfo
  {author} {\bibfnamefont {M.}~\bibnamefont {Lebrat}}, \bibinfo {author}
  {\bibfnamefont {T.}~\bibnamefont {Uehlinger}}, \bibinfo {author}
  {\bibfnamefont {D.}~\bibnamefont {Greif}}, \ and\ \bibinfo {author}
  {\bibfnamefont {T.}~\bibnamefont {Esslinger}},\ }\href
  {http://dx.doi.org/10.1038/nature13915} {\bibfield  {journal} {\bibinfo
  {journal} {Nature}\ }\textbf {\bibinfo {volume} {515}},\ \bibinfo {pages}
  {237} (\bibinfo {year} {2014})}\BibitemShut {NoStop}%
\bibitem [{\citenamefont {Greiner}\ \emph {et~al.}(2002)\citenamefont
  {Greiner}, \citenamefont {Mandel}, \citenamefont {Esslinger}, \citenamefont
  {Hansch},\ and\ \citenamefont {Bloch}}]{Greiner02}%
  \BibitemOpen
  \bibfield  {author} {\bibinfo {author} {\bibfnamefont {M.}~\bibnamefont
  {Greiner}}, \bibinfo {author} {\bibfnamefont {O.}~\bibnamefont {Mandel}},
  \bibinfo {author} {\bibfnamefont {T.}~\bibnamefont {Esslinger}}, \bibinfo
  {author} {\bibfnamefont {T.~W.}\ \bibnamefont {Hansch}}, \ and\ \bibinfo
  {author} {\bibfnamefont {I.}~\bibnamefont {Bloch}},\ }\href {\doibase
  10.1038/415039a} {\bibfield  {journal} {\bibinfo  {journal} {Nature}\
  }\textbf {\bibinfo {volume} {415}},\ \bibinfo {pages} {39} (\bibinfo {year}
  {2002})}\BibitemShut {NoStop}%
\bibitem [{\citenamefont {Courteille}\ \emph {et~al.}(1998)\citenamefont
  {Courteille}, \citenamefont {Freeland}, \citenamefont {Heinzen},
  \citenamefont {van Abeelen},\ and\ \citenamefont {Verhaar}}]{Courteille98}%
  \BibitemOpen
  \bibfield  {author} {\bibinfo {author} {\bibfnamefont {P.}~\bibnamefont
  {Courteille}}, \bibinfo {author} {\bibfnamefont {R.~S.}\ \bibnamefont
  {Freeland}}, \bibinfo {author} {\bibfnamefont {D.~J.}\ \bibnamefont
  {Heinzen}}, \bibinfo {author} {\bibfnamefont {F.~A.}\ \bibnamefont {van
  Abeelen}}, \ and\ \bibinfo {author} {\bibfnamefont {B.~J.}\ \bibnamefont
  {Verhaar}},\ }\href {\doibase 10.1103/PhysRevLett.81.69} {\bibfield
  {journal} {\bibinfo  {journal} {Phys. Rev. Lett.}\ }\textbf {\bibinfo
  {volume} {81}},\ \bibinfo {pages} {69} (\bibinfo {year} {1998})}\BibitemShut
  {NoStop}%
\bibitem [{\citenamefont {Inouye}\ \emph {et~al.}(1998)\citenamefont {Inouye},
  \citenamefont {Andrews}, \citenamefont {Stenger}, \citenamefont {Miesner},
  \citenamefont {Stamper-Kurn},\ and\ \citenamefont {Ketterle}}]{Inouye98}%
  \BibitemOpen
  \bibfield  {author} {\bibinfo {author} {\bibfnamefont {S.}~\bibnamefont
  {Inouye}}, \bibinfo {author} {\bibfnamefont {M.~R.}\ \bibnamefont {Andrews}},
  \bibinfo {author} {\bibfnamefont {J.}~\bibnamefont {Stenger}}, \bibinfo
  {author} {\bibfnamefont {H.-J.}\ \bibnamefont {Miesner}}, \bibinfo {author}
  {\bibfnamefont {D.~M.}\ \bibnamefont {Stamper-Kurn}}, \ and\ \bibinfo
  {author} {\bibfnamefont {W.}~\bibnamefont {Ketterle}},\ }\href {\doibase
  10.1038/32354} {\bibfield  {journal} {\bibinfo  {journal} {Nature}\ }\textbf
  {\bibinfo {volume} {392}},\ \bibinfo {pages} {151} (\bibinfo {year}
  {1998})}\BibitemShut {NoStop}%
\bibitem [{\citenamefont {Bloch}\ \emph {et~al.}(2012)\citenamefont {Bloch},
  \citenamefont {Dalibard},\ and\ \citenamefont {Nascimbene}}]{Bloch12}%
  \BibitemOpen
  \bibfield  {author} {\bibinfo {author} {\bibfnamefont {I.}~\bibnamefont
  {Bloch}}, \bibinfo {author} {\bibfnamefont {J.}~\bibnamefont {Dalibard}}, \
  and\ \bibinfo {author} {\bibfnamefont {S.}~\bibnamefont {Nascimbene}},\
  }\href {\doibase 10.1038/nphys2259} {\bibfield  {journal} {\bibinfo
  {journal} {Nat Phys}\ }\textbf {\bibinfo {volume} {8}},\ \bibinfo {pages}
  {267} (\bibinfo {year} {2012})}\BibitemShut {NoStop}%
\bibitem [{\citenamefont {Anderson}\ and\ \citenamefont
  {Kasevich}(1998)}]{Anderson98}%
  \BibitemOpen
  \bibfield  {author} {\bibinfo {author} {\bibfnamefont {B.~P.}\ \bibnamefont
  {Anderson}}\ and\ \bibinfo {author} {\bibfnamefont {M.~A.}\ \bibnamefont
  {Kasevich}},\ }\href {\doibase 10.1126/science.282.5394.1686} {\bibfield
  {journal} {\bibinfo  {journal} {Science}\ }\textbf {\bibinfo {volume}
  {282}},\ \bibinfo {pages} {1686} (\bibinfo {year} {1998})}\BibitemShut
  {NoStop}%
\bibitem [{\citenamefont {Steinhauer}(2016)}]{Steinhauer16}%
  \BibitemOpen
  \bibfield  {author} {\bibinfo {author} {\bibfnamefont {J.}~\bibnamefont
  {Steinhauer}},\ }\href {http://dx.doi.org/10.1038/nphys3863} {\bibfield
  {journal} {\bibinfo  {journal} {Nat Phys}\ }\textbf {\bibinfo {volume}
  {12}},\ \bibinfo {pages} {959} (\bibinfo {year} {2016})}\BibitemShut
  {NoStop}%
\bibitem [{\citenamefont {Sala}\ \emph {et~al.}(2013)\citenamefont {Sala},
  \citenamefont {F\"orster},\ and\ \citenamefont {Saenz}}]{Sala13}%
  \BibitemOpen
  \bibfield  {author} {\bibinfo {author} {\bibfnamefont {S.}~\bibnamefont
  {Sala}}, \bibinfo {author} {\bibfnamefont {J.}~\bibnamefont {F\"orster}}, \
  and\ \bibinfo {author} {\bibfnamefont {A.}~\bibnamefont {Saenz}},\ }\href
  {https://arxiv.org/abs/1311.2304} {\bibfield  {journal} {\bibinfo  {journal}
  {arXiv:1311.2304}\ } (\bibinfo {year} {2013})}\BibitemShut {NoStop}%
\bibitem [{\citenamefont {L\"uhmann}\ \emph {et~al.}(2015)\citenamefont
  {L\"uhmann}, \citenamefont {Weitenberg},\ and\ \citenamefont
  {Sengstock}}]{Sengstock15}%
  \BibitemOpen
  \bibfield  {author} {\bibinfo {author} {\bibfnamefont {D.-S.}\ \bibnamefont
  {L\"uhmann}}, \bibinfo {author} {\bibfnamefont {C.}~\bibnamefont
  {Weitenberg}}, \ and\ \bibinfo {author} {\bibfnamefont {K.}~\bibnamefont
  {Sengstock}},\ }\href {\doibase 10.1103/PhysRevX.5.031016} {\bibfield
  {journal} {\bibinfo  {journal} {Phys. Rev. X}\ }\textbf {\bibinfo {volume}
  {5}},\ \bibinfo {pages} {031016} (\bibinfo {year} {2015})}\BibitemShut
  {NoStop}%
\bibitem [{\citenamefont {Bloch}\ \emph {et~al.}(2008)\citenamefont {Bloch},
  \citenamefont {Dalibard},\ and\ \citenamefont {Zwerger}}]{Bloch08}%
  \BibitemOpen
  \bibfield  {author} {\bibinfo {author} {\bibfnamefont {I.}~\bibnamefont
  {Bloch}}, \bibinfo {author} {\bibfnamefont {J.}~\bibnamefont {Dalibard}}, \
  and\ \bibinfo {author} {\bibfnamefont {W.}~\bibnamefont {Zwerger}},\ }\href
  {\doibase 10.1103/RevModPhys.80.885} {\bibfield  {journal} {\bibinfo
  {journal} {Rev. Mod. Phys.}\ }\textbf {\bibinfo {volume} {80}},\ \bibinfo
  {pages} {885} (\bibinfo {year} {2008})}\BibitemShut {NoStop}%
\bibitem [{\citenamefont {Lieb}\ and\ \citenamefont
  {Liniger}(1963)}]{Liniger63}%
  \BibitemOpen
  \bibfield  {author} {\bibinfo {author} {\bibfnamefont {E.~H.}\ \bibnamefont
  {Lieb}}\ and\ \bibinfo {author} {\bibfnamefont {W.}~\bibnamefont {Liniger}},\
  }\href {\doibase 10.1103/PhysRev.130.1605} {\bibfield  {journal} {\bibinfo
  {journal} {Phys. Rev.}\ }\textbf {\bibinfo {volume} {130}},\ \bibinfo {pages}
  {1605} (\bibinfo {year} {1963})}\BibitemShut {NoStop}%
\bibitem [{\citenamefont {Lieb}(1963)}]{Lieb63}%
  \BibitemOpen
  \bibfield  {author} {\bibinfo {author} {\bibfnamefont {E.~H.}\ \bibnamefont
  {Lieb}},\ }\href {\doibase 10.1103/PhysRev.130.1616} {\bibfield  {journal}
  {\bibinfo  {journal} {Phys. Rev.}\ }\textbf {\bibinfo {volume} {130}},\
  \bibinfo {pages} {1616} (\bibinfo {year} {1963})}\BibitemShut {NoStop}%
\bibitem [{\citenamefont {Girardeau}(1960)}]{Girardeau60}%
  \BibitemOpen
  \bibfield  {author} {\bibinfo {author} {\bibfnamefont {M.}~\bibnamefont
  {Girardeau}},\ }\href {\doibase http://dx.doi.org/10.1063/1.1703687}
  {\bibfield  {journal} {\bibinfo  {journal} {Journal of Mathematical Physics}\
  }\textbf {\bibinfo {volume} {1}},\ \bibinfo {pages} {516} (\bibinfo {year}
  {1960})}\BibitemShut {NoStop}%
\bibitem [{\citenamefont {Yukalov}\ and\ \citenamefont
  {Girardeau}(2005)}]{Yukalov05}%
  \BibitemOpen
  \bibfield  {author} {\bibinfo {author} {\bibfnamefont {V.~I.}\ \bibnamefont
  {Yukalov}}\ and\ \bibinfo {author} {\bibfnamefont {M.~D.}\ \bibnamefont
  {Girardeau}},\ }\href {http://stacks.iop.org/1612-202X/2/i=8/a=001}
  {\bibfield  {journal} {\bibinfo  {journal} {Laser Physics Letters}\ }\textbf
  {\bibinfo {volume} {2}},\ \bibinfo {pages} {375} (\bibinfo {year}
  {2005})}\BibitemShut {NoStop}%
\bibitem [{\citenamefont {Calogero}(1969)}]{Calogero69}%
  \BibitemOpen
  \bibfield  {author} {\bibinfo {author} {\bibfnamefont {F.}~\bibnamefont
  {Calogero}},\ }\href {\doibase http://dx.doi.org/10.1063/1.1664820}
  {\bibfield  {journal} {\bibinfo  {journal} {Journal of Mathematical Physics}\
  }\textbf {\bibinfo {volume} {10}},\ \bibinfo {pages} {2191} (\bibinfo {year}
  {1969})}\BibitemShut {NoStop}%
\bibitem [{\citenamefont {Cohen}\ and\ \citenamefont {Lee}(1985)}]{Cohen85}%
  \BibitemOpen
  \bibfield  {author} {\bibinfo {author} {\bibfnamefont {L.}~\bibnamefont
  {Cohen}}\ and\ \bibinfo {author} {\bibfnamefont {C.}~\bibnamefont {Lee}},\
  }\href {\doibase http://dx.doi.org/10.1063/1.526688} {\bibfield  {journal}
  {\bibinfo  {journal} {Journal of Mathematical Physics}\ }\textbf {\bibinfo
  {volume} {26}},\ \bibinfo {pages} {3105} (\bibinfo {year}
  {1985})}\BibitemShut {NoStop}%
\bibitem [{\citenamefont {Yan}(2003)}]{Yan03}%
  \BibitemOpen
  \bibfield  {author} {\bibinfo {author} {\bibfnamefont {J.}~\bibnamefont
  {Yan}},\ }\href {\doibase 10.1023/A:1026029104217} {\bibfield  {journal}
  {\bibinfo  {journal} {Journal of Statistical Physics}\ }\textbf {\bibinfo
  {volume} {113}},\ \bibinfo {pages} {623} (\bibinfo {year}
  {2003})}\BibitemShut {NoStop}%
\bibitem [{\citenamefont {Lode}\ \emph
  {et~al.}(2012{\natexlab{a}})\citenamefont {Lode}, \citenamefont {Sakmann},
  \citenamefont {Alon}, \citenamefont {Cederbaum},\ and\ \citenamefont
  {Streltsov}}]{Lode12}%
  \BibitemOpen
  \bibfield  {author} {\bibinfo {author} {\bibfnamefont {A.~U.~J.}\
  \bibnamefont {Lode}}, \bibinfo {author} {\bibfnamefont {K.}~\bibnamefont
  {Sakmann}}, \bibinfo {author} {\bibfnamefont {O.~E.}\ \bibnamefont {Alon}},
  \bibinfo {author} {\bibfnamefont {L.~S.}\ \bibnamefont {Cederbaum}}, \ and\
  \bibinfo {author} {\bibfnamefont {A.~I.}\ \bibnamefont {Streltsov}},\ }\href
  {\doibase 10.1103/PhysRevA.86.063606} {\bibfield  {journal} {\bibinfo
  {journal} {Phys. Rev. A}\ }\textbf {\bibinfo {volume} {86}},\ \bibinfo
  {pages} {063606} (\bibinfo {year} {2012}{\natexlab{a}})}\BibitemShut
  {NoStop}%
\bibitem [{\citenamefont {Baym}\ and\ \citenamefont {Pethick}(1996)}]{Baym96}%
  \BibitemOpen
  \bibfield  {author} {\bibinfo {author} {\bibfnamefont {G.}~\bibnamefont
  {Baym}}\ and\ \bibinfo {author} {\bibfnamefont {C.~J.}\ \bibnamefont
  {Pethick}},\ }\href {\doibase 10.1103/PhysRevLett.76.6} {\bibfield  {journal}
  {\bibinfo  {journal} {Phys. Rev. Lett.}\ }\textbf {\bibinfo {volume} {76}},\
  \bibinfo {pages} {6} (\bibinfo {year} {1996})}\BibitemShut {NoStop}%
\bibitem [{\citenamefont {Bogoliubov}(1947)}]{Bogoliubov47}%
  \BibitemOpen
  \bibfield  {author} {\bibinfo {author} {\bibfnamefont {N.~N.}\ \bibnamefont
  {Bogoliubov}},\ }\href@noop {} {\bibfield  {journal} {\bibinfo  {journal}
  {Journal of Physics}\ }\textbf {\bibinfo {volume} {11}},\ \bibinfo {pages}
  {23} (\bibinfo {year} {1947})}\BibitemShut {NoStop}%
\bibitem [{\citenamefont {Lee}\ and\ \citenamefont {Yang}(1957)}]{Lee57}%
  \BibitemOpen
  \bibfield  {author} {\bibinfo {author} {\bibfnamefont {T.~D.}\ \bibnamefont
  {Lee}}\ and\ \bibinfo {author} {\bibfnamefont {C.~N.}\ \bibnamefont {Yang}},\
  }\href {\doibase 10.1103/PhysRev.105.1119} {\bibfield  {journal} {\bibinfo
  {journal} {Phys. Rev.}\ }\textbf {\bibinfo {volume} {105}},\ \bibinfo {pages}
  {1119} (\bibinfo {year} {1957})}\BibitemShut {NoStop}%
\bibitem [{\citenamefont {Lee}\ \emph {et~al.}(1957)\citenamefont {Lee},
  \citenamefont {Huang},\ and\ \citenamefont {Yang}}]{Lee57_2}%
  \BibitemOpen
  \bibfield  {author} {\bibinfo {author} {\bibfnamefont {T.~D.}\ \bibnamefont
  {Lee}}, \bibinfo {author} {\bibfnamefont {K.}~\bibnamefont {Huang}}, \ and\
  \bibinfo {author} {\bibfnamefont {C.~N.}\ \bibnamefont {Yang}},\ }\href
  {\doibase 10.1103/PhysRev.106.1135} {\bibfield  {journal} {\bibinfo
  {journal} {Phys. Rev.}\ }\textbf {\bibinfo {volume} {106}},\ \bibinfo {pages}
  {1135} (\bibinfo {year} {1957})}\BibitemShut {NoStop}%
\bibitem [{\citenamefont {Fisher}\ \emph {et~al.}(1989)\citenamefont {Fisher},
  \citenamefont {Weichman}, \citenamefont {Grinstein},\ and\ \citenamefont
  {Fisher}}]{Fisher89}%
  \BibitemOpen
  \bibfield  {author} {\bibinfo {author} {\bibfnamefont {M.~P.~A.}\
  \bibnamefont {Fisher}}, \bibinfo {author} {\bibfnamefont {P.~B.}\
  \bibnamefont {Weichman}}, \bibinfo {author} {\bibfnamefont {G.}~\bibnamefont
  {Grinstein}}, \ and\ \bibinfo {author} {\bibfnamefont {D.~S.}\ \bibnamefont
  {Fisher}},\ }\href {\doibase 10.1103/PhysRevB.40.546} {\bibfield  {journal}
  {\bibinfo  {journal} {Phys. Rev. B}\ }\textbf {\bibinfo {volume} {40}},\
  \bibinfo {pages} {546} (\bibinfo {year} {1989})}\BibitemShut {NoStop}%
\bibitem [{\citenamefont {Dutta}\ \emph {et~al.}(2015)\citenamefont {Dutta},
  \citenamefont {Gajda}, \citenamefont {Hauke}, \citenamefont {Lewenstein},
  \citenamefont {L\"uhmann}, \citenamefont {Malomed}, \citenamefont
  {Sowi\'nski},\ and\ \citenamefont {Zakrzewski}}]{Dutta15}%
  \BibitemOpen
  \bibfield  {author} {\bibinfo {author} {\bibfnamefont {O.}~\bibnamefont
  {Dutta}}, \bibinfo {author} {\bibfnamefont {M.}~\bibnamefont {Gajda}},
  \bibinfo {author} {\bibfnamefont {P.}~\bibnamefont {Hauke}}, \bibinfo
  {author} {\bibfnamefont {M.}~\bibnamefont {Lewenstein}}, \bibinfo {author}
  {\bibfnamefont {D.-S.}\ \bibnamefont {L\"uhmann}}, \bibinfo {author}
  {\bibfnamefont {B.~A.}\ \bibnamefont {Malomed}}, \bibinfo {author}
  {\bibfnamefont {T.}~\bibnamefont {Sowi\'nski}}, \ and\ \bibinfo {author}
  {\bibfnamefont {J.}~\bibnamefont {Zakrzewski}},\ }\href
  {http://stacks.iop.org/0034-4885/78/i=6/a=066001} {\bibfield  {journal}
  {\bibinfo  {journal} {Reports on Progress in Physics}\ }\textbf {\bibinfo
  {volume} {78}},\ \bibinfo {pages} {066001} (\bibinfo {year}
  {2015})}\BibitemShut {NoStop}%
\bibitem [{\citenamefont {White}\ and\ \citenamefont {Noack}(1992)}]{White92}%
  \BibitemOpen
  \bibfield  {author} {\bibinfo {author} {\bibfnamefont {S.~R.}\ \bibnamefont
  {White}}\ and\ \bibinfo {author} {\bibfnamefont {R.~M.}\ \bibnamefont
  {Noack}},\ }\href {\doibase 10.1103/PhysRevLett.68.3487} {\bibfield
  {journal} {\bibinfo  {journal} {Phys. Rev. Lett.}\ }\textbf {\bibinfo
  {volume} {68}},\ \bibinfo {pages} {3487} (\bibinfo {year}
  {1992})}\BibitemShut {NoStop}%
\bibitem [{\citenamefont {White}(1992)}]{White92_2}%
  \BibitemOpen
  \bibfield  {author} {\bibinfo {author} {\bibfnamefont {S.~R.}\ \bibnamefont
  {White}},\ }\href {\doibase 10.1103/PhysRevLett.69.2863} {\bibfield
  {journal} {\bibinfo  {journal} {Phys. Rev. Lett.}\ }\textbf {\bibinfo
  {volume} {69}},\ \bibinfo {pages} {2863} (\bibinfo {year}
  {1992})}\BibitemShut {NoStop}%
\bibitem [{\citenamefont {White}(1993)}]{White93}%
  \BibitemOpen
  \bibfield  {author} {\bibinfo {author} {\bibfnamefont {S.~R.}\ \bibnamefont
  {White}},\ }\href {\doibase 10.1103/PhysRevB.48.10345} {\bibfield  {journal}
  {\bibinfo  {journal} {Phys. Rev. B}\ }\textbf {\bibinfo {volume} {48}},\
  \bibinfo {pages} {10345} (\bibinfo {year} {1993})}\BibitemShut {NoStop}%
\bibitem [{\citenamefont {Schollw\"ock}(2005)}]{Schollwock05}%
  \BibitemOpen
  \bibfield  {author} {\bibinfo {author} {\bibfnamefont {U.}~\bibnamefont
  {Schollw\"ock}},\ }\href {\doibase 10.1103/RevModPhys.77.259} {\bibfield
  {journal} {\bibinfo  {journal} {Rev. Mod. Phys.}\ }\textbf {\bibinfo {volume}
  {77}},\ \bibinfo {pages} {259} (\bibinfo {year} {2005})}\BibitemShut
  {NoStop}%
\bibitem [{\citenamefont {Vidal}(2004)}]{Vidal04}%
  \BibitemOpen
  \bibfield  {author} {\bibinfo {author} {\bibfnamefont {G.}~\bibnamefont
  {Vidal}},\ }\href {\doibase 10.1103/PhysRevLett.93.040502} {\bibfield
  {journal} {\bibinfo  {journal} {Phys. Rev. Lett.}\ }\textbf {\bibinfo
  {volume} {93}},\ \bibinfo {pages} {040502} (\bibinfo {year}
  {2004})}\BibitemShut {NoStop}%
\bibitem [{\citenamefont {Daley}\ \emph {et~al.}(2004)\citenamefont {Daley},
  \citenamefont {Kollath}, \citenamefont {Schollw\"ock},\ and\ \citenamefont
  {Vidal}}]{Daley04}%
  \BibitemOpen
  \bibfield  {author} {\bibinfo {author} {\bibfnamefont {A.~J.}\ \bibnamefont
  {Daley}}, \bibinfo {author} {\bibfnamefont {C.}~\bibnamefont {Kollath}},
  \bibinfo {author} {\bibfnamefont {U.}~\bibnamefont {Schollw\"ock}}, \ and\
  \bibinfo {author} {\bibfnamefont {G.}~\bibnamefont {Vidal}},\ }\href
  {http://stacks.iop.org/1742-5468/2004/i=04/a=P04005} {\bibfield  {journal}
  {\bibinfo  {journal} {Journal of Statistical Mechanics: Theory and
  Experiment}\ }\textbf {\bibinfo {volume} {2004}},\ \bibinfo {pages} {P04005}
  (\bibinfo {year} {2004})}\BibitemShut {NoStop}%
\bibitem [{\citenamefont {White}\ and\ \citenamefont
  {Feiguin}(2004)}]{White04}%
  \BibitemOpen
  \bibfield  {author} {\bibinfo {author} {\bibfnamefont {S.~R.}\ \bibnamefont
  {White}}\ and\ \bibinfo {author} {\bibfnamefont {A.~E.}\ \bibnamefont
  {Feiguin}},\ }\href {\doibase 10.1103/PhysRevLett.93.076401} {\bibfield
  {journal} {\bibinfo  {journal} {Phys. Rev. Lett.}\ }\textbf {\bibinfo
  {volume} {93}},\ \bibinfo {pages} {076401} (\bibinfo {year}
  {2004})}\BibitemShut {NoStop}%
\bibitem [{\citenamefont {McMillan}(1965)}]{McMillan65}%
  \BibitemOpen
  \bibfield  {author} {\bibinfo {author} {\bibfnamefont {W.~L.}\ \bibnamefont
  {McMillan}},\ }\href {\doibase 10.1103/PhysRev.138.A442} {\bibfield
  {journal} {\bibinfo  {journal} {Phys. Rev.}\ }\textbf {\bibinfo {volume}
  {138}},\ \bibinfo {pages} {A442} (\bibinfo {year} {1965})}\BibitemShut
  {NoStop}%
\bibitem [{\citenamefont {Anderson}(1975)}]{Anderson75}%
  \BibitemOpen
  \bibfield  {author} {\bibinfo {author} {\bibfnamefont {J.~B.}\ \bibnamefont
  {Anderson}},\ }\href {\doibase http://dx.doi.org/10.1063/1.431514} {\bibfield
   {journal} {\bibinfo  {journal} {The Journal of Chemical Physics}\ }\textbf
  {\bibinfo {volume} {63}},\ \bibinfo {pages} {1499} (\bibinfo {year}
  {1975})}\BibitemShut {NoStop}%
\bibitem [{\citenamefont {Reynolds}\ \emph {et~al.}(1982)\citenamefont
  {Reynolds}, \citenamefont {Ceperley}, \citenamefont {Alder},\ and\
  \citenamefont {Lester}}]{Reynolds82}%
  \BibitemOpen
  \bibfield  {author} {\bibinfo {author} {\bibfnamefont {P.~J.}\ \bibnamefont
  {Reynolds}}, \bibinfo {author} {\bibfnamefont {D.~M.}\ \bibnamefont
  {Ceperley}}, \bibinfo {author} {\bibfnamefont {B.~J.}\ \bibnamefont {Alder}},
  \ and\ \bibinfo {author} {\bibfnamefont {W.~A.}\ \bibnamefont {Lester}},\
  }\href {\doibase http://dx.doi.org/10.1063/1.443766} {\bibfield  {journal}
  {\bibinfo  {journal} {The Journal of Chemical Physics}\ }\textbf {\bibinfo
  {volume} {77}},\ \bibinfo {pages} {5593} (\bibinfo {year}
  {1982})}\BibitemShut {NoStop}%
\bibitem [{\citenamefont {Blume}\ and\ \citenamefont
  {Greene}(2001)}]{Blume2001}%
  \BibitemOpen
  \bibfield  {author} {\bibinfo {author} {\bibfnamefont {D.}~\bibnamefont
  {Blume}}\ and\ \bibinfo {author} {\bibfnamefont {C.~H.}\ \bibnamefont
  {Greene}},\ }\href {\doibase 10.1103/PhysRevA.63.063601} {\bibfield
  {journal} {\bibinfo  {journal} {Phys. Rev. A}\ }\textbf {\bibinfo {volume}
  {63}},\ \bibinfo {pages} {063601} (\bibinfo {year} {2001})}\BibitemShut
  {NoStop}%
\bibitem [{\citenamefont {Astrakharchik}\ \emph {et~al.}(2004)\citenamefont
  {Astrakharchik}, \citenamefont {Blume}, \citenamefont {Giorgini},\ and\
  \citenamefont {Granger}}]{Astrakharchik04}%
  \BibitemOpen
  \bibfield  {author} {\bibinfo {author} {\bibfnamefont {G.~E.}\ \bibnamefont
  {Astrakharchik}}, \bibinfo {author} {\bibfnamefont {D.}~\bibnamefont
  {Blume}}, \bibinfo {author} {\bibfnamefont {S.}~\bibnamefont {Giorgini}}, \
  and\ \bibinfo {author} {\bibfnamefont {B.~E.}\ \bibnamefont {Granger}},\
  }\href {http://stacks.iop.org/0953-4075/37/i=7/a=066} {\bibfield  {journal}
  {\bibinfo  {journal} {Journal of Physics B: Atomic, Molecular and Optical
  Physics}\ }\textbf {\bibinfo {volume} {37}},\ \bibinfo {pages} {S205}
  (\bibinfo {year} {2004})}\BibitemShut {NoStop}%
\bibitem [{\citenamefont {Bijl}(1940)}]{Bijl40}%
  \BibitemOpen
  \bibfield  {author} {\bibinfo {author} {\bibfnamefont {A.}~\bibnamefont
  {Bijl}},\ }\href {\doibase http://dx.doi.org/10.1016/0031-8914(40)90166-5}
  {\bibfield  {journal} {\bibinfo  {journal} {Physica}\ }\textbf {\bibinfo
  {volume} {7}},\ \bibinfo {pages} {869 } (\bibinfo {year} {1940})}\BibitemShut
  {NoStop}%
\bibitem [{\citenamefont {Jastrow}(1955)}]{Jastrow55}%
  \BibitemOpen
  \bibfield  {author} {\bibinfo {author} {\bibfnamefont {R.}~\bibnamefont
  {Jastrow}},\ }\href {\doibase 10.1103/PhysRev.98.1479} {\bibfield  {journal}
  {\bibinfo  {journal} {Phys. Rev.}\ }\textbf {\bibinfo {volume} {98}},\
  \bibinfo {pages} {1479} (\bibinfo {year} {1955})}\BibitemShut {NoStop}%
\bibitem [{\citenamefont {Esry}(1997)}]{Esry97}%
  \BibitemOpen
  \bibfield  {author} {\bibinfo {author} {\bibfnamefont {B.~D.}\ \bibnamefont
  {Esry}},\ }\href {\doibase 10.1103/PhysRevA.55.1147} {\bibfield  {journal}
  {\bibinfo  {journal} {Phys. Rev. A}\ }\textbf {\bibinfo {volume} {55}},\
  \bibinfo {pages} {1147} (\bibinfo {year} {1997})}\BibitemShut {NoStop}%
\bibitem [{\citenamefont {Streltsov}\ \emph {et~al.}(2006)\citenamefont
  {Streltsov}, \citenamefont {Alon},\ and\ \citenamefont
  {Cederbaum}}]{Streltsov06}%
  \BibitemOpen
  \bibfield  {author} {\bibinfo {author} {\bibfnamefont {A.~I.}\ \bibnamefont
  {Streltsov}}, \bibinfo {author} {\bibfnamefont {O.~E.}\ \bibnamefont {Alon}},
  \ and\ \bibinfo {author} {\bibfnamefont {L.~S.}\ \bibnamefont {Cederbaum}},\
  }\href {\doibase 10.1103/PhysRevA.73.063626} {\bibfield  {journal} {\bibinfo
  {journal} {Phys. Rev. A}\ }\textbf {\bibinfo {volume} {73}},\ \bibinfo
  {pages} {063626} (\bibinfo {year} {2006})}\BibitemShut {NoStop}%
\bibitem [{\citenamefont {L\"owdin}(1955)}]{Lowdin55}%
  \BibitemOpen
  \bibfield  {author} {\bibinfo {author} {\bibfnamefont {P.-O.}\ \bibnamefont
  {L\"owdin}},\ }\href {\doibase 10.1103/PhysRev.97.1474} {\bibfield  {journal}
  {\bibinfo  {journal} {Phys. Rev.}\ }\textbf {\bibinfo {volume} {97}},\
  \bibinfo {pages} {1474} (\bibinfo {year} {1955})}\BibitemShut {NoStop}%
\bibitem [{\citenamefont {Coester}(1958)}]{Coester58}%
  \BibitemOpen
  \bibfield  {author} {\bibinfo {author} {\bibfnamefont {F.}~\bibnamefont
  {Coester}},\ }\href {\doibase http://dx.doi.org/10.1016/0029-5582(58)90280-3}
  {\bibfield  {journal} {\bibinfo  {journal} {Nuclear Physics}\ }\textbf
  {\bibinfo {volume} {7}},\ \bibinfo {pages} {421 } (\bibinfo {year}
  {1958})}\BibitemShut {NoStop}%
\bibitem [{\citenamefont {Coester}\ and\ \citenamefont
  {K\"ummel}(1960)}]{Coester60}%
  \BibitemOpen
  \bibfield  {author} {\bibinfo {author} {\bibfnamefont {F.}~\bibnamefont
  {Coester}}\ and\ \bibinfo {author} {\bibfnamefont {H.}~\bibnamefont
  {K\"ummel}},\ }\href {\doibase
  http://dx.doi.org/10.1016/0029-5582(60)90140-1} {\bibfield  {journal}
  {\bibinfo  {journal} {Nuclear Physics}\ }\textbf {\bibinfo {volume} {17}},\
  \bibinfo {pages} {477 } (\bibinfo {year} {1960})}\BibitemShut {NoStop}%
\bibitem [{\citenamefont {\v{C}\'i\v{z}ek}(1966)}]{Cizek66}%
  \BibitemOpen
  \bibfield  {author} {\bibinfo {author} {\bibfnamefont {J.}~\bibnamefont
  {\v{C}\'i\v{z}ek}},\ }\href {\doibase http://dx.doi.org/10.1063/1.1727484}
  {\bibfield  {journal} {\bibinfo  {journal} {The Journal of Chemical Physics}\
  }\textbf {\bibinfo {volume} {45}},\ \bibinfo {pages} {4256} (\bibinfo {year}
  {1966})}\BibitemShut {NoStop}%
\bibitem [{\citenamefont {Cederbaum}\ \emph {et~al.}(2006)\citenamefont
  {Cederbaum}, \citenamefont {Alon},\ and\ \citenamefont
  {Streltsov}}]{Cederbaum06}%
  \BibitemOpen
  \bibfield  {author} {\bibinfo {author} {\bibfnamefont {L.~S.}\ \bibnamefont
  {Cederbaum}}, \bibinfo {author} {\bibfnamefont {O.~E.}\ \bibnamefont {Alon}},
  \ and\ \bibinfo {author} {\bibfnamefont {A.~I.}\ \bibnamefont {Streltsov}},\
  }\href {\doibase 10.1103/PhysRevA.73.043609} {\bibfield  {journal} {\bibinfo
  {journal} {Phys. Rev. A}\ }\textbf {\bibinfo {volume} {73}},\ \bibinfo
  {pages} {043609} (\bibinfo {year} {2006})}\BibitemShut {NoStop}%
\bibitem [{\citenamefont {Lysaght}\ \emph {et~al.}(2009)\citenamefont
  {Lysaght}, \citenamefont {van~der Hart},\ and\ \citenamefont
  {Burke}}]{Lysaght09}%
  \BibitemOpen
  \bibfield  {author} {\bibinfo {author} {\bibfnamefont {M.~A.}\ \bibnamefont
  {Lysaght}}, \bibinfo {author} {\bibfnamefont {H.~W.}\ \bibnamefont {van~der
  Hart}}, \ and\ \bibinfo {author} {\bibfnamefont {P.~G.}\ \bibnamefont
  {Burke}},\ }\href {\doibase 10.1103/PhysRevA.79.053411} {\bibfield  {journal}
  {\bibinfo  {journal} {Phys. Rev. A}\ }\textbf {\bibinfo {volume} {79}},\
  \bibinfo {pages} {053411} (\bibinfo {year} {2009})}\BibitemShut {NoStop}%
\bibitem [{\citenamefont {Hochstuhl}\ and\ \citenamefont
  {Bonitz}(2012)}]{Hochstuhl12}%
  \BibitemOpen
  \bibfield  {author} {\bibinfo {author} {\bibfnamefont {D.}~\bibnamefont
  {Hochstuhl}}\ and\ \bibinfo {author} {\bibfnamefont {M.}~\bibnamefont
  {Bonitz}},\ }\href {\doibase 10.1103/PhysRevA.86.053424} {\bibfield
  {journal} {\bibinfo  {journal} {Phys. Rev. A}\ }\textbf {\bibinfo {volume}
  {86}},\ \bibinfo {pages} {053424} (\bibinfo {year} {2012})}\BibitemShut
  {NoStop}%
\bibitem [{\citenamefont {Pabst}\ \emph {et~al.}(2012)\citenamefont {Pabst},
  \citenamefont {Greenman}, \citenamefont {Mazziotti},\ and\ \citenamefont
  {Santra}}]{Pabst12}%
  \BibitemOpen
  \bibfield  {author} {\bibinfo {author} {\bibfnamefont {S.}~\bibnamefont
  {Pabst}}, \bibinfo {author} {\bibfnamefont {L.}~\bibnamefont {Greenman}},
  \bibinfo {author} {\bibfnamefont {D.~A.}\ \bibnamefont {Mazziotti}}, \ and\
  \bibinfo {author} {\bibfnamefont {R.}~\bibnamefont {Santra}},\ }\href
  {\doibase 10.1103/PhysRevA.85.023411} {\bibfield  {journal} {\bibinfo
  {journal} {Phys. Rev. A}\ }\textbf {\bibinfo {volume} {85}},\ \bibinfo
  {pages} {023411} (\bibinfo {year} {2012})}\BibitemShut {NoStop}%
\bibitem [{\citenamefont {Kvaal}(2012)}]{Kvaal12}%
  \BibitemOpen
  \bibfield  {author} {\bibinfo {author} {\bibfnamefont {S.}~\bibnamefont
  {Kvaal}},\ }\href {\doibase http://dx.doi.org/10.1063/1.4718427} {\bibfield
  {journal} {\bibinfo  {journal} {The Journal of Chemical Physics}\ }\textbf
  {\bibinfo {volume} {136}},\ \bibinfo {eid} {194109} (\bibinfo {year}
  {2012}),\ http://dx.doi.org/10.1063/1.4718427}\BibitemShut {NoStop}%
\bibitem [{\citenamefont {Sato}\ and\ \citenamefont {Ishikawa}(2013)}]{Sato13}%
  \BibitemOpen
  \bibfield  {author} {\bibinfo {author} {\bibfnamefont {T.}~\bibnamefont
  {Sato}}\ and\ \bibinfo {author} {\bibfnamefont {K.~L.}\ \bibnamefont
  {Ishikawa}},\ }\href {\doibase 10.1103/PhysRevA.88.023402} {\bibfield
  {journal} {\bibinfo  {journal} {Phys. Rev. A}\ }\textbf {\bibinfo {volume}
  {88}},\ \bibinfo {pages} {023402} (\bibinfo {year} {2013})}\BibitemShut
  {NoStop}%
\bibitem [{\citenamefont {Bauch}\ \emph {et~al.}(2014)\citenamefont {Bauch},
  \citenamefont {S\o{}rensen},\ and\ \citenamefont {Madsen}}]{Bauch14}%
  \BibitemOpen
  \bibfield  {author} {\bibinfo {author} {\bibfnamefont {S.}~\bibnamefont
  {Bauch}}, \bibinfo {author} {\bibfnamefont {L.~K.}\ \bibnamefont
  {S\o{}rensen}}, \ and\ \bibinfo {author} {\bibfnamefont {L.~B.}\ \bibnamefont
  {Madsen}},\ }\href {\doibase 10.1103/PhysRevA.90.062508} {\bibfield
  {journal} {\bibinfo  {journal} {Phys. Rev. A}\ }\textbf {\bibinfo {volume}
  {90}},\ \bibinfo {pages} {062508} (\bibinfo {year} {2014})}\BibitemShut
  {NoStop}%
\bibitem [{\citenamefont {Popmintchev}\ \emph {et~al.}(2010)\citenamefont
  {Popmintchev}, \citenamefont {Chen}, \citenamefont {Arpin}, \citenamefont
  {Murnane},\ and\ \citenamefont {Kapteyn}}]{Popmintchev10}%
  \BibitemOpen
  \bibfield  {author} {\bibinfo {author} {\bibfnamefont {T.}~\bibnamefont
  {Popmintchev}}, \bibinfo {author} {\bibfnamefont {M.-C.}\ \bibnamefont
  {Chen}}, \bibinfo {author} {\bibfnamefont {P.}~\bibnamefont {Arpin}},
  \bibinfo {author} {\bibfnamefont {M.~M.}\ \bibnamefont {Murnane}}, \ and\
  \bibinfo {author} {\bibfnamefont {H.~C.}\ \bibnamefont {Kapteyn}},\ }\href
  {\doibase 10.1038/nphoton.2010.256} {\bibfield  {journal} {\bibinfo
  {journal} {Nat Photon}\ }\textbf {\bibinfo {volume} {4}},\ \bibinfo {pages}
  {822} (\bibinfo {year} {2010})}\BibitemShut {NoStop}%
\bibitem [{\citenamefont {Calegari}\ \emph {et~al.}(2014)\citenamefont
  {Calegari}, \citenamefont {Ayuso}, \citenamefont {Trabattoni}, \citenamefont
  {Belshaw}, \citenamefont {De~Camillis}, \citenamefont {Anumula},
  \citenamefont {Frassetto}, \citenamefont {Poletto}, \citenamefont {Palacios},
  \citenamefont {Decleva}, \citenamefont {Greenwood}, \citenamefont
  {Mart{\'\i}n},\ and\ \citenamefont {Nisoli}}]{Calegari14}%
  \BibitemOpen
  \bibfield  {author} {\bibinfo {author} {\bibfnamefont {F.}~\bibnamefont
  {Calegari}}, \bibinfo {author} {\bibfnamefont {D.}~\bibnamefont {Ayuso}},
  \bibinfo {author} {\bibfnamefont {A.}~\bibnamefont {Trabattoni}}, \bibinfo
  {author} {\bibfnamefont {L.}~\bibnamefont {Belshaw}}, \bibinfo {author}
  {\bibfnamefont {S.}~\bibnamefont {De~Camillis}}, \bibinfo {author}
  {\bibfnamefont {S.}~\bibnamefont {Anumula}}, \bibinfo {author} {\bibfnamefont
  {F.}~\bibnamefont {Frassetto}}, \bibinfo {author} {\bibfnamefont
  {L.}~\bibnamefont {Poletto}}, \bibinfo {author} {\bibfnamefont
  {A.}~\bibnamefont {Palacios}}, \bibinfo {author} {\bibfnamefont
  {P.}~\bibnamefont {Decleva}}, \bibinfo {author} {\bibfnamefont {J.~B.}\
  \bibnamefont {Greenwood}}, \bibinfo {author} {\bibfnamefont {F.}~\bibnamefont
  {Mart{\'\i}n}}, \ and\ \bibinfo {author} {\bibfnamefont {M.}~\bibnamefont
  {Nisoli}},\ }\href {\doibase 10.1126/science.1254061} {\bibfield  {journal}
  {\bibinfo  {journal} {Science}\ }\textbf {\bibinfo {volume} {346}},\ \bibinfo
  {pages} {336} (\bibinfo {year} {2014})}\BibitemShut {NoStop}%
\bibitem [{\citenamefont {Kraus}\ \emph {et~al.}(2015)\citenamefont {Kraus},
  \citenamefont {Mignolet}, \citenamefont {Baykusheva}, \citenamefont
  {Rupenyan}, \citenamefont {Horn{\'y}}, \citenamefont {Penka}, \citenamefont
  {Grassi}, \citenamefont {Tolstikhin}, \citenamefont {Schneider},
  \citenamefont {Jensen}, \citenamefont {Madsen}, \citenamefont {Bandrauk},
  \citenamefont {Remacle},\ and\ \citenamefont {W{\"o}rner}}]{Kraus15}%
  \BibitemOpen
  \bibfield  {author} {\bibinfo {author} {\bibfnamefont {P.~M.}\ \bibnamefont
  {Kraus}}, \bibinfo {author} {\bibfnamefont {B.}~\bibnamefont {Mignolet}},
  \bibinfo {author} {\bibfnamefont {D.}~\bibnamefont {Baykusheva}}, \bibinfo
  {author} {\bibfnamefont {A.}~\bibnamefont {Rupenyan}}, \bibinfo {author}
  {\bibfnamefont {L.}~\bibnamefont {Horn{\'y}}}, \bibinfo {author}
  {\bibfnamefont {E.~F.}\ \bibnamefont {Penka}}, \bibinfo {author}
  {\bibfnamefont {G.}~\bibnamefont {Grassi}}, \bibinfo {author} {\bibfnamefont
  {O.~I.}\ \bibnamefont {Tolstikhin}}, \bibinfo {author} {\bibfnamefont
  {J.}~\bibnamefont {Schneider}}, \bibinfo {author} {\bibfnamefont
  {F.}~\bibnamefont {Jensen}}, \bibinfo {author} {\bibfnamefont {L.~B.}\
  \bibnamefont {Madsen}}, \bibinfo {author} {\bibfnamefont {A.~D.}\
  \bibnamefont {Bandrauk}}, \bibinfo {author} {\bibfnamefont {F.}~\bibnamefont
  {Remacle}}, \ and\ \bibinfo {author} {\bibfnamefont {H.~J.}\ \bibnamefont
  {W{\"o}rner}},\ }\href {\doibase 10.1126/science.aab2160} {\bibfield
  {journal} {\bibinfo  {journal} {Science}\ }\textbf {\bibinfo {volume}
  {350}},\ \bibinfo {pages} {790} (\bibinfo {year} {2015})}\BibitemShut
  {NoStop}%
\bibitem [{\citenamefont {Zanghellini}\ \emph {et~al.}(2003)\citenamefont
  {Zanghellini}, \citenamefont {Kitzler}, \citenamefont {Fabian}, \citenamefont
  {Brabec},\ and\ \citenamefont {Scrinzi}}]{Zanghellini03}%
  \BibitemOpen
  \bibfield  {author} {\bibinfo {author} {\bibfnamefont {J.}~\bibnamefont
  {Zanghellini}}, \bibinfo {author} {\bibfnamefont {M.}~\bibnamefont
  {Kitzler}}, \bibinfo {author} {\bibfnamefont {C.}~\bibnamefont {Fabian}},
  \bibinfo {author} {\bibfnamefont {T.}~\bibnamefont {Brabec}}, \ and\ \bibinfo
  {author} {\bibfnamefont {A.}~\bibnamefont {Scrinzi}},\ }\href@noop {}
  {\bibfield  {journal} {\bibinfo  {journal} {Laser Physics}\ }\textbf
  {\bibinfo {volume} {13}},\ \bibinfo {pages} {1064} (\bibinfo {year}
  {2003})}\BibitemShut {NoStop}%
\bibitem [{\citenamefont {Kato}\ and\ \citenamefont {Kono}(2004)}]{Kato04}%
  \BibitemOpen
  \bibfield  {author} {\bibinfo {author} {\bibfnamefont {T.}~\bibnamefont
  {Kato}}\ and\ \bibinfo {author} {\bibfnamefont {H.}~\bibnamefont {Kono}},\
  }\href {\doibase http://dx.doi.org/10.1016/j.cplett.2004.05.106} {\bibfield
  {journal} {\bibinfo  {journal} {Chemical Physics Letters}\ }\textbf {\bibinfo
  {volume} {392}},\ \bibinfo {pages} {533 } (\bibinfo {year}
  {2004})}\BibitemShut {NoStop}%
\bibitem [{\citenamefont {Nest}\ \emph {et~al.}(2005)\citenamefont {Nest},
  \citenamefont {Klamroth},\ and\ \citenamefont {Saalfrank}}]{Nest05}%
  \BibitemOpen
  \bibfield  {author} {\bibinfo {author} {\bibfnamefont {M.}~\bibnamefont
  {Nest}}, \bibinfo {author} {\bibfnamefont {T.}~\bibnamefont {Klamroth}}, \
  and\ \bibinfo {author} {\bibfnamefont {P.}~\bibnamefont {Saalfrank}},\ }\href
  {\doibase http://dx.doi.org/10.1063/1.1862243} {\bibfield  {journal}
  {\bibinfo  {journal} {The Journal of Chemical Physics}\ }\textbf {\bibinfo
  {volume} {122}},\ \bibinfo {eid} {124102} (\bibinfo {year} {2005}),\
  http://dx.doi.org/10.1063/1.1862243}\BibitemShut {NoStop}%
\bibitem [{\citenamefont {Haxton}\ \emph {et~al.}(2011)\citenamefont {Haxton},
  \citenamefont {Lawler},\ and\ \citenamefont {McCurdy}}]{Haxton11}%
  \BibitemOpen
  \bibfield  {author} {\bibinfo {author} {\bibfnamefont {D.~J.}\ \bibnamefont
  {Haxton}}, \bibinfo {author} {\bibfnamefont {K.~V.}\ \bibnamefont {Lawler}},
  \ and\ \bibinfo {author} {\bibfnamefont {C.~W.}\ \bibnamefont {McCurdy}},\
  }\href {\doibase 10.1103/PhysRevA.83.063416} {\bibfield  {journal} {\bibinfo
  {journal} {Phys. Rev. A}\ }\textbf {\bibinfo {volume} {83}},\ \bibinfo
  {pages} {063416} (\bibinfo {year} {2011})}\BibitemShut {NoStop}%
\bibitem [{\citenamefont {Meyer}\ \emph {et~al.}(1990)\citenamefont {Meyer},
  \citenamefont {Manthe},\ and\ \citenamefont {Cederbaum}}]{Meyer90}%
  \BibitemOpen
  \bibfield  {author} {\bibinfo {author} {\bibfnamefont {H.-D.}\ \bibnamefont
  {Meyer}}, \bibinfo {author} {\bibfnamefont {U.}~\bibnamefont {Manthe}}, \
  and\ \bibinfo {author} {\bibfnamefont {L.}~\bibnamefont {Cederbaum}},\ }\href
  {\doibase http://dx.doi.org/10.1016/0009-2614(90)87014-I} {\bibfield
  {journal} {\bibinfo  {journal} {Chemical Physics Letters}\ }\textbf {\bibinfo
  {volume} {165}},\ \bibinfo {pages} {73 } (\bibinfo {year}
  {1990})}\BibitemShut {NoStop}%
\bibitem [{\citenamefont {Beck}\ \emph {et~al.}(2000)\citenamefont {Beck},
  \citenamefont {J\"ackle}, \citenamefont {Worth},\ and\ \citenamefont
  {Meyer}}]{Beck00}%
  \BibitemOpen
  \bibfield  {author} {\bibinfo {author} {\bibfnamefont {M.}~\bibnamefont
  {Beck}}, \bibinfo {author} {\bibfnamefont {A.}~\bibnamefont {J\"ackle}},
  \bibinfo {author} {\bibfnamefont {G.}~\bibnamefont {Worth}}, \ and\ \bibinfo
  {author} {\bibfnamefont {H.-D.}\ \bibnamefont {Meyer}},\ }\href {\doibase
  http://dx.doi.org/10.1016/S0370-1573(99)00047-2} {\bibfield  {journal}
  {\bibinfo  {journal} {Physics Reports}\ }\textbf {\bibinfo {volume} {324}},\
  \bibinfo {pages} {1 } (\bibinfo {year} {2000})}\BibitemShut {NoStop}%
\bibitem [{\citenamefont {Alon}\ \emph {et~al.}(2008)\citenamefont {Alon},
  \citenamefont {Streltsov},\ and\ \citenamefont {Cederbaum}}]{Alon08}%
  \BibitemOpen
  \bibfield  {author} {\bibinfo {author} {\bibfnamefont {O.~E.}\ \bibnamefont
  {Alon}}, \bibinfo {author} {\bibfnamefont {A.~I.}\ \bibnamefont {Streltsov}},
  \ and\ \bibinfo {author} {\bibfnamefont {L.~S.}\ \bibnamefont {Cederbaum}},\
  }\href {\doibase 10.1103/PhysRevA.77.033613} {\bibfield  {journal} {\bibinfo
  {journal} {Phys. Rev. A}\ }\textbf {\bibinfo {volume} {77}},\ \bibinfo
  {pages} {033613} (\bibinfo {year} {2008})}\BibitemShut {NoStop}%
\bibitem [{\citenamefont {Alon}\ \emph
  {et~al.}(2007{\natexlab{a}})\citenamefont {Alon}, \citenamefont {Streltsov},\
  and\ \citenamefont {Cederbaum}}]{Alon07}%
  \BibitemOpen
  \bibfield  {author} {\bibinfo {author} {\bibfnamefont {O.~E.}\ \bibnamefont
  {Alon}}, \bibinfo {author} {\bibfnamefont {A.~I.}\ \bibnamefont {Streltsov}},
  \ and\ \bibinfo {author} {\bibfnamefont {L.~S.}\ \bibnamefont {Cederbaum}},\
  }\href {\doibase 10.1103/PhysRevA.76.062501} {\bibfield  {journal} {\bibinfo
  {journal} {Phys. Rev. A}\ }\textbf {\bibinfo {volume} {76}},\ \bibinfo
  {pages} {062501} (\bibinfo {year} {2007}{\natexlab{a}})}\BibitemShut
  {NoStop}%
\bibitem [{\citenamefont {Alon}\ \emph {et~al.}(2012)\citenamefont {Alon},
  \citenamefont {Streltsov}, \citenamefont {Sakmann}, \citenamefont {Lode},
  \citenamefont {Grond},\ and\ \citenamefont {Cederbaum}}]{Alon12}%
  \BibitemOpen
  \bibfield  {author} {\bibinfo {author} {\bibfnamefont {O.~E.}\ \bibnamefont
  {Alon}}, \bibinfo {author} {\bibfnamefont {A.~I.}\ \bibnamefont {Streltsov}},
  \bibinfo {author} {\bibfnamefont {K.}~\bibnamefont {Sakmann}}, \bibinfo
  {author} {\bibfnamefont {A.~U.}\ \bibnamefont {Lode}}, \bibinfo {author}
  {\bibfnamefont {J.}~\bibnamefont {Grond}}, \ and\ \bibinfo {author}
  {\bibfnamefont {L.~S.}\ \bibnamefont {Cederbaum}},\ }\href {\doibase
  http://dx.doi.org/10.1016/j.chemphys.2011.09.026} {\bibfield  {journal}
  {\bibinfo  {journal} {Chemical Physics}\ }\textbf {\bibinfo {volume} {401}},\
  \bibinfo {pages} {2 } (\bibinfo {year} {2012})},\ \bibinfo {note} {recent
  advances in electron correlation methods and applications}\BibitemShut
  {NoStop}%
\bibitem [{\citenamefont {Wang}\ and\ \citenamefont {Thoss}(2003)}]{Wang03}%
  \BibitemOpen
  \bibfield  {author} {\bibinfo {author} {\bibfnamefont {H.}~\bibnamefont
  {Wang}}\ and\ \bibinfo {author} {\bibfnamefont {M.}~\bibnamefont {Thoss}},\
  }\href {\doibase http://dx.doi.org/10.1063/1.1580111} {\bibfield  {journal}
  {\bibinfo  {journal} {The Journal of Chemical Physics}\ }\textbf {\bibinfo
  {volume} {119}},\ \bibinfo {pages} {1289} (\bibinfo {year}
  {2003})}\BibitemShut {NoStop}%
\bibitem [{\citenamefont {Manthe}(2008)}]{Manthe08}%
  \BibitemOpen
  \bibfield  {author} {\bibinfo {author} {\bibfnamefont {U.}~\bibnamefont
  {Manthe}},\ }\href {\doibase http://dx.doi.org/10.1063/1.2902982} {\bibfield
  {journal} {\bibinfo  {journal} {The Journal of Chemical Physics}\ }\textbf
  {\bibinfo {volume} {128}},\ \bibinfo {eid} {164116} (\bibinfo {year}
  {2008}),\ http://dx.doi.org/10.1063/1.2902982}\BibitemShut {NoStop}%
\bibitem [{\citenamefont {Vendrell}\ and\ \citenamefont
  {Meyer}(2011)}]{Vendrell11}%
  \BibitemOpen
  \bibfield  {author} {\bibinfo {author} {\bibfnamefont {O.}~\bibnamefont
  {Vendrell}}\ and\ \bibinfo {author} {\bibfnamefont {H.-D.}\ \bibnamefont
  {Meyer}},\ }\href {\doibase http://dx.doi.org/10.1063/1.3535541} {\bibfield
  {journal} {\bibinfo  {journal} {The Journal of Chemical Physics}\ }\textbf
  {\bibinfo {volume} {134}},\ \bibinfo {eid} {044135} (\bibinfo {year}
  {2011}),\ http://dx.doi.org/10.1063/1.3535541}\BibitemShut {NoStop}%
\bibitem [{\citenamefont {Kr\"onke}\ \emph {et~al.}(2013)\citenamefont
  {Kr\"onke}, \citenamefont {Cao}, \citenamefont {Vendrell},\ and\
  \citenamefont {Schmelcher}}]{Kronke13}%
  \BibitemOpen
  \bibfield  {author} {\bibinfo {author} {\bibfnamefont {S.}~\bibnamefont
  {Kr\"onke}}, \bibinfo {author} {\bibfnamefont {L.}~\bibnamefont {Cao}},
  \bibinfo {author} {\bibfnamefont {O.}~\bibnamefont {Vendrell}}, \ and\
  \bibinfo {author} {\bibfnamefont {P.}~\bibnamefont {Schmelcher}},\ }\href
  {http://stacks.iop.org/1367-2630/15/i=6/a=063018} {\bibfield  {journal}
  {\bibinfo  {journal} {New Journal of Physics}\ }\textbf {\bibinfo {volume}
  {15}},\ \bibinfo {pages} {063018} (\bibinfo {year} {2013})}\BibitemShut
  {NoStop}%
\bibitem [{\citenamefont {Cao}\ \emph {et~al.}(2013)\citenamefont {Cao},
  \citenamefont {Kr\"onke}, \citenamefont {Vendrell},\ and\ \citenamefont
  {Schmelcher}}]{Cao13}%
  \BibitemOpen
  \bibfield  {author} {\bibinfo {author} {\bibfnamefont {L.}~\bibnamefont
  {Cao}}, \bibinfo {author} {\bibfnamefont {S.}~\bibnamefont {Kr\"onke}},
  \bibinfo {author} {\bibfnamefont {O.}~\bibnamefont {Vendrell}}, \ and\
  \bibinfo {author} {\bibfnamefont {P.}~\bibnamefont {Schmelcher}},\ }\href
  {\doibase http://dx.doi.org/10.1063/1.4821350} {\bibfield  {journal}
  {\bibinfo  {journal} {The Journal of Chemical Physics}\ }\textbf {\bibinfo
  {volume} {139}},\ \bibinfo {eid} {134103} (\bibinfo {year} {2013}),\
  http://dx.doi.org/10.1063/1.4821350}\BibitemShut {NoStop}%
\bibitem [{\citenamefont {Schmitz}\ \emph {et~al.}(2013)\citenamefont
  {Schmitz}, \citenamefont {Kr\"onke}, \citenamefont {Cao},\ and\ \citenamefont
  {Schmelcher}}]{Schmitz13}%
  \BibitemOpen
  \bibfield  {author} {\bibinfo {author} {\bibfnamefont {R.}~\bibnamefont
  {Schmitz}}, \bibinfo {author} {\bibfnamefont {S.}~\bibnamefont {Kr\"onke}},
  \bibinfo {author} {\bibfnamefont {L.}~\bibnamefont {Cao}}, \ and\ \bibinfo
  {author} {\bibfnamefont {P.}~\bibnamefont {Schmelcher}},\ }\href {\doibase
  10.1103/PhysRevA.88.043601} {\bibfield  {journal} {\bibinfo  {journal} {Phys.
  Rev. A}\ }\textbf {\bibinfo {volume} {88}},\ \bibinfo {pages} {043601}
  (\bibinfo {year} {2013})}\BibitemShut {NoStop}%
\bibitem [{\citenamefont {Sakmann}\ \emph {et~al.}(2009)\citenamefont
  {Sakmann}, \citenamefont {Streltsov}, \citenamefont {Alon},\ and\
  \citenamefont {Cederbaum}}]{Sakmann09}%
  \BibitemOpen
  \bibfield  {author} {\bibinfo {author} {\bibfnamefont {K.}~\bibnamefont
  {Sakmann}}, \bibinfo {author} {\bibfnamefont {A.~I.}\ \bibnamefont
  {Streltsov}}, \bibinfo {author} {\bibfnamefont {O.~E.}\ \bibnamefont {Alon}},
  \ and\ \bibinfo {author} {\bibfnamefont {L.~S.}\ \bibnamefont {Cederbaum}},\
  }\href {\doibase 10.1103/PhysRevLett.103.220601} {\bibfield  {journal}
  {\bibinfo  {journal} {Phys. Rev. Lett.}\ }\textbf {\bibinfo {volume} {103}},\
  \bibinfo {pages} {220601} (\bibinfo {year} {2009})}\BibitemShut {NoStop}%
\bibitem [{\citenamefont {Sakmann}\ \emph {et~al.}(2014)\citenamefont
  {Sakmann}, \citenamefont {Streltsov}, \citenamefont {Alon},\ and\
  \citenamefont {Cederbaum}}]{Sakmann14}%
  \BibitemOpen
  \bibfield  {author} {\bibinfo {author} {\bibfnamefont {K.}~\bibnamefont
  {Sakmann}}, \bibinfo {author} {\bibfnamefont {A.~I.}\ \bibnamefont
  {Streltsov}}, \bibinfo {author} {\bibfnamefont {O.~E.}\ \bibnamefont {Alon}},
  \ and\ \bibinfo {author} {\bibfnamefont {L.~S.}\ \bibnamefont {Cederbaum}},\
  }\href {\doibase 10.1103/PhysRevA.89.023602} {\bibfield  {journal} {\bibinfo
  {journal} {Phys. Rev. A}\ }\textbf {\bibinfo {volume} {89}},\ \bibinfo
  {pages} {023602} (\bibinfo {year} {2014})}\BibitemShut {NoStop}%
\bibitem [{\citenamefont {Sakmann}\ \emph {et~al.}(2008)\citenamefont
  {Sakmann}, \citenamefont {Streltsov}, \citenamefont {Alon},\ and\
  \citenamefont {Cederbaum}}]{Sakmann08}%
  \BibitemOpen
  \bibfield  {author} {\bibinfo {author} {\bibfnamefont {K.}~\bibnamefont
  {Sakmann}}, \bibinfo {author} {\bibfnamefont {A.~I.}\ \bibnamefont
  {Streltsov}}, \bibinfo {author} {\bibfnamefont {O.~E.}\ \bibnamefont {Alon}},
  \ and\ \bibinfo {author} {\bibfnamefont {L.~S.}\ \bibnamefont {Cederbaum}},\
  }\href {\doibase 10.1103/PhysRevA.78.023615} {\bibfield  {journal} {\bibinfo
  {journal} {Phys. Rev. A}\ }\textbf {\bibinfo {volume} {78}},\ \bibinfo
  {pages} {023615} (\bibinfo {year} {2008})}\BibitemShut {NoStop}%
\bibitem [{\citenamefont {Klaiman}\ and\ \citenamefont
  {Alon}(2015)}]{Klaiman15}%
  \BibitemOpen
  \bibfield  {author} {\bibinfo {author} {\bibfnamefont {S.}~\bibnamefont
  {Klaiman}}\ and\ \bibinfo {author} {\bibfnamefont {O.~E.}\ \bibnamefont
  {Alon}},\ }\href {\doibase 10.1103/PhysRevA.91.063613} {\bibfield  {journal}
  {\bibinfo  {journal} {Phys. Rev. A}\ }\textbf {\bibinfo {volume} {91}},\
  \bibinfo {pages} {063613} (\bibinfo {year} {2015})}\BibitemShut {NoStop}%
\bibitem [{\citenamefont {Streltsov}\ \emph {et~al.}(2008)\citenamefont
  {Streltsov}, \citenamefont {Alon},\ and\ \citenamefont
  {Cederbaum}}]{Streltsov08}%
  \BibitemOpen
  \bibfield  {author} {\bibinfo {author} {\bibfnamefont {A.~I.}\ \bibnamefont
  {Streltsov}}, \bibinfo {author} {\bibfnamefont {O.~E.}\ \bibnamefont {Alon}},
  \ and\ \bibinfo {author} {\bibfnamefont {L.~S.}\ \bibnamefont {Cederbaum}},\
  }\href {\doibase 10.1103/PhysRevLett.100.130401} {\bibfield  {journal}
  {\bibinfo  {journal} {Phys. Rev. Lett.}\ }\textbf {\bibinfo {volume} {100}},\
  \bibinfo {pages} {130401} (\bibinfo {year} {2008})}\BibitemShut {NoStop}%
\bibitem [{\citenamefont {Streltsov}\ \emph {et~al.}(2009)\citenamefont
  {Streltsov}, \citenamefont {Alon},\ and\ \citenamefont
  {Cederbaum}}]{Streltsov09}%
  \BibitemOpen
  \bibfield  {author} {\bibinfo {author} {\bibfnamefont {A.~I.}\ \bibnamefont
  {Streltsov}}, \bibinfo {author} {\bibfnamefont {O.~E.}\ \bibnamefont {Alon}},
  \ and\ \bibinfo {author} {\bibfnamefont {L.~S.}\ \bibnamefont {Cederbaum}},\
  }\href {\doibase 10.1103/PhysRevA.80.043616} {\bibfield  {journal} {\bibinfo
  {journal} {Phys. Rev. A}\ }\textbf {\bibinfo {volume} {80}},\ \bibinfo
  {pages} {043616} (\bibinfo {year} {2009})}\BibitemShut {NoStop}%
\bibitem [{\citenamefont {Streltsova}\ \emph {et~al.}(2014)\citenamefont
  {Streltsova}, \citenamefont {Alon}, \citenamefont {Cederbaum},\ and\
  \citenamefont {Streltsov}}]{Streltsova14}%
  \BibitemOpen
  \bibfield  {author} {\bibinfo {author} {\bibfnamefont {O.~I.}\ \bibnamefont
  {Streltsova}}, \bibinfo {author} {\bibfnamefont {O.~E.}\ \bibnamefont
  {Alon}}, \bibinfo {author} {\bibfnamefont {L.~S.}\ \bibnamefont {Cederbaum}},
  \ and\ \bibinfo {author} {\bibfnamefont {A.~I.}\ \bibnamefont {Streltsov}},\
  }\href {\doibase 10.1103/PhysRevA.89.061602} {\bibfield  {journal} {\bibinfo
  {journal} {Phys. Rev. A}\ }\textbf {\bibinfo {volume} {89}},\ \bibinfo
  {pages} {061602} (\bibinfo {year} {2014})}\BibitemShut {NoStop}%
\bibitem [{\citenamefont {Lode}\ \emph
  {et~al.}(2012{\natexlab{b}})\citenamefont {Lode}, \citenamefont {Streltsov},
  \citenamefont {Sakmann}, \citenamefont {Alon},\ and\ \citenamefont
  {Cederbaum}}]{Lode12_2}%
  \BibitemOpen
  \bibfield  {author} {\bibinfo {author} {\bibfnamefont {A.~U.}\ \bibnamefont
  {Lode}}, \bibinfo {author} {\bibfnamefont {A.~I.}\ \bibnamefont {Streltsov}},
  \bibinfo {author} {\bibfnamefont {K.}~\bibnamefont {Sakmann}}, \bibinfo
  {author} {\bibfnamefont {O.~E.}\ \bibnamefont {Alon}}, \ and\ \bibinfo
  {author} {\bibfnamefont {L.~S.}\ \bibnamefont {Cederbaum}},\ }\href {\doibase
  10.1073/pnas.1201345109} {\bibfield  {journal} {\bibinfo  {journal}
  {Proceedings of the National Academy of Sciences}\ }\textbf {\bibinfo
  {volume} {109}},\ \bibinfo {pages} {13521} (\bibinfo {year}
  {2012}{\natexlab{b}})},\ \Eprint
  {http://arxiv.org/abs/http://www.pnas.org/content/109/34/13521.full.pdf}
  {http://www.pnas.org/content/109/34/13521.full.pdf} \BibitemShut {NoStop}%
\bibitem [{\citenamefont {Beinke}\ \emph {et~al.}(2015)\citenamefont {Beinke},
  \citenamefont {Klaiman}, \citenamefont {Cederbaum}, \citenamefont
  {Streltsov},\ and\ \citenamefont {Alon}}]{Beinke15}%
  \BibitemOpen
  \bibfield  {author} {\bibinfo {author} {\bibfnamefont {R.}~\bibnamefont
  {Beinke}}, \bibinfo {author} {\bibfnamefont {S.}~\bibnamefont {Klaiman}},
  \bibinfo {author} {\bibfnamefont {L.~S.}\ \bibnamefont {Cederbaum}}, \bibinfo
  {author} {\bibfnamefont {A.~I.}\ \bibnamefont {Streltsov}}, \ and\ \bibinfo
  {author} {\bibfnamefont {O.~E.}\ \bibnamefont {Alon}},\ }\href {\doibase
  10.1103/PhysRevA.92.043627} {\bibfield  {journal} {\bibinfo  {journal} {Phys.
  Rev. A}\ }\textbf {\bibinfo {volume} {92}},\ \bibinfo {pages} {043627}
  (\bibinfo {year} {2015})}\BibitemShut {NoStop}%
\bibitem [{\citenamefont {Olsen}\ \emph {et~al.}(1988)\citenamefont {Olsen},
  \citenamefont {Roos}, \citenamefont {J\o{}rgensen},\ and\ \citenamefont
  {Jensen}}]{Olsen88}%
  \BibitemOpen
  \bibfield  {author} {\bibinfo {author} {\bibfnamefont {J.}~\bibnamefont
  {Olsen}}, \bibinfo {author} {\bibfnamefont {B.~O.}\ \bibnamefont {Roos}},
  \bibinfo {author} {\bibfnamefont {P.}~\bibnamefont {J\o{}rgensen}}, \ and\
  \bibinfo {author} {\bibfnamefont {H.~J.~A.}\ \bibnamefont {Jensen}},\ }\href
  {\doibase http://dx.doi.org/10.1063/1.455063} {\bibfield  {journal} {\bibinfo
   {journal} {The Journal of Chemical Physics}\ }\textbf {\bibinfo {volume}
  {89}},\ \bibinfo {pages} {2185} (\bibinfo {year} {1988})}\BibitemShut
  {NoStop}%
\bibitem [{\citenamefont {Miyagi}\ and\ \citenamefont {Madsen}(2013)}]{Haru13}%
  \BibitemOpen
  \bibfield  {author} {\bibinfo {author} {\bibfnamefont {H.}~\bibnamefont
  {Miyagi}}\ and\ \bibinfo {author} {\bibfnamefont {L.~B.}\ \bibnamefont
  {Madsen}},\ }\href {\doibase 10.1103/PhysRevA.87.062511} {\bibfield
  {journal} {\bibinfo  {journal} {Phys. Rev. A}\ }\textbf {\bibinfo {volume}
  {87}},\ \bibinfo {pages} {062511} (\bibinfo {year} {2013})}\BibitemShut
  {NoStop}%
\bibitem [{\citenamefont {Miyagi}\ and\ \citenamefont
  {Madsen}(2014)}]{Haru14_1}%
  \BibitemOpen
  \bibfield  {author} {\bibinfo {author} {\bibfnamefont {H.}~\bibnamefont
  {Miyagi}}\ and\ \bibinfo {author} {\bibfnamefont {L.~B.}\ \bibnamefont
  {Madsen}},\ }\href {\doibase 10.1103/PhysRevA.89.063416} {\bibfield
  {journal} {\bibinfo  {journal} {Phys. Rev. A}\ }\textbf {\bibinfo {volume}
  {89}},\ \bibinfo {pages} {063416} (\bibinfo {year} {2014})}\BibitemShut
  {NoStop}%
\bibitem [{\citenamefont {Miyagi}\ and\ \citenamefont
  {Bojer~Madsen}(2014)}]{Haru14_2}%
  \BibitemOpen
  \bibfield  {author} {\bibinfo {author} {\bibfnamefont {H.}~\bibnamefont
  {Miyagi}}\ and\ \bibinfo {author} {\bibfnamefont {L.}~\bibnamefont
  {Bojer~Madsen}},\ }\href {\doibase http://dx.doi.org/10.1063/1.4872005}
  {\bibfield  {journal} {\bibinfo  {journal} {The Journal of Chemical Physics}\
  }\textbf {\bibinfo {volume} {140}},\ \bibinfo {eid} {164309} (\bibinfo {year}
  {2014}),\ http://dx.doi.org/10.1063/1.4872005}\BibitemShut {NoStop}%
\bibitem [{\citenamefont {Kramer}\ and\ \citenamefont
  {Saraceno}(1981)}]{Kramer81}%
  \BibitemOpen
  \bibfield  {author} {\bibinfo {author} {\bibfnamefont {P.}~\bibnamefont
  {Kramer}}\ and\ \bibinfo {author} {\bibfnamefont {M.}~\bibnamefont
  {Saraceno}},\ }\href {\doibase 10.1007/3-540-10579-4} {\emph {\bibinfo
  {title} {Geometry of the Time-Dependent Variational Principle in Quantum
  Mechanics}}}\ (\bibinfo  {publisher} {Springer-Verlag Berlin Heidelberg},\
  \bibinfo {year} {1981})\BibitemShut {NoStop}%
\bibitem [{\citenamefont {Cederbaum}(2016)}]{Cederbaum16}%
  \BibitemOpen
  \bibfield  {author} {\bibinfo {author} {\bibfnamefont {L.~S.}\ \bibnamefont
  {Cederbaum}},\ }\href {\doibase 10.1021/acs.jpca.5b09444} {\bibfield
  {journal} {\bibinfo  {journal} {The Journal of Physical Chemistry A}\
  }\textbf {\bibinfo {volume} {120}},\ \bibinfo {pages} {3009} (\bibinfo {year}
  {2016})},\ \bibinfo {note} {pMID: 26594868},\ \Eprint
  {http://arxiv.org/abs/http://dx.doi.org/10.1021/acs.jpca.5b09444}
  {http://dx.doi.org/10.1021/acs.jpca.5b09444} \BibitemShut {NoStop}%
\bibitem [{\citenamefont {Broeckhove}\ \emph {et~al.}(1988)\citenamefont
  {Broeckhove}, \citenamefont {Lathouwers}, \citenamefont {Kesteloot},\ and\
  \citenamefont {Leuven}}]{Broeckhove88}%
  \BibitemOpen
  \bibfield  {author} {\bibinfo {author} {\bibfnamefont {J.}~\bibnamefont
  {Broeckhove}}, \bibinfo {author} {\bibfnamefont {L.}~\bibnamefont
  {Lathouwers}}, \bibinfo {author} {\bibfnamefont {E.}~\bibnamefont
  {Kesteloot}}, \ and\ \bibinfo {author} {\bibfnamefont {P.~V.}\ \bibnamefont
  {Leuven}},\ }\href {\doibase http://dx.doi.org/10.1016/0009-2614(88)80380-4}
  {\bibfield  {journal} {\bibinfo  {journal} {Chemical Physics Letters}\
  }\textbf {\bibinfo {volume} {149}},\ \bibinfo {pages} {547 } (\bibinfo {year}
  {1988})}\BibitemShut {NoStop}%
\bibitem [{\citenamefont {Caillat}\ \emph {et~al.}(2005)\citenamefont
  {Caillat}, \citenamefont {Zanghellini}, \citenamefont {Kitzler},
  \citenamefont {Koch}, \citenamefont {Kreuzer},\ and\ \citenamefont
  {Scrinzi}}]{Caillat05}%
  \BibitemOpen
  \bibfield  {author} {\bibinfo {author} {\bibfnamefont {J.}~\bibnamefont
  {Caillat}}, \bibinfo {author} {\bibfnamefont {J.}~\bibnamefont
  {Zanghellini}}, \bibinfo {author} {\bibfnamefont {M.}~\bibnamefont
  {Kitzler}}, \bibinfo {author} {\bibfnamefont {O.}~\bibnamefont {Koch}},
  \bibinfo {author} {\bibfnamefont {W.}~\bibnamefont {Kreuzer}}, \ and\
  \bibinfo {author} {\bibfnamefont {A.}~\bibnamefont {Scrinzi}},\ }\href
  {\doibase 10.1103/PhysRevA.71.012712} {\bibfield  {journal} {\bibinfo
  {journal} {Phys. Rev. A}\ }\textbf {\bibinfo {volume} {71}},\ \bibinfo
  {pages} {012712} (\bibinfo {year} {2005})}\BibitemShut {NoStop}%
\bibitem [{\citenamefont {Haxton}\ and\ \citenamefont
  {McCurdy}(2015)}]{Haxton15}%
  \BibitemOpen
  \bibfield  {author} {\bibinfo {author} {\bibfnamefont {D.~J.}\ \bibnamefont
  {Haxton}}\ and\ \bibinfo {author} {\bibfnamefont {C.~W.}\ \bibnamefont
  {McCurdy}},\ }\href {\doibase 10.1103/PhysRevA.91.012509} {\bibfield
  {journal} {\bibinfo  {journal} {Phys. Rev. A}\ }\textbf {\bibinfo {volume}
  {91}},\ \bibinfo {pages} {012509} (\bibinfo {year} {2015})}\BibitemShut
  {NoStop}%
\bibitem [{\citenamefont {Alon}\ \emph
  {et~al.}(2007{\natexlab{b}})\citenamefont {Alon}, \citenamefont {Streltsov},\
  and\ \citenamefont {Cederbaum}}]{Alon07_2}%
  \BibitemOpen
  \bibfield  {author} {\bibinfo {author} {\bibfnamefont {O.~E.}\ \bibnamefont
  {Alon}}, \bibinfo {author} {\bibfnamefont {A.~I.}\ \bibnamefont {Streltsov}},
  \ and\ \bibinfo {author} {\bibfnamefont {L.~S.}\ \bibnamefont {Cederbaum}},\
  }\href {\doibase http://dx.doi.org/10.1063/1.2771159} {\bibfield  {journal}
  {\bibinfo  {journal} {The Journal of Chemical Physics}\ }\textbf {\bibinfo
  {volume} {127}},\ \bibinfo {eid} {154103} (\bibinfo {year}
  {2007}{\natexlab{b}}),\ http://dx.doi.org/10.1063/1.2771159}\BibitemShut
  {NoStop}%
\bibitem [{\citenamefont {G\"orlitz}\ \emph {et~al.}(2001)\citenamefont
  {G\"orlitz}, \citenamefont {Vogels}, \citenamefont {Leanhardt}, \citenamefont
  {Raman}, \citenamefont {Gustavson}, \citenamefont {Abo-Shaeer}, \citenamefont
  {Chikkatur}, \citenamefont {Gupta}, \citenamefont {Inouye}, \citenamefont
  {Rosenband},\ and\ \citenamefont {Ketterle}}]{Gorlitz01}%
  \BibitemOpen
  \bibfield  {author} {\bibinfo {author} {\bibfnamefont {A.}~\bibnamefont
  {G\"orlitz}}, \bibinfo {author} {\bibfnamefont {J.~M.}\ \bibnamefont
  {Vogels}}, \bibinfo {author} {\bibfnamefont {A.~E.}\ \bibnamefont
  {Leanhardt}}, \bibinfo {author} {\bibfnamefont {C.}~\bibnamefont {Raman}},
  \bibinfo {author} {\bibfnamefont {T.~L.}\ \bibnamefont {Gustavson}}, \bibinfo
  {author} {\bibfnamefont {J.~R.}\ \bibnamefont {Abo-Shaeer}}, \bibinfo
  {author} {\bibfnamefont {A.~P.}\ \bibnamefont {Chikkatur}}, \bibinfo {author}
  {\bibfnamefont {S.}~\bibnamefont {Gupta}}, \bibinfo {author} {\bibfnamefont
  {S.}~\bibnamefont {Inouye}}, \bibinfo {author} {\bibfnamefont
  {T.}~\bibnamefont {Rosenband}}, \ and\ \bibinfo {author} {\bibfnamefont
  {W.}~\bibnamefont {Ketterle}},\ }\href {\doibase
  10.1103/PhysRevLett.87.130402} {\bibfield  {journal} {\bibinfo  {journal}
  {Phys. Rev. Lett.}\ }\textbf {\bibinfo {volume} {87}},\ \bibinfo {pages}
  {130402} (\bibinfo {year} {2001})}\BibitemShut {NoStop}%
\bibitem [{\citenamefont {Moritz}\ \emph {et~al.}(2003)\citenamefont {Moritz},
  \citenamefont {St\"oferle}, \citenamefont {K\"ohl},\ and\ \citenamefont
  {Esslinger}}]{Moritz03}%
  \BibitemOpen
  \bibfield  {author} {\bibinfo {author} {\bibfnamefont {H.}~\bibnamefont
  {Moritz}}, \bibinfo {author} {\bibfnamefont {T.}~\bibnamefont {St\"oferle}},
  \bibinfo {author} {\bibfnamefont {M.}~\bibnamefont {K\"ohl}}, \ and\ \bibinfo
  {author} {\bibfnamefont {T.}~\bibnamefont {Esslinger}},\ }\href {\doibase
  10.1103/PhysRevLett.91.250402} {\bibfield  {journal} {\bibinfo  {journal}
  {Phys. Rev. Lett.}\ }\textbf {\bibinfo {volume} {91}},\ \bibinfo {pages}
  {250402} (\bibinfo {year} {2003})}\BibitemShut {NoStop}%
\bibitem [{\citenamefont {Tolra}\ \emph {et~al.}(2004)\citenamefont {Tolra},
  \citenamefont {O'Hara}, \citenamefont {Huckans}, \citenamefont {Phillips},
  \citenamefont {Rolston},\ and\ \citenamefont {Porto}}]{Laburthe04}%
  \BibitemOpen
  \bibfield  {author} {\bibinfo {author} {\bibfnamefont {B.~L.}\ \bibnamefont
  {Tolra}}, \bibinfo {author} {\bibfnamefont {K.~M.}\ \bibnamefont {O'Hara}},
  \bibinfo {author} {\bibfnamefont {J.~H.}\ \bibnamefont {Huckans}}, \bibinfo
  {author} {\bibfnamefont {W.~D.}\ \bibnamefont {Phillips}}, \bibinfo {author}
  {\bibfnamefont {S.~L.}\ \bibnamefont {Rolston}}, \ and\ \bibinfo {author}
  {\bibfnamefont {J.~V.}\ \bibnamefont {Porto}},\ }\href {\doibase
  10.1103/PhysRevLett.92.190401} {\bibfield  {journal} {\bibinfo  {journal}
  {Phys. Rev. Lett.}\ }\textbf {\bibinfo {volume} {92}},\ \bibinfo {pages}
  {190401} (\bibinfo {year} {2004})}\BibitemShut {NoStop}%
\bibitem [{\citenamefont {Kinoshita}\ \emph {et~al.}(2005)\citenamefont
  {Kinoshita}, \citenamefont {Wenger},\ and\ \citenamefont
  {Weiss}}]{Kinoshita05}%
  \BibitemOpen
  \bibfield  {author} {\bibinfo {author} {\bibfnamefont {T.}~\bibnamefont
  {Kinoshita}}, \bibinfo {author} {\bibfnamefont {T.}~\bibnamefont {Wenger}}, \
  and\ \bibinfo {author} {\bibfnamefont {D.~S.}\ \bibnamefont {Weiss}},\ }\href
  {\doibase 10.1103/PhysRevLett.95.190406} {\bibfield  {journal} {\bibinfo
  {journal} {Phys. Rev. Lett.}\ }\textbf {\bibinfo {volume} {95}},\ \bibinfo
  {pages} {190406} (\bibinfo {year} {2005})}\BibitemShut {NoStop}%
\bibitem [{\citenamefont {Hofferberth}\ \emph {et~al.}(2007)\citenamefont
  {Hofferberth}, \citenamefont {Lesanovsky}, \citenamefont {Fischer},
  \citenamefont {Schumm},\ and\ \citenamefont {Schmiedmayer}}]{Hofferberth07}%
  \BibitemOpen
  \bibfield  {author} {\bibinfo {author} {\bibfnamefont {S.}~\bibnamefont
  {Hofferberth}}, \bibinfo {author} {\bibfnamefont {I.}~\bibnamefont
  {Lesanovsky}}, \bibinfo {author} {\bibfnamefont {B.}~\bibnamefont {Fischer}},
  \bibinfo {author} {\bibfnamefont {T.}~\bibnamefont {Schumm}}, \ and\ \bibinfo
  {author} {\bibfnamefont {J.}~\bibnamefont {Schmiedmayer}},\ }\href {\doibase
  10.1038/nature06149} {\bibfield  {journal} {\bibinfo  {journal} {Nature}\
  }\textbf {\bibinfo {volume} {449}},\ \bibinfo {pages} {324} (\bibinfo {year}
  {2007})}\BibitemShut {NoStop}%
\bibitem [{\citenamefont {Olshanii}(1998)}]{Olshanii98}%
  \BibitemOpen
  \bibfield  {author} {\bibinfo {author} {\bibfnamefont {M.}~\bibnamefont
  {Olshanii}},\ }\href {\doibase 10.1103/PhysRevLett.81.938} {\bibfield
  {journal} {\bibinfo  {journal} {Phys. Rev. Lett.}\ }\textbf {\bibinfo
  {volume} {81}},\ \bibinfo {pages} {938} (\bibinfo {year} {1998})}\BibitemShut
  {NoStop}%
\bibitem [{\citenamefont {Press}\ \emph {et~al.}(1993)\citenamefont {Press},
  \citenamefont {Teukolsky}, \citenamefont {Vetterling},\ and\ \citenamefont
  {Flannery}}]{NumRecipe}%
  \BibitemOpen
  \bibfield  {author} {\bibinfo {author} {\bibfnamefont {W.~H.}\ \bibnamefont
  {Press}}, \bibinfo {author} {\bibfnamefont {S.~A.}\ \bibnamefont
  {Teukolsky}}, \bibinfo {author} {\bibfnamefont {W.~T.}\ \bibnamefont
  {Vetterling}}, \ and\ \bibinfo {author} {\bibfnamefont {B.~P.}\ \bibnamefont
  {Flannery}},\ }\href@noop {} {\emph {\bibinfo {title} {Numerical Recipes in
  FORTRAN; The Art of Scientific Computing}}},\ \bibinfo {edition} {2nd}\ ed.\
  (\bibinfo  {publisher} {Cambridge University Press},\ \bibinfo {address} {New
  York, NY, USA},\ \bibinfo {year} {1993})\BibitemShut {NoStop}%
\bibitem [{\citenamefont {Worth}\ \emph {et~al.}(2007)\citenamefont {Worth},
  \citenamefont {Beck}, \citenamefont {J\"ackle},\ and\ \citenamefont
  {Meyer}}]{MCTDH}%
  \BibitemOpen
  \bibfield  {author} {\bibinfo {author} {\bibfnamefont {G.~A.}\ \bibnamefont
  {Worth}}, \bibinfo {author} {\bibfnamefont {M.~H.}\ \bibnamefont {Beck}},
  \bibinfo {author} {\bibfnamefont {A.}~\bibnamefont {J\"ackle}}, \ and\
  \bibinfo {author} {\bibfnamefont {H.-D.}\ \bibnamefont {Meyer}},\ }\href
  {http://mctdh.uni-hd.de} {\  (\bibinfo {year} {2007})},\ \bibinfo {note} {the
  MCTDH package, version 8.4, University of Heidelberg, Heidelberg,
  Germany}\BibitemShut {NoStop}%
\bibitem [{\citenamefont {Kosloff}\ and\ \citenamefont
  {Tal-Ezer}(1986)}]{Kosloff86}%
  \BibitemOpen
  \bibfield  {author} {\bibinfo {author} {\bibfnamefont {R.}~\bibnamefont
  {Kosloff}}\ and\ \bibinfo {author} {\bibfnamefont {H.}~\bibnamefont
  {Tal-Ezer}},\ }\href {\doibase
  http://dx.doi.org/10.1016/0009-2614(86)80262-7} {\bibfield  {journal}
  {\bibinfo  {journal} {Chemical Physics Letters}\ }\textbf {\bibinfo {volume}
  {127}},\ \bibinfo {pages} {223 } (\bibinfo {year} {1986})}\BibitemShut
  {NoStop}%
\bibitem [{\citenamefont {Szabo}\ and\ \citenamefont
  {Ostlund}(1996)}]{Szabo96}%
  \BibitemOpen
  \bibfield  {author} {\bibinfo {author} {\bibfnamefont {A.}~\bibnamefont
  {Szabo}}\ and\ \bibinfo {author} {\bibfnamefont {N.}~\bibnamefont
  {Ostlund}},\ }\href@noop {} {\emph {\bibinfo {title} {Modern Quantum
  Chemistry}}}\ (\bibinfo  {publisher} {Dover Publications},\ \bibinfo
  {address} {Inc., 31 East 2nd Street, Mineola, N.Y. 11501},\ \bibinfo {year}
  {1996})\BibitemShut {NoStop}%
\bibitem [{\citenamefont {Penrose}\ and\ \citenamefont
  {Onsager}(1956)}]{Penrose56}%
  \BibitemOpen
  \bibfield  {author} {\bibinfo {author} {\bibfnamefont {O.}~\bibnamefont
  {Penrose}}\ and\ \bibinfo {author} {\bibfnamefont {L.}~\bibnamefont
  {Onsager}},\ }\href {\doibase 10.1103/PhysRev.104.576} {\bibfield  {journal}
  {\bibinfo  {journal} {Phys. Rev.}\ }\textbf {\bibinfo {volume} {104}},\
  \bibinfo {pages} {576} (\bibinfo {year} {1956})}\BibitemShut {NoStop}%
\bibitem [{\citenamefont {Nozi\`eres}\ and\ \citenamefont
  {Saint~James}(1982)}]{Nozieres82}%
  \BibitemOpen
  \bibfield  {author} {\bibinfo {author} {\bibfnamefont {P.}~\bibnamefont
  {Nozi\`eres}}\ and\ \bibinfo {author} {\bibfnamefont {D.}~\bibnamefont
  {Saint~James}},\ }\href {\doibase 10.1051/jphys:019820043070113300}
  {\bibfield  {journal} {\bibinfo  {journal} {J. Phys. France}\ }\textbf
  {\bibinfo {volume} {43}},\ \bibinfo {pages} {1133} (\bibinfo {year}
  {1982})}\BibitemShut {NoStop}%
\bibitem [{\citenamefont {Hilligs\o{}e}\ and\ \citenamefont
  {M\o{}lmer}(2005)}]{Hilligsoe05}%
  \BibitemOpen
  \bibfield  {author} {\bibinfo {author} {\bibfnamefont {K.~M.}\ \bibnamefont
  {Hilligs\o{}e}}\ and\ \bibinfo {author} {\bibfnamefont {K.}~\bibnamefont
  {M\o{}lmer}},\ }\href {\doibase 10.1103/PhysRevA.71.041602} {\bibfield
  {journal} {\bibinfo  {journal} {Phys. Rev. A}\ }\textbf {\bibinfo {volume}
  {71}},\ \bibinfo {pages} {041602} (\bibinfo {year} {2005})}\BibitemShut
  {NoStop}%
\bibitem [{\citenamefont {Bonneau}\ \emph {et~al.}(2013)\citenamefont
  {Bonneau}, \citenamefont {Ruaudel}, \citenamefont {Lopes}, \citenamefont
  {Jaskula}, \citenamefont {Aspect}, \citenamefont {Boiron},\ and\
  \citenamefont {Westbrook}}]{Bonneau13}%
  \BibitemOpen
  \bibfield  {author} {\bibinfo {author} {\bibfnamefont {M.}~\bibnamefont
  {Bonneau}}, \bibinfo {author} {\bibfnamefont {J.}~\bibnamefont {Ruaudel}},
  \bibinfo {author} {\bibfnamefont {R.}~\bibnamefont {Lopes}}, \bibinfo
  {author} {\bibfnamefont {J.-C.}\ \bibnamefont {Jaskula}}, \bibinfo {author}
  {\bibfnamefont {A.}~\bibnamefont {Aspect}}, \bibinfo {author} {\bibfnamefont
  {D.}~\bibnamefont {Boiron}}, \ and\ \bibinfo {author} {\bibfnamefont {C.~I.}\
  \bibnamefont {Westbrook}},\ }\href {\doibase 10.1103/PhysRevA.87.061603}
  {\bibfield  {journal} {\bibinfo  {journal} {Phys. Rev. A}\ }\textbf {\bibinfo
  {volume} {87}},\ \bibinfo {pages} {061603} (\bibinfo {year}
  {2013})}\BibitemShut {NoStop}%
\bibitem [{\citenamefont {Khaykovich}\ \emph {et~al.}(2002)\citenamefont
  {Khaykovich}, \citenamefont {Schreck}, \citenamefont {Ferrari}, \citenamefont
  {Bourdel}, \citenamefont {Cubizolles}, \citenamefont {Carr}, \citenamefont
  {Castin},\ and\ \citenamefont {Salomon}}]{Khaykovich02}%
  \BibitemOpen
  \bibfield  {author} {\bibinfo {author} {\bibfnamefont {L.}~\bibnamefont
  {Khaykovich}}, \bibinfo {author} {\bibfnamefont {F.}~\bibnamefont {Schreck}},
  \bibinfo {author} {\bibfnamefont {G.}~\bibnamefont {Ferrari}}, \bibinfo
  {author} {\bibfnamefont {T.}~\bibnamefont {Bourdel}}, \bibinfo {author}
  {\bibfnamefont {J.}~\bibnamefont {Cubizolles}}, \bibinfo {author}
  {\bibfnamefont {L.~D.}\ \bibnamefont {Carr}}, \bibinfo {author}
  {\bibfnamefont {Y.}~\bibnamefont {Castin}}, \ and\ \bibinfo {author}
  {\bibfnamefont {C.}~\bibnamefont {Salomon}},\ }\href
  {http://search.ebscohost.com/login.aspx?direct=true&db=afh&AN=6878651&site=e%
host-live} {\bibfield  {journal} {\bibinfo  {journal} {Science}\ }\textbf
  {\bibinfo {volume} {296}},\ \bibinfo {pages} {1290 } (\bibinfo {year}
  {2002})}\BibitemShut {NoStop}%
\bibitem [{\citenamefont {Strecker}\ \emph {et~al.}(2002)\citenamefont
  {Strecker}, \citenamefont {Partridge}, \citenamefont {Truscott},\ and\
  \citenamefont {Hulet}}]{Strecker02}%
  \BibitemOpen
  \bibfield  {author} {\bibinfo {author} {\bibfnamefont {K.~E.}\ \bibnamefont
  {Strecker}}, \bibinfo {author} {\bibfnamefont {G.~B.}\ \bibnamefont
  {Partridge}}, \bibinfo {author} {\bibfnamefont {A.~G.}\ \bibnamefont
  {Truscott}}, \ and\ \bibinfo {author} {\bibfnamefont {R.~G.}\ \bibnamefont
  {Hulet}},\ }\href {\doibase 10.1038/nature747} {\bibfield  {journal}
  {\bibinfo  {journal} {Nature}\ }\textbf {\bibinfo {volume} {417}},\ \bibinfo
  {pages} {150} (\bibinfo {year} {2002})}\BibitemShut {NoStop}%
\bibitem [{\citenamefont {Burger}\ \emph {et~al.}(1999)\citenamefont {Burger},
  \citenamefont {Bongs}, \citenamefont {Dettmer}, \citenamefont {Ertmer},
  \citenamefont {Sengstock}, \citenamefont {Sanpera}, \citenamefont
  {Shlyapnikov},\ and\ \citenamefont {Lewenstein}}]{Burger99}%
  \BibitemOpen
  \bibfield  {author} {\bibinfo {author} {\bibfnamefont {S.}~\bibnamefont
  {Burger}}, \bibinfo {author} {\bibfnamefont {K.}~\bibnamefont {Bongs}},
  \bibinfo {author} {\bibfnamefont {S.}~\bibnamefont {Dettmer}}, \bibinfo
  {author} {\bibfnamefont {W.}~\bibnamefont {Ertmer}}, \bibinfo {author}
  {\bibfnamefont {K.}~\bibnamefont {Sengstock}}, \bibinfo {author}
  {\bibfnamefont {A.}~\bibnamefont {Sanpera}}, \bibinfo {author} {\bibfnamefont
  {G.~V.}\ \bibnamefont {Shlyapnikov}}, \ and\ \bibinfo {author} {\bibfnamefont
  {M.}~\bibnamefont {Lewenstein}},\ }\href {\doibase
  10.1103/PhysRevLett.83.5198} {\bibfield  {journal} {\bibinfo  {journal}
  {Phys. Rev. Lett.}\ }\textbf {\bibinfo {volume} {83}},\ \bibinfo {pages}
  {5198} (\bibinfo {year} {1999})}\BibitemShut {NoStop}%
\bibitem [{\citenamefont {Denschlag}\ \emph {et~al.}(2000)\citenamefont
  {Denschlag}, \citenamefont {Simsarian}, \citenamefont {Feder}, \citenamefont
  {Clark}, \citenamefont {Collins}, \citenamefont {Cubizolles}, \citenamefont
  {Deng}, \citenamefont {Hagley}, \citenamefont {Helmerson}, \citenamefont
  {Reinhardt}, \citenamefont {Rolston}, \citenamefont {Schneider},\ and\
  \citenamefont {Phillips}}]{Denschlag97}%
  \BibitemOpen
  \bibfield  {author} {\bibinfo {author} {\bibfnamefont {J.}~\bibnamefont
  {Denschlag}}, \bibinfo {author} {\bibfnamefont {J.~E.}\ \bibnamefont
  {Simsarian}}, \bibinfo {author} {\bibfnamefont {D.~L.}\ \bibnamefont
  {Feder}}, \bibinfo {author} {\bibfnamefont {C.~W.}\ \bibnamefont {Clark}},
  \bibinfo {author} {\bibfnamefont {L.~A.}\ \bibnamefont {Collins}}, \bibinfo
  {author} {\bibfnamefont {J.}~\bibnamefont {Cubizolles}}, \bibinfo {author}
  {\bibfnamefont {L.}~\bibnamefont {Deng}}, \bibinfo {author} {\bibfnamefont
  {E.~W.}\ \bibnamefont {Hagley}}, \bibinfo {author} {\bibfnamefont
  {K.}~\bibnamefont {Helmerson}}, \bibinfo {author} {\bibfnamefont {W.~P.}\
  \bibnamefont {Reinhardt}}, \bibinfo {author} {\bibfnamefont {S.~L.}\
  \bibnamefont {Rolston}}, \bibinfo {author} {\bibfnamefont {B.~I.}\
  \bibnamefont {Schneider}}, \ and\ \bibinfo {author} {\bibfnamefont {W.~D.}\
  \bibnamefont {Phillips}},\ }\href {\doibase 10.1126/science.287.5450.97}
  {\bibfield  {journal} {\bibinfo  {journal} {Science}\ }\textbf {\bibinfo
  {volume} {287}},\ \bibinfo {pages} {97} (\bibinfo {year} {2000})}\BibitemShut
  {NoStop}%
\bibitem [{\citenamefont {Aidelsburger}\ \emph {et~al.}(2013)\citenamefont
  {Aidelsburger}, \citenamefont {Atala}, \citenamefont {Lohse}, \citenamefont
  {Barreiro}, \citenamefont {Paredes},\ and\ \citenamefont
  {Bloch}}]{Aidelsburger13}%
  \BibitemOpen
  \bibfield  {author} {\bibinfo {author} {\bibfnamefont {M.}~\bibnamefont
  {Aidelsburger}}, \bibinfo {author} {\bibfnamefont {M.}~\bibnamefont {Atala}},
  \bibinfo {author} {\bibfnamefont {M.}~\bibnamefont {Lohse}}, \bibinfo
  {author} {\bibfnamefont {J.~T.}\ \bibnamefont {Barreiro}}, \bibinfo {author}
  {\bibfnamefont {B.}~\bibnamefont {Paredes}}, \ and\ \bibinfo {author}
  {\bibfnamefont {I.}~\bibnamefont {Bloch}},\ }\href {\doibase
  10.1103/PhysRevLett.111.185301} {\bibfield  {journal} {\bibinfo  {journal}
  {Phys. Rev. Lett.}\ }\textbf {\bibinfo {volume} {111}},\ \bibinfo {pages}
  {185301} (\bibinfo {year} {2013})}\BibitemShut {NoStop}%
\bibitem [{\citenamefont {Goldman}\ and\ \citenamefont
  {Dalibard}(2014)}]{Goldman14}%
  \BibitemOpen
  \bibfield  {author} {\bibinfo {author} {\bibfnamefont {N.}~\bibnamefont
  {Goldman}}\ and\ \bibinfo {author} {\bibfnamefont {J.}~\bibnamefont
  {Dalibard}},\ }\href {\doibase 10.1103/PhysRevX.4.031027} {\bibfield
  {journal} {\bibinfo  {journal} {Phys. Rev. X}\ }\textbf {\bibinfo {volume}
  {4}},\ \bibinfo {pages} {031027} (\bibinfo {year} {2014})}\BibitemShut
  {NoStop}%
\bibitem [{\citenamefont {Streltsov}\ \emph {et~al.}(2010)\citenamefont
  {Streltsov}, \citenamefont {Alon},\ and\ \citenamefont
  {Cederbaum}}]{Streltsov10}%
  \BibitemOpen
  \bibfield  {author} {\bibinfo {author} {\bibfnamefont {A.~I.}\ \bibnamefont
  {Streltsov}}, \bibinfo {author} {\bibfnamefont {O.~E.}\ \bibnamefont {Alon}},
  \ and\ \bibinfo {author} {\bibfnamefont {L.~S.}\ \bibnamefont {Cederbaum}},\
  }\href {\doibase 10.1103/PhysRevA.81.022124} {\bibfield  {journal} {\bibinfo
  {journal} {Phys. Rev. A}\ }\textbf {\bibinfo {volume} {81}},\ \bibinfo
  {pages} {022124} (\bibinfo {year} {2010})}\BibitemShut {NoStop}%
\bibitem [{\citenamefont {Streltsov}\ \emph {et~al.}(2011)\citenamefont
  {Streltsov}, \citenamefont {Sakmann}, \citenamefont {Alon},\ and\
  \citenamefont {Cederbaum}}]{Streltsov11}%
  \BibitemOpen
  \bibfield  {author} {\bibinfo {author} {\bibfnamefont {A.~I.}\ \bibnamefont
  {Streltsov}}, \bibinfo {author} {\bibfnamefont {K.}~\bibnamefont {Sakmann}},
  \bibinfo {author} {\bibfnamefont {O.~E.}\ \bibnamefont {Alon}}, \ and\
  \bibinfo {author} {\bibfnamefont {L.~S.}\ \bibnamefont {Cederbaum}},\ }\href
  {\doibase 10.1103/PhysRevA.83.043604} {\bibfield  {journal} {\bibinfo
  {journal} {Phys. Rev. A}\ }\textbf {\bibinfo {volume} {83}},\ \bibinfo
  {pages} {043604} (\bibinfo {year} {2011})}\BibitemShut {NoStop}%
\bibitem [{\citenamefont {Fasshauer}\ and\ \citenamefont
  {Lode}(2016)}]{Fasshauer16}%
  \BibitemOpen
  \bibfield  {author} {\bibinfo {author} {\bibfnamefont {E.}~\bibnamefont
  {Fasshauer}}\ and\ \bibinfo {author} {\bibfnamefont {A.~U.~J.}\ \bibnamefont
  {Lode}},\ }\href {\doibase 10.1103/PhysRevA.93.033635} {\bibfield  {journal}
  {\bibinfo  {journal} {Phys. Rev. A}\ }\textbf {\bibinfo {volume} {93}},\
  \bibinfo {pages} {033635} (\bibinfo {year} {2016})}\BibitemShut {NoStop}%
\bibitem [{\citenamefont {Anderson}\ \emph {et~al.}(1999)\citenamefont
  {Anderson}, \citenamefont {Bai}, \citenamefont {Bischof}, \citenamefont
  {Blackford}, \citenamefont {Demmel}, \citenamefont {Dongarra}, \citenamefont
  {Croz}, \citenamefont {Greenbaum}, \citenamefont {Hammarling}, \citenamefont
  {McKenney},\ and\ \citenamefont {Sorensen}}]{lapack}%
  \BibitemOpen
  \bibfield  {author} {\bibinfo {author} {\bibfnamefont {E.}~\bibnamefont
  {Anderson}}, \bibinfo {author} {\bibfnamefont {Z.}~\bibnamefont {Bai}},
  \bibinfo {author} {\bibfnamefont {C.}~\bibnamefont {Bischof}}, \bibinfo
  {author} {\bibfnamefont {S.}~\bibnamefont {Blackford}}, \bibinfo {author}
  {\bibfnamefont {J.}~\bibnamefont {Demmel}}, \bibinfo {author} {\bibfnamefont
  {J.}~\bibnamefont {Dongarra}}, \bibinfo {author} {\bibfnamefont {J.~D.}\
  \bibnamefont {Croz}}, \bibinfo {author} {\bibfnamefont {A.}~\bibnamefont
  {Greenbaum}}, \bibinfo {author} {\bibfnamefont {S.}~\bibnamefont
  {Hammarling}}, \bibinfo {author} {\bibfnamefont {A.}~\bibnamefont
  {McKenney}}, \ and\ \bibinfo {author} {\bibfnamefont {D.}~\bibnamefont
  {Sorensen}},\ }\href@noop {} {\emph {\bibinfo {title} {{LAPACK} Users'
  Guide}}},\ \bibinfo {edition} {3rd}\ ed.\ (\bibinfo  {publisher} {Society for
  Industrial and Applied Mathematics},\ \bibinfo {address} {Philadelphia, PA},\
  \bibinfo {year} {1999})\BibitemShut {NoStop}%
\bibitem [{\citenamefont {Beck}\ and\ \citenamefont {Meyer}(1997)}]{Beck97}%
  \BibitemOpen
  \bibfield  {author} {\bibinfo {author} {\bibfnamefont {M.}~\bibnamefont
  {Beck}}\ and\ \bibinfo {author} {\bibfnamefont {H.-D.}\ \bibnamefont
  {Meyer}},\ }\href {\doibase 10.1007/s004600050342} {\bibfield  {journal}
  {\bibinfo  {journal} {Zeitschrift f{\"u}r Physik D Atoms, Molecules and
  Clusters}\ }\textbf {\bibinfo {volume} {42}},\ \bibinfo {pages} {113}
  (\bibinfo {year} {1997})}\BibitemShut {NoStop}%
\bibitem [{\citenamefont {Heather}\ and\ \citenamefont
  {Light}(1983)}]{Heather83}%
  \BibitemOpen
  \bibfield  {author} {\bibinfo {author} {\bibfnamefont {R.~W.}\ \bibnamefont
  {Heather}}\ and\ \bibinfo {author} {\bibfnamefont {J.~C.}\ \bibnamefont
  {Light}},\ }\href {\doibase http://dx.doi.org/10.1063/1.445574} {\bibfield
  {journal} {\bibinfo  {journal} {The Journal of Chemical Physics}\ }\textbf
  {\bibinfo {volume} {79}},\ \bibinfo {pages} {147} (\bibinfo {year}
  {1983})}\BibitemShut {NoStop}%
\end{thebibliography}%

\onecolumngrid
\newpage


  \section*{TABLES}
 
\begin{table}[h!]
 \fontsize{8}{10}\selectfont
\caption{Ground-state energy [in units of $E_{0}$ see text after Eq. (\ref{H_relax})] of $100$ bosons trapped in a 1D harmonic potential interacting through a contact potential with a strength $\lambda = 0.01$ and $0.1$. The TD-RASSCF-B calculations were performed with a single ${\cal P}_{1}$ orbital, $M_{1}=1$ and $M_{2}=M-1$ ${\cal P}_{2}$ orbitals, with $M$ the total number of orbitals. The results were obtained using the general RAS scheme, which includes both even and odd excitations. The excitation schemes are indicated with the usual notations -S, -SD, $\cdots$, up to $N_{\text{max}}=9$ and the RAS schemes are labeled by the value of $N_{\text{max}}$ for larger excitations, e.g. -10, -20. In addition MCTDHB calculations were carried out to compare the accuracy of the TD-RASSCF-B method and the efficiency. The number of configurations used in the wavefunction expansion are indicated in parentheses. The result obtained with a single orbital is equivalent to the GP method. To highlight the difference between the TD-RASSCF-B and MCTDHB results, the digits that differ are underlined.}
  \begin{tabular}{p{1.8cm} p{0.1cm} p{1.8cm} p{1.8cm} p{1.8cm} p{1.8cm} p{1.8cm} p{1.8cm} p{1.8cm} p{1.8cm} p{0.1cm}} 
  \hline \hline
  \multicolumn{1}{c}{ } & \multicolumn{8}{c}{ Orbitals } &    \\
       \cline{2-11}
    \centering Method &&\centering $1$ &\centering $2$  &\centering $3$ &\centering $4$ &\centering $5$ &\centering $6$  &\centering $7$ &\centering $8$ &\\
    \hline 
    \centering $\lambda=0.01$ & &  &  & & &  & &  & \\  
    \cline{1-2}
    \centering MCTDHB && \centering 68.76816487 & \centering 68.75335446 & \centering  68.74538390  &  \centering  68.74152088  & \centering 68.73891122 & \centering - & \centering - & \centering - & \\
         && \centering (1) & \centering (101) & \centering  (5151)  &  \centering  (176851)  & \centering (4598126) & \centering (96560646) & \centering (1705904746) & \centering (26075972546) & \\
    \centering          -SD && \centering    -              & \centering 68.753\underline{55024} & \centering  68.745\underline{65678} &  \centering  68.741\underline{84781}  & \centering 68.73\underline{926413} & \centering 68.73761231 & \centering 68.736360917& \centering 68.73545355 & \\
          &&  & \centering (3) & \centering  (6)  &  \centering  (10)  & \centering (15) & \centering (21) & \centering (28) & \centering (36) & \\
    \centering     -SDTQ && \centering   -               & \centering 68.75335\underline{660} & \centering  68.74538\underline{672}  &  \centering  68.74152\underline{449}  & \centering 68.73891\underline{508} & \centering 68.73724366 & \centering 68.73598073 & \centering 68.73506372 & \\
          &&  & \centering (5) & \centering  (15)  &  \centering  (35)  & \centering (70) & \centering (126) & \centering (210) & \centering (330) & \\
    \centering     -SDTQ56 && \centering   -               & \centering 68.7533544\underline{8} & \centering  68.7453839\underline{3}  &  \centering  68.741520\underline{92}  & \centering 68.738911\underline{26} & \centering 68.73723959 & \centering 68.73597655 & \centering 68.73505943 & \\
          &&  & \centering (7) & \centering  (28)  &  \centering  (84)  & \centering (210) & \centering (462) & \centering (924) & \centering (1716) & \\
    \centering     -SDTQ5678 && \centering   -               & \centering 68.75335446 & \centering  68.74538390 &  \centering  68.74152088  & \centering 68.73891122 & \centering 68.73723955 & \centering 68.73597651 & \centering 68.73505938 & \\
          &&  & \centering (9) & \centering  (45)  &  \centering  (165)  & \centering (495) & \centering (1287) & \centering (3003) & \centering (6435) & \\
    \centering     -10 && \centering   -               & \centering 68.75335446 & \centering  68.74538390 &  \centering  68.74152088  & \centering 68.73891122 & \centering 68.73723955 & \centering 68.73597651 & \centering 68.73505938 & \\
          &&  & \centering (11) & \centering  (66)  &  \centering  (286)  & \centering (1001) & \centering (3003) & \centering (8008) & \centering (19448) & \\

    \centering $\lambda=0.1$  & &  &  & & &  & &  & \\  
    \cline{1-2}
    \centering MCTDHB && \centering 193.5509587 & \centering 193.0154216 & \centering  192.6308389  &  \centering  192.3920265  & \centering 192.2138048 & \centering - & \centering - & \centering - & \\
             && \centering (1) & \centering (101) & \centering  (5151)  &  \centering  (176851)  & \centering (4598126) & \centering (96560646) & \centering (1705904746) & \centering (26075972546) & \\
    \centering         -SDT && \centering    -              & \centering 193.0\underline{783470} & \centering  192.\underline{7594351} &  \centering  192.\underline{5665491}  & \centering 192.\underline{4187690} & \centering 192.3153377 & \centering 192.2315684& \centering 192.1681159 & \\
          &&  & \centering (4) & \centering  (10)  &  \centering  (20)  & \centering (35) & \centering (56) & \centering (84) & \centering (120) & \\
    \centering     -SDTQ5 && \centering   -               & \centering 193.0\underline{310688} & \centering  192.6\underline{591396}  &  \centering  192.\underline{4322339}  & \centering 192.2\underline{605383} & \centering 192.1396169 & \centering 192.0434375 & \centering 191.9701535 & \\
          &&  & \centering (6) & \centering  (21)  &  \centering  (56)  & \centering (126) & \centering (252) & \centering (462) & \centering (792) & \\
    \centering     -SDTQ567 && \centering   -               & \centering 193.01\underline{91782} & \centering  192.63\underline{70892}  &  \centering  192.\underline{4017271}  & \centering 192.2\underline{250102} & \centering 192.1000115 & \centering 192.0013608 & \centering 191.9259461 & \\
          &&  & \centering (8) & \centering  (36)  &  \centering  (120)  & \centering (330) & \centering (792) & \centering (1716) & \centering (3432) & \\
    \centering     -SDTQ56789 && \centering   -               & \centering 193.01\underline{62747} & \centering  192.63\underline{21926} &  \centering  192.39\underline{43825}  & \centering 192.21\underline{65110} & \centering 192.0903987 & \centering 191.9911974 & \centering 191.9152346 & \\
          &&  & \centering (10) & \centering  (55)  &  \centering  (220)  & \centering (715) & \centering (2002) & \centering (5005) & \centering (11440) & \\
    \centering     -10 && \centering   -               & \centering 193.015\underline{8613} & \centering  192.63\underline{15747} &  \centering  192.39\underline{33448}  & \centering 192.21\underline{53160} & \centering 192.0890211 & \centering 191.9894655 & \centering 191.9133983 & \\
          &&  & \centering (11) & \centering  (66)  &  \centering  (286)  & \centering (1001) & \centering (3003) & \centering (8008) & \centering (19448) & \\     
    \centering     -15 && \centering   -               & \centering 193.01542\underline{91} & \centering  192.6308\underline{506} &  \centering  192.3920\underline{555}  & \centering 192.2138\underline{378} & \centering 192.0872814 & \centering 191.9879161 & \centering 191.9117429 & \\
          &&  & \centering (16) & \centering  (136)  &  \centering  (816)  & \centering (3876) & \centering (15504) & \centering (54264) & \centering (170544) & \\     
    \centering     -20 && \centering   -               & \centering 193.01542\underline{20} & \centering  192.63083\underline{92} &  \centering  192.39202\underline{71}  & \centering 192.21380\underline{55} & \centering 192.0872403 & \centering 191.9878729 & \centering - & \\
          &&  & \centering (21) & \centering  (231)  &  \centering  (1771)  & \centering (10626) & \centering (53130) & \centering (230230) & \centering (888030) & \\     
    \centering     -23 && \centering   -               & \centering 193.0154216 & \centering  192.6308389 &  \centering  192.3920265  & \centering 192.213804\underline{9} & \centering 192.0872393 & \centering 191.9878720 & \centering - & \\
          &&  & \centering (24) & \centering  (300)  &  \centering  (2600)  & \centering (17550) & \centering (98280) & \centering (475020) & \centering (2035800) & \\     
    \centering     -25 && \centering   -               & \centering 193.0154216 & \centering  192.6308389 &  \centering  192.3920265  & \centering 192.2138048 & \centering 192.0872393 & \centering 191.9878719 & \centering - & \\
          &&  & \centering (26) & \centering  (351)  &  \centering  (3276)  & \centering (23751) & \centering (142506) & \centering (736281) & \centering (3365856) & \\     
    \hline \hline
 \end{tabular}

 \end{table}
\newpage
 
\begin{table}[h!]
 \fontsize{8}{10}\selectfont
\caption{Same as Table I but for ground-state energies obtained using RAS schemes with \textit{only} even excitations. The excitation schemes are indicated with the notations -D, -DQ, $\cdots$, up to $N_{\text{max}}=8$ and the RAS schemes are labeled by the value of $N_{\text{max}}$ for larger excitations, e.g. -10, -20. }
  \begin{tabular}{p{1.8cm} p{0.1cm} p{1.8cm} p{1.8cm} p{1.8cm} p{1.8cm} p{1.8cm} p{1.8cm} p{1.8cm} p{1.8cm} p{0.1cm}} 
  \hline \hline
  \multicolumn{1}{c}{ } & \multicolumn{8}{c}{ Orbitals } &    \\
       \cline{2-11}
    \centering Method &&\centering $1$ &\centering $2$  &\centering $3$ &\centering $4$ &\centering $5$ &\centering $6$  &\centering $7$ &\centering $8$ &\\
    \hline 
    \centering $\lambda=0.01$ & &  &  & & &  & &  & \\  
    \cline{1-2}
    \centering MCTDHB && \centering 68.76816487 & \centering 68.75335446 & \centering  68.74538390  &  \centering  68.74152088  & \centering 68.73891122 & \centering - & \centering - & \centering - & \\
         && \centering (1) & \centering (101) & \centering  (5151)  &  \centering  (176851)  & \centering (4598126) & \centering (96560646) & \centering (1705904746) & \centering (26075972546) & \\
    \centering          -D && \centering    -              & \centering 68.753\underline{55024} & \centering  68.745\underline{65780} &  \centering  68.741\underline{84905}  & \centering 68.73\underline{92653} & \centering 68.73761353 & \centering 68.73636218 & \centering 68.73545481 & \\
          &&  & \centering (2) & \centering  (4)  &  \centering  (7)  & \centering (11) & \centering (16) & \centering (22) & \centering (29) & \\
    \centering     -DQ && \centering   -               & \centering 68.75335\underline{660} & \centering  68.74538\underline{974}  &  \centering  68.7415\underline{3257}  & \centering 68.7389\underline{2541} & \centering 68.73725598 & \centering 68.73599408 & \centering 68.73507803 & \\
          &&  & \centering (3) & \centering  (9)  &  \centering  (22)  & \centering (46) & \centering (86) & \centering (148) & \centering (239) & \\
    \centering     -DQ6 && \centering   -               & \centering 68.7533544\underline{8} & \centering  68.74538\underline{700}  &  \centering  68.74152\underline{917}  & \centering 68.7389\underline{2180} & \centering 68.73725217 & \centering 68.73599018 & \centering 68.73507404 & \\
          &&  & \centering (4) & \centering  (16)  &  \centering  (50)  & \centering (130) & \centering (296) & \centering (610) & \centering (1163) & \\
    \centering     -DQ68 && \centering   -               & \centering 68.75335446 & \centering  68.74538\underline{697} &  \centering  68.74152\underline{914}  & \centering 68.7389\underline{2177} & \centering 68.73725213 & \centering 68.73599014 & \centering 68.73507400 & \\
          &&  & \centering (5) & \centering  (25)  &  \centering  (95)  & \centering (295) & \centering (791) & \centering (1897) & \centering (4166) & \\
    \centering     -10 && \centering   -               & \centering 68.75335446 & \centering  68.74538\underline{697} &  \centering  68.74152\underline{914}  & \centering 68.7389\underline{2177} & \centering 68.73725213 & \centering 68.73599014 & \centering 68.73507400 & \\
          &&  & \centering (6) & \centering  (36)  &  \centering  (161)  & \centering (581) & \centering (1792) & \centering (4900) & \centering (12174) & \\

    \centering $\lambda=0.1$  & &  &  & & &  & &  & \\  
    \cline{1-2}
    \centering MCTDHB && \centering 193.5509587 & \centering 193.0154216 & \centering  192.6308389  &  \centering  192.3920265  & \centering 192.2138048 & \centering - & \centering - & \centering - & \\
             && \centering (1) & \centering (101) & \centering  (5151)  &  \centering  (176851)  & \centering (4598126) & \centering (96560646) & \centering (1705904746) & \centering (26075972546) & \\
    \centering         -D && \centering    -              & \centering 193.\underline{1320396} & \centering  192.\underline{8258797} &  \centering  192.\underline{6453389}  & \centering 192.\underline{5031705} & \centering 192.4050140 & \centering 192.3243169 & \centering 192.2636708 & \\
          &&  & \centering (2) & \centering  (4)  &  \centering  (7)  & \centering (11) & \centering (16) & \centering (22) & \centering (29) & \\
    \centering     -DQ && \centering   -               & \centering 193.0\underline{449780} & \centering  192.6\underline{788633}  &  \centering  192.\underline{4619220}  & \centering 192.2\underline{951569} & \centering 192.1793448 & \centering 192.0861393 & \centering 192.0157234 & \\
          &&  & \centering (3) & \centering  (9)  &  \centering  (22)  & \centering (46) & \centering (86) & \centering (148) & \centering (239) & \\
    \centering     -DQ6 && \centering   -               & \centering 193.0\underline{227865} & \centering  192.6\underline{439550}  &  \centering  192.\underline{4175380}  & \centering 192.2\underline{457797} & \centering 192.1259116 & \centering 192.0303689 & \centering 191.9579281 & \\
          &&  & \centering (4) & \centering  (16)  &  \centering  (50)  & \centering (130) & \centering (296) & \centering (610) & \centering (1163) & \\
    \centering     -DQ68 && \centering   -               & \centering 193.01\underline{71558} & \centering  192.63\underline{56968} &  \centering  192.\underline{4065816}  & \centering 192.2\underline{338183} & \centering 192.1129247 & \centering 192.0169500 & \centering 191.9440484 & \\
          &&  & \centering (5) & \centering  (25)  &  \centering  (95)  & \centering (295) & \centering (791) & \centering (1897) & \centering (4166) & \\
          &&  & \centering (6) & \centering  (36)  &  \centering  (161)  & \centering (581) & \centering (1792) & \centering (4900) & \centering (12174) & \\     
    \centering     -20 && \centering   -               & \centering 193.0154216 & \centering  192.63\underline{33324} &  \centering  192.\underline{4031759}  & \centering 192.2\underline{301918} & \centering 192.1089342 & \centering 192.0128774 & \centering 191.9398292 & \\
          &&  & \centering (11) & \centering  (121)  &  \centering  (946)  & \centering (5786) & \centering (29458) & \centering (129844) & \centering (508937) & \\     
    \centering     -30 && \centering   -               & \centering 193.0154216 & \centering  192.63\underline{33323} &  \centering  192.\underline{4031756}  & \centering 192.2\underline{301915} & \centering 192.1089338 & \centering 192.0128771 & \centering - & \\
          &&  & \centering (16) & \centering  (256)  &  \centering  (2856)  & \centering (24616) & \centering (174624) & \centering (1061208) & \centering (5678340) & \\     
    \hline \hline
 \end{tabular}

 \end{table}

  \newpage
  
  \section*{FIGURES}

\end{document}